\documentclass[pdflatex,sn-mathphys-num]{sn-jnl}

\usepackage[T1]{fontenc}
\usepackage[utf8]{inputenc}
\usepackage{lmodern}
\usepackage[english]{babel}
\usepackage[autostyle]{csquotes}
\usepackage{amsmath,amssymb}
\usepackage{graphicx}
\usepackage{tikz}
\usetikzlibrary{positioning,arrows.meta,calc}
\usepackage{booktabs}
\usepackage{array}
\usepackage{float}
\makeatletter
\let\table\tableorg
\let\endtable\endtableorg
\let\sidewaystable\sidewaystableorg
\let\endsidewaystable\endsidewaystableorg
\makeatother
\usepackage{tabularx}
\usepackage{longtable}
\usepackage{pdflscape}

\newcommand{\abbildung}[1]{%
  \IfFileExists{#1.pdf}{\includegraphics[width=\linewidth]{#1.pdf}}%
  {\fbox{\parbox[c][4cm][c]{0.9\linewidth}{\centering\ttfamily\detokenize{#1} missing}}}}

\newsavebox\breitebox
\newcommand{\maxbreite}[1]{\sbox\breitebox{#1}\ifdim\wd\breitebox>\linewidth
\resizebox{\linewidth}{!}{\usebox\breitebox}\else\usebox\breitebox\fi}

\newenvironment{fragen}{\begin{list}{}{\settowidth{\labelwidth}{(1a)}%
  \setlength{\leftmargin}{\dimexpr\labelwidth+\labelsep\relax}%
  \setlength{\itemsep}{1pt}\setlength{\topsep}{3pt}}}{\end{list}}

\makeatletter
\newcommand{\FloatBarrier}{\par\begingroup\let\@elt\relax
  \edef\@tempa{\@botlist\@deferlist\@dbldeferlist}%
  \ifx\@tempa\@empty\else\newpage\fi\endgroup}
\makeatother

\newcommand{\pf}{\pi_F}
\newcommand{\pfnull}{\pi_{F,0}}

\newcommand{\zDS}{30}
\newcommand{\zDupDS}{4}
\newcommand{\zDupMax}{708}
\newcommand{\zICLauc}{0.958}
\newcommand{\zICLasrEins}{53}

\newcommand{\zPFNauc}{0.962}
\newcommand{\zPFNasrEins}{65}

\newcommand{\zCATasrEins}{52}

\newcommand{\zZiele}{1,480}

\newcommand{\zICLangegriffen}{1,307}
\newcommand{\zPFNangegriffen}{1,316}
\newcommand{\zTFMaucMax}{0.010}
\newcommand{\zPFNidentMax}{$8.4 \cdot 10^{-5}$}

\newcommand{\zMnullUeberZeilen}{9}

\newcommand{\zMstartGekippt}{0.82}
\newcommand{\zMstartGehalten}{0.75}

\newcommand{\zMschubGekippt}{0.23}
\newcommand{\zMschubGehalten}{0.15}

\newcommand{\zMauc}{0.95}

\newcommand{\zMcosZgekippt}{0.78}

\newcommand{\zMcosZgekipptFuenf}{0.51}

\newcommand{\zMcosTmin}{0.26}
\newcommand{\zMcosFmin}{0.27}

\newcommand{\zHolmAPEins}{< 0.001}

\newcommand{\zHolmAPZwei}{= 0.022}

\newcommand{\zHolmArmMax}{< 0.001}

\newcommand{\zRFasrEins}{98.7}

\newcommand{\zLRasrEins}{1.8}

\newcommand{\zLRauc}{0.885}

\newcommand{\zDupSauber}{2}

\newcommand{\zOhneCreditMax}{0.005}

\newcommand{\zFeatMin}{4}
\newcommand{\zFeatMax}{100}
\newcommand{\zRowsMin}{540}
\newcommand{\zRowsMax}{566,602}
\newcommand{\zMinorityMinPct}{0.2}
\newcommand{\zCtxMin}{432}
\newcommand{\zGiftAnteilMax}{1.2}
\newcommand{\zCreditZiele}{30}
\newcommand{\zDupZiele}{17}
\newcommand{\zVorhersagen}{16,280}

\newcommand{\zICLhaltenFuenf}{14}
\newcommand{\zICLhaltenDupFuenf}{11}
\newcommand{\zPFNhaltenFuenf}{12}
\newcommand{\zPFNhaltenDupFuenf}{10}
\newcommand{\zHaltenPiTOhneMax}{0.63}
\newcommand{\zHaltenPiTVielMin}{0.99}
\newcommand{\zDupGekipptFuenfMax}{3}

\newcommand{\zDiffPFNICLDrei}{3}

\newcommand{\zDiffPFNICL}{12}
\newcommand{\zRhoTFMEins}{0.80}
\newcommand{\zDaucMittelMax}{0.001}
\newcommand{\zICLkorrigiertEins}{79}
\newcommand{\zPFNkorrigiertEins}{84}

\newcommand{\zMcosMuAlleMin}{0.998}

\newcommand{\zPFNohneFalsch}{7}
\newcommand{\zPFNgiftEins}{31}
\newcommand{\zPFNechtEins}{45}

\newcommand{\zPFNwahrEins}{23}

\newcommand{\zPFNgiftDrei}{72}
\newcommand{\zPFNechtDrei}{24}

\newcommand{\zPFNkontrFalschDrei}{1}
\newcommand{\zPFNwahrDrei}{4}

\newcommand{\zPFNaFnullGekipptPct}{11}
\newcommand{\zPFNaFnullGehaltenPct}{3}

\newcommand{\zPFNstaerksteEins}{83}

\newcommand{\zRhoCATICLEins}{0.10}
\newcommand{\zRhoCATPFNEins}{0.21}
\newcommand{\zRhoCATICLZwei}{0.43}
\newcommand{\zRhoCATPFNZwei}{0.33}
\newcommand{\zRhoCATICLDrei}{0.61}
\newcommand{\zRhoCATPFNDrei}{0.56}
\newcommand{\zRFhaltenEins}{16}
\newcommand{\zLRdeltaPiEins}{0.003}
\newcommand{\zPFNaPueberZielEins}{11}
\newcommand{\zPFNaPueberZielDrei}{87}
\newcommand{\zHolmGesamtAPZwei}{0.066}

\newcommand{\zKoppICLohneMittel}{1.1}
\newcommand{\zKoppICLgekipptMittel}{20.7}
\newcommand{\zKoppICLgehaltenMittel}{6.6}
\newcommand{\zKoppICLkontrMittel}{0.4}

\title[Targeted in-context poisoning of tabular foundation models]{Targeted In-Context Poisoning of Tabular Foundation Models: An Empirical Study}

\author*[1]{\fnm{Adem} \sur{Alparslan}}\email{adem.alparslan@fom.de}

\affil*[1]{\orgname{FOM University of Applied Sciences for Economics and Management}, \orgaddress{\city{Essen}, \country{Germany}}}

\abstract{Tabular foundation models (TFMs) predict the label of a query row conditioned on a context of labelled rows, without training on them. Drawn from operational data, the context becomes an attack surface. We study targeted in-context poisoning in its simplest form: an attacker with write access to the context's source places exact copies of a target row there with the wrong label. On \zDS\ OpenML datasets for binary classification, one such poison row flips on average \zICLasrEins\,\% (TabICL) and \zPFNasrEins\,\% (TabPFN) of the predictions correct without attack, and three poison rows flip almost all. The same copy with the true label flips almost none, and performance on the remaining rows hardly changes. Predictions flip by two routes: in TabICL, the target row's embedding rotates towards the mean of the wrong class; in TabPFN, the poison row pulls genuine rows of the wrong class along. The TFMs thus infer soundly, but from forged evidence. Predictions of trained models flip as well: almost always for a random forest, for CatBoost as often as for the TFMs, hardly for a logistic regression. New with TFMs is not the kind of vulnerability but the timing: the data take effect at the moment of prediction, not through training, and a poison row in a live context can be deleted afterwards. Three established requirements must therefore cover the context: access control for its source, traceability per prediction, which can also deter, and validation at prediction time. A detection mechanism shows how such validation can work.}


\begin{document}
\maketitle

\section{Introduction}
\label{sec:einleitung}

Tabular foundation models (TFMs) carry the in-context learning (ICL) of
language models over to tables~\citep{brown2020, muller2022,
hollmann2025, qu2025, vanbreugel2024}. They predict the label of a new
row, the query row, from the labelled rows in their context. This happens
in a single forward pass and without training on the task. Instead, they
are trained in advance, mostly on a large number of synthetic tables.
These are drawn from a distribution over datasets, the prior, which is
defined by random structural causal models~\citep{hollmann2025, qu2026,
grinsztajn2026, jaeger2026}. Read in Bayesian terms, TFMs thereby
approximate the posterior predictive distribution that follows from this
prior and the context~\citep{muller2022, hollmann2025}. For small and
medium-sized datasets, TFMs are reported to match or even exceed the
performance of tree-based models and deep learning
models~\citep{qu2026, grinsztajn2026, kurashkin2026}.

In operation, the context can come directly from operational
data~\citep{sap2026}. Data platforms, for example, allow TFMs to be run
directly on the table in which the data
reside~\citep{kongdas2026, googlecloud2026}. Whoever can write to these
data changes the input of a TFM and thus the basis of its next
prediction, without waiting for a training run. This makes the context a
potential attack surface.

We examine an attack on this surface in its simplest targeted form and
call it targeted in-context poisoning. The attacker wants
a particular query row, the target row, to receive the wrong label, and
writes one to five exact copies of the target row with the wrong label
in place of as many context rows. Such a copy is called a poison row,
the remaining context rows are called genuine context rows. A prediction
flips if the model predicts the target row correctly without attack and
wrongly under attack. No context row can be closer to the target row in
its features than its exact copy. If a model relies on similar rows,
this should show most clearly for the exact copy. This is why we choose
this form. A TFM, however, compares rows in its representation, which it
forms from the features and the labels of the context, and there the
poison row need not be close to the target row. Whether and how often
the attack works is therefore an empirical question.

The exact copy is at the same time the most conspicuous form of the
attack, and this is intended. As a duplicate of the target row, it can
be found if the context is checked for duplicates of the query row
before every prediction. The study thus does not look for an attack
method that outwits a check of the context. Instead, it aims to raise
awareness that the context of a TFM can be an attack surface through
which its predictions can be steered. Nor does it pass judgement on the
performance of the TFMs examined.

For trained models, the manipulation of the training data, known as data
poisoning, is well studied~\citep{biggio2012, xiao2012, shafahi2018,
goldblum2023, cina2023, vassilev2025}. For TFMs, in contrast, the study
of their vulnerability is still at an early stage. \citet{djilani2026}
study evasion attacks: they perturb the query row slightly and show that
even small, structured perturbations lower the accuracy of TFMs
considerably, while the context remains unchanged. \citet{bilos2026}
corrupt the context as a whole, for example by flipping the labels of
its most central rows, and measure the loss in mean accuracy. Their
attacks target the mechanism by which the models predict, not individual
predictions, and lower the performance of the model overall.

Two studies come close to the attack examined here. \citet{anwar2025}
manipulate individual context rows of a transformer that learns in
context: for single-layer linear transformers that learn linear
regression in context, they prove that a single altered context row can
force an arbitrary prediction. \citet{wahdany2026} place copies of a
query row with the wrong label in the context of a TFM, but use them for
membership inference: how many copies are needed to flip the prediction
reveals whether the row is already in the context.

As a poisoning attack that redirects the prediction for a chosen query
row, the manipulation of the context of a TFM has, to our knowledge, not
been studied so far. It is analogous to the
poisoning of demonstrations for language
models~\citep{he2025} and of retrieval corpora~\citep{zou2025}. As in
indirect prompt injection~\citep{greshake2023}, the attacker places the
input in data that the model reads only at prediction time. In the
taxonomy of the National Institute of Standards and Technology (NIST),
the attack is a targeted poisoning attack: it is not meant to lower the
performance of the model overall, but to redirect the prediction for
individual target rows chosen by the attacker~\citep{vassilev2025}. It
requires control over the context, which here takes the place of the
training data. The attacker can thus alter rows of the context and, in
particular, flip their labels (label flipping). For trained models, this
form of attack is considered effective~\citep{vassilev2025}. Whether
label flipping is equally effective against a TFM is what this study
examines.

We run the attack on two current TFMs of different architecture,
TabICLv2~\citep{qu2026} and TabPFN-3.5~\citep{jaeger2026}, referred to
below as TabICL and TabPFN. The basis is \zDS\ public datasets for
binary classification on which both TFMs classify well. Besides the
attack arm, in which the context contains the poison rows, there is a
control arm: there the same copies carry the true label and are called
control rows. Comparing the two arms is meant to separate the effect of
the wrong label from the effect that a copy of the target row in the
context has by itself.

The findings on data poisoning concern models that are trained
on the poisoned data. Whether flipping is a peculiarity of TFMs is shown
by three such models, trained on the same poisoned context: a random
forest, CatBoost~\citep{prokhorenkova2018} and a logistic regression.
The questions of this study are nevertheless addressed to the TFMs
alone, because the datasets are selected by their performance.

Three research questions guide the study. The first asks about the
effect of the wrong label and, to this end, compares the attack arm with
the control arm. It is posed for each of the two TFMs:
\begin{fragen}
\item[(1a)] Do one to five poison rows flip more predictions in TabICL
that are correct without attack than as many control rows?
\item[(1b)] Do they also flip more such predictions in TabPFN than as
many control rows?
\end{fragen}

The first question settles whether and how often predictions flip, but
not what the poison row changes inside a TFM. For this, we look in each
model at the place where it forms its decision.

In TabICL, an MLP forms the decision from the embedding of the target
row. It cannot be read off there directly. \citet{bilos2026} report a
readout rule that largely reproduces the predictions of TabICL on
classification tasks, and \citet{kurashkin2026} adopt it in their survey:
for each class the centroid is formed as the mean of the embeddings of
its context rows, and the target row receives the class of the nearest
centroid. This rule is
reported for contexts without attack. If a poison row only shifted the
centroid, it should have little effect, because it enters the centroid
of the wrong class as only one of many rows. If the rule also holds
under attack, a flipped target row lies closer to the centroid of the
wrong class and a held one closer to the centroid of the true class. The
second question therefore has two parts:
\begin{fragen}
\item[(2a)] In TabICL, does the embedding of a flipped target row lie
closer to the centroid of the wrong class under attack?
\item[(2b)] Does the embedding of a held target row lie closer to the
centroid of the true class under attack?
\end{fragen}

The decision of TabPFN is a weighted vote over the labels of the context
rows: the model forms its prediction in a decoder through attention over
these labels, designed as a soft nearest-neighbour retrieval over the
context~\citep{grinsztajn2026}. Unlike in TabICL, the decision can be
read off directly here, from the weights that the target row places on
the context rows. The prediction flips exactly when the rows with the
wrong label together receive more than half of the weight. If the attack
flips a prediction, it is plausible that the poison rows, as exact copies
of the target row, alone receive more than half.
The third question reads:
\begin{fragen}
\item[(3)] In TabPFN, do the poison rows receive more than half of the
weight for flipped target rows?
\end{fragen}

With the answers to these questions and the conclusions drawn from them,
the study makes three contributions. First, it shows that the context of
a TFM is an attack surface and can be manipulated in a targeted way.
Second, it describes what happens inside two TFMs of different
architecture when their context is poisoned. Third, it argues that three established requirements for secure
operation must cover the context of a TFM: access control, traceability
and validation.

The study is structured as follows: Section~\ref{sec:setup} describes
the experimental setup, and Section~\ref{sec:ergebnisse} reports first
the effect of the attack, also on trained models, and then what happens
inside the two TFMs. Section~\ref{sec:diskussion} answers the research questions,
interprets flipping as a sound inference from forged evidence, derives
implications for operation and names limitations, and
Section~\ref{sec:fazit} concludes.

\section{Experimental setup}
\label{sec:setup}

\subsection{Threat model, attack arm and control arm}
\label{sec:aufbau}

We examine the attack in binary classification. The goal of an attacker
is that a particular query row, the target row, receives the label the
attacker wants, while all other predictions remain unchanged, so that
the attack does not show in the overall performance of the model. In the experiment, this is always the label opposite to the true one,
which we call the wrong label. The feature values of the target row are known
to the attacker before the prediction, for example because the attacker
submits the query row. Knowledge of the model or its weights is not
required, and the number of poison rows is chosen without observing
predictions of the model.

For the attack to work, the poison rows must reach the context. We
therefore assume that the attacker can write and overwrite entries of
the source from which the context is formed. We further assume that
these entries reach the context of the target row before its prediction, for example
because the context comprises the whole source. Access to the model
itself is not required.

To flip the prediction of the target row, the attacker replaces $b$ of
the $n$ context rows with poison rows. A poison row is an exact copy of
the target row: it agrees with it in every feature value but carries the
wrong label. The number $b$ of replaced rows is the budget, with
$b \in \{1, 2, 3, 4, 5\}$. The remaining $n - b$ context rows stay
unchanged. Replacing the rows instead of adding them is a design
decision. It fixes the context size $n$ and thus the number of rows the
TFM reads. For both TFMs it also fixes the scaling of the queries in the
attention, which depends on the context length~\citep{qu2026,
grinsztajn2026}. Replacing uses the fact that the attacker can overwrite
existing entries.

The rows to be replaced are drawn at random from the whole context, so
the attack does not systematically remove genuine rows near the target
row. All models see the same poisoned context for the same target row. The budgets are nested: the rows replaced at budget
$b > 1$ include those replaced at $b - 1$, so that two budgets differ
only in the poison rows added.

Every target row is predicted under three conditions: once without
attack and, for every budget, once in the attack arm and once in the
control arm, eleven times in total. In the attack arm, the context
contains the poison rows. In the control arm, control rows replace the
same context rows as in the attack arm. A control row is the same copy
of the target row with the true label. The two arms thus differ only in
the label of the copies. A difference between them is therefore due to
the label, not to the copy itself.

\subsection{Data}
\label{sec:daten}

The \zDS\ datasets for binary classification come from OpenML
\citep{vanschoren2014} (Appendix~\ref{app:daten}). All have exclusively
numerical features. All models thus receive the same input, and no
model-specific handling of categorical features confounds the results.

On every dataset, both TFMs reach an AUC of at least 0.8 without attack,
averaged over the five test folds (see below). The study thus excludes
tasks on which a TFM classifies poorly. The threshold
of 0.8 is a pragmatic choice. Both criteria, exclusively numerical features and an
AUC of both TFMs of at least 0.8, are tailored to the TFMs. The datasets
do not come from a fixed benchmark suite. We collected 32 datasets for binary
classification with exclusively numerical features from OpenML. Two of them fail the AUC threshold
(Appendix~\ref{app:daten}). The selection was based solely on the data
and the performance without attack, not on the effect of the attack. The
remaining datasets range from \zRowsMin\ to \zRowsMax\ rows, from
\zFeatMin\ to \zFeatMax\ features, and from a balanced class
distribution to a minority class of \zMinorityMinPct\,\%.

Every dataset is reduced to a stratified random sample of at most 3,000
rows. This lies in the range of small and medium-sized tables in which
TFMs are considered particularly strong \citep{hollmann2025,
erickson2025}. The features are passed to the models unchanged. Only for
the logistic regression are they imputed and standardised
(Appendix~\ref{app:referenz}).

Every sample is split into five stratified folds. Each fold serves once
as test rows, the other four as context. The context thus contains at
most 2,400 rows and fewer for small datasets, \zCtxMin\ in the smallest.
Even five poison rows therefore make up at most \zGiftAnteilMax\,\% of
the context. From the test rows of every fold, up to five target rows
per class are drawn at random, so that both classes are equally
represented. The exception is \texttt{creditcard}: each of its test
folds contains only one row of the minority class, so that this dataset
has \zCreditZiele\ target rows, 25 of the majority class and 5 of the
minority class. This gives \zZiele\ target rows, 50 for each of the
other datasets. Even without attack, the context contains exact
duplicates with the true label in \zDupDS\ of the \zDS\ datasets, for
\zDupZiele\ target rows in total, and \zDupMax\ for one target row. They
remain in the context. Among the target rows that hold even at five
poison rows, most have such a duplicate (Section~\ref{sec:budget}).

Without attack, all target rows of a fold share the same context. In the
attack arm and in the control arm, every target row receives its own
context, which contains only its own copies.

\subsection{Quantifying the effect of the attack}
\label{sec:wirkung}

Table~\ref{tab:masse} summarises the quantities of this study. For every
target row, $\pfnull$ and $\pf$ denote the predicted probability of the
wrong label without and under attack. We evaluate only target rows for which the model predicts the true label
without attack, as a rule $\pfnull < 0.5$. For the others, a wrong
prediction could not be attributed to the attack. The evaluated rows are
called the attacked target rows of the model: \zICLangegriffen\ of the \zZiele\ target rows for TabICL and
\zPFNangegriffen\ for TabPFN (Appendix~\ref{app:guete}). Without attack,
the two TFMs reach a mean AUC over the datasets of \zICLauc\ (TabICL)
and \zPFNauc\ (TabPFN). An attacked target row is called flipped if the
model predicts the wrong label for it under attack, as a rule $\pf > 0.5$, above the decision threshold, and held
otherwise.

Every attack is directed at a single target row and succeeds if this row
flips.
To measure how often it succeeds, the attack is carried out for
every target row separately. The attack success rate (ASR)
is the share of successes among these individual attacks
\citep{shafahi2018}, per budget, dataset and model. The target rows are
drawn balanced by class (Section~\ref{sec:daten}), but the attacked ones
no longer are, because the models predict rows of the minority class
correctly less often (Appendix~\ref{app:guete}). The ASR of a dataset
therefore weights the majority class somewhat more strongly.

Every model has its own attacked target rows. The two TFMs are therefore
compared per dataset, not per target row
(Table~\ref{tab:app_asr_ds_b}).

Beyond the target row, every model in every arm also predicts the
remaining test rows of its fold with the context of this target row. We
record the AUC on them to check whether the attack stays targeted, that
is, does not change the performance on the remaining rows. It is
computed for the context of every attacked target row, averaged per
dataset and then over the datasets.

\begin{table}[!t]
\centering\scriptsize
\caption{Notation. The index $0$ marks the run without attack. $T$ and $F$ mark the true and the wrong label of the target row. The predicted label is the one with the larger probability, class 1 at exactly 0.5. We call 0.5 the decision threshold.}
\label{tab:masse}
\begin{tabularx}{\textwidth}{@{}>{\raggedright\arraybackslash}p{2.6cm} >{\raggedright\arraybackslash}X@{}}
\toprule
Quantity & Meaning   \\
\midrule
\multicolumn{2}{@{}l}{All models: output} \\
$\pfnull$, $\pf$ & predicted probability $\pfnull$ of the wrong label without attack and predicted probability $\pf$ under attack \\
$\pi_T$ & predicted probability of the true label, $\pi_T = 1 - \pf$ \\
ASR & attack success rate: share of the attacked target rows that flip under attack, per budget, dataset and model. In the text in per cent, in the tables as a share \\
AUC & performance on the test rows of the fold without attack, and on the remaining test rows under attack and in the control arm \\
\midrule
\multicolumn{2}{@{}l}{TabICL: relative closeness of the target row to the class centroids} \\
$z$, $z_0$ & embedding $z$ of the target row in the respective arm and embedding $z_0$ without attack, each taken after the last ICL block and the final layer normalisation, the input of the MLP \\
$\mu_T$, $\mu_F$ & class centroids: mean $\mu_T$ of the embeddings of the context rows with the true label of the target row and mean $\mu_F$ of those with the wrong label, formed in the respective arm. $\mu_{T,0}$ and $\mu_{F,0}$ are the same centroids without attack \\
$m$ & relative closeness of the target row to the two class centroids, an approximation of the decision: $m = \cos(z, \mu_F)/\cos(z, \mu_T)$, formed per member and averaged over the members. At $m > 1$ the target row is closer in angle to the centroid of the wrong class \\
$m_0$ & $m$ without attack \\

$\cos(z, z_0)$ & cosine between the embedding of the target row under and without attack, 1 without movement \\
$\cos(\mu_T, \mu_{T,0})$, $\cos(\mu_F, \mu_{F,0})$ & cosine $\cos(\mu_T, \mu_{T,0})$ between the centroid of the true class under and without attack and cosine $\cos(\mu_F, \mu_{F,0})$ for the centroid of the wrong class, 1 without movement \\
$\nu_F$ & number of rows with the wrong label among the 25 nearest genuine context rows of the target row at the input of the MLP, by cosine, excluding the poison rows or control rows, averaged over the members. At $\nu_F > 12.5$ the wrong class holds the majority \\
\midrule
\multicolumn{2}{@{}l}{TabPFN: weights in the decoder} \\
$a_j$ & weight of context row $j$, $a_j = \tfrac{1}{H}\sum_{h} \operatorname{softmax}_j\bigl(q^{(h)\top} k_j^{(h)}/\sqrt{d}\bigr)$, $\sum_j a_j = 1$, with the query $q^{(h)}$ of the target row, scaled depending on the context length, and the key $k_j^{(h)}$ of context row $j$ in head $h$, $H = 6$ heads and dimension $d = 64$, then averaged over the members \\
$a_T$, $a_F$ & weight $a_T$ of the true label and weight $a_F$ of the wrong label: each the sum of the weights of all context rows with that label. $a_F$ is the decision itself, $\pf = a_F$ \\
$a_{F,G}$, $a_{F,P}$ & under attack the two parts of $a_F$: weight $a_{F,G}$ of the genuine rows of the wrong class and weight $a_{F,P}$ of the poison rows, $a_F = a_{F,G} + a_{F,P}$ \\
$a_{T,G}$, $a_{T,C}$ & in the control arm the two parts of $a_T$: weight $a_{T,G}$ of the genuine rows of the true class and weight $a_{T,C}$ of the control rows, $a_T = a_{T,G} + a_{T,C}$ \\
$a_{F,0}$, $a_{T,0}$ & weight $a_{F,0}$ of the wrong label and weight $a_{T,0}$ of the true label without attack \\
\bottomrule
\end{tabularx}
\end{table}

\subsection{Models and readout sites}
\label{sec:modelle}

TabICL \citep{qu2026} and TabPFN \citep{jaeger2026} are pretrained on
synthetic tables. They process the context and the query row, here the
target row, in roughly three steps.
First, they embed every row into a vector, each column conditioned on the
context~\citep{qu2026, grinsztajn2026}. The vectors of the context rows
contain their labels, the vector of the query row contains none. Then all
rows pass through the ICL blocks, 12 in TabICL and 24 in TabPFN, in which
every row attends to the context rows. Finally, TabICL forms the
prediction with an MLP on the embedding of the query
row~\citep{qu2026}, TabPFN with an attention-based retrieval decoder~\citep{grinsztajn2026}: the query row distributes weights over
the context rows, a softmax normalises them to sum to one, and the labels
of the context rows enter the prediction with these weights.

Both models are run as an ensemble of eight members, as in their default
setting. Each member sees the table with its own preprocessing of the
features \citep{qu2025, grinsztajn2026}, its own column order and its own
class encoding, so its embeddings lie in a space of their own. The
following quantities are therefore computed per member and then averaged
over the members. All settings of both models stay at their defaults.

For TabICL, the decision of the MLP cannot be read off directly and is
therefore approximated by the position of the target row relative to the
class centroids. The embedding $z$ of the target row is the output of the
last ICL block after the final layer normalisation and the only input of
the MLP~\citep{qu2026}. The class centroids $\mu_T$ and $\mu_F$ are the
means of the embeddings of the context rows with the true and with the
wrong label of the target row. They are recomputed for each target row and
in each arm from the respective context, so a poison row enters $\mu_F$
and a control row enters $\mu_T$.
The relative closeness
\[
  m = \frac{\cos(z, \mu_F)}{\cos(z, \mu_T)}
\]
divides the cosine similarity of the target row to the centroid of the
wrong class by that to the centroid of the true class. It is greater than
one when the target row is closer in angle to the centroid of the wrong
class. The rule $m > 1$ thus corresponds to the readout rule over the
nearest centroid reported for TabICL~\citep{bilos2026}. Both cosine
values are positive in every measurement and in every member, so $m$ is
well defined (Appendix~\ref{app:icl}). Without attack, the relative
closeness is called $m_0$. Whether the target row or the class centroids
move is shown by $\cos(z, z_0)$, $\cos(\mu_T, \mu_{T,0})$ and
$\cos(\mu_F, \mu_{F,0})$. They measure how far the embedding of the
target row and the two class centroids rotate relative to the context
without attack. A value of one means no rotation. While $m$
describes the position of the target row relative to the means of whole
classes, $\nu_F$ shows its immediate surroundings: it counts how many of
the 25 nearest genuine context rows of the target row at the input of the
MLP carry the wrong label, by cosine and excluding the poison rows or control
rows. The
wrong class holds the majority among these neighbours when more than half
of them carry the wrong label. Because $\nu_F$ is averaged over the members of the ensemble and
is therefore not an integer, the threshold lies at
$\nu_F > 12.5$.

For TabPFN, we read off the weights $a_j$ that the target row places on
the context rows in the decoder, after the scaling of the query
\citep{grinsztajn2026, jaeger2026} and averaged over the heads. With
these weights the model votes. In contrast to TabICL, the decision of
TabPFN is thus measured directly. Summed by the label of the rows, they
give the weight $a_T$ of the true class and $a_F$ of the wrong class,
without attack $a_{T,0}$ and $a_{F,0}$. The decoder outputs the one-hot
labels averaged with these weights as probabilities, and TabPFN averages
these probabilities over the eight members of the ensemble. Hence
$\pf = a_F$ up to numerical precision, with a largest deviation of
\zPFNidentMax\ (Table~\ref{tab:app_e_vote_1}). Under attack, the weight
$a_F$ of the wrong class splits into the weight $a_{F,G}$ of the genuine
rows with the wrong label and the weight $a_{F,P}$ of the poison rows:
\[
  a_F = a_{F,G} + a_{F,P}.
\]
In the control arm, the weight $a_T$ of the true class is composed of the
weight $a_{T,G}$ of the genuine rows with the true label and the weight
$a_{T,C}$ of the control rows:
\[
  a_T = a_{T,G} + a_{T,C}.
\]
In addition, we determine whether one of the poison rows is the strongest
row, that is, carries the largest weight in every member, even where it
alone receives less than half.

The trained models are a random forest, CatBoost
\citep{prokhorenkova2018} and a logistic regression, likewise at their
default settings, the latter with up to 5,000 iterations
(Appendix~\ref{app:referenz}). They are trained on the same context
including the copies, anew for every target row, every budget and every
arm, and once per fold without attack. In these models, the poisoning thus acts
through training. Without attack they
likewise reach an average AUC above 0.8, but a lower one than the TFMs
(Appendix~\ref{app:trainiert_guete}).

\subsection{Statistical analysis}
\label{sec:statistik}

The unit of analysis is the dataset. Each model makes \zVorhersagen\
predictions, eleven per target row. These predictions are not
independent: the target rows of a dataset come from the same
distribution, and their contexts derive from the same fold context. Every quantity
is therefore first averaged per dataset over its target rows from all
five folds and then reported over the \zDS\ datasets. The standard
deviations are given in the appendix. Where a count pools the target
rows of all datasets, this is noted at that point. Groups such as
flipped and held target rows are formed per dataset and budget. A
dataset belongs to a group as soon as it has one target row in it, so
that its mean can rest on few target rows. Two groups are compared
paired over the datasets that have both.

Each research question from (1a) to (3) is assigned a test
(Table~\ref{tab:tests} in Appendix~\ref{app:tests}). All tests are
Wilcoxon signed-rank tests over the datasets~\citep{demsar2006}. For
(1a) and (1b), the test is a paired comparison of the attack arm with
the control arm. For (2a) and (2b), it tests $m$ against 1, because that is the
threshold of the readout rule. For (3), it tests $a_{F,P}$ against 0.5, because that is the
decision threshold: if the poison rows receive more than half of the
weight, they flip the prediction alone. All tests are two-sided, although the
questions are directional, and are thus conservative. Datasets with a
difference of zero are excluded from a test. They occur only in tests
(1a) and (1b) at one poison row, on datasets where neither a poison row
nor a control row flips a target row. A result is called significant at
$p < 0.05$. Because each test answers a question of its own, it forms a
family of its own with its five budgets, within which we correct
according to Holm~\citep{demsar2006}. Every reported $p$ is
Holm-corrected unless it is explicitly marked as uncorrected. If one
corrects over all 25 tests instead, every significant result remains,
except for test~(3) at two poison rows
($p = \zHolmGesamtAPZwei$ with Holm over all 25 tests). The
test tables in Appendix~\ref{app:tests} (Tables~\ref{tab:test_a},
\ref{tab:test_d} and~\ref{tab:test_e}) give as effect size the mean
difference and the number of datasets on each side, with the excluded
datasets in the column \enquote{equal}.

For all other quantities we run no tests and report them descriptively.
These include the difference between the two TFMs, in TabICL the
movement of the target row and of the class centroids and the
neighbourhood $\nu_F$, in TabPFN the weight $a_{F,G}$, and the results
of the trained models. For each dataset, the AUC with which $m$ and
$\nu_F$ separate flipped from held target rows shows how well they
capture the decision. We interpret causally only what the experiment
manipulates, that is, the budget $b$ and the label of the copies in the
two arms. We describe what happens inside the models without explaining
it causally, because the experiment does not intervene in them.

\section{Results}
\label{sec:ergebnisse}

\subsection{Effect of the attack}
\label{sec:budget}

\begin{figure}[tb]
\centering
\includegraphics[width=0.75\linewidth]{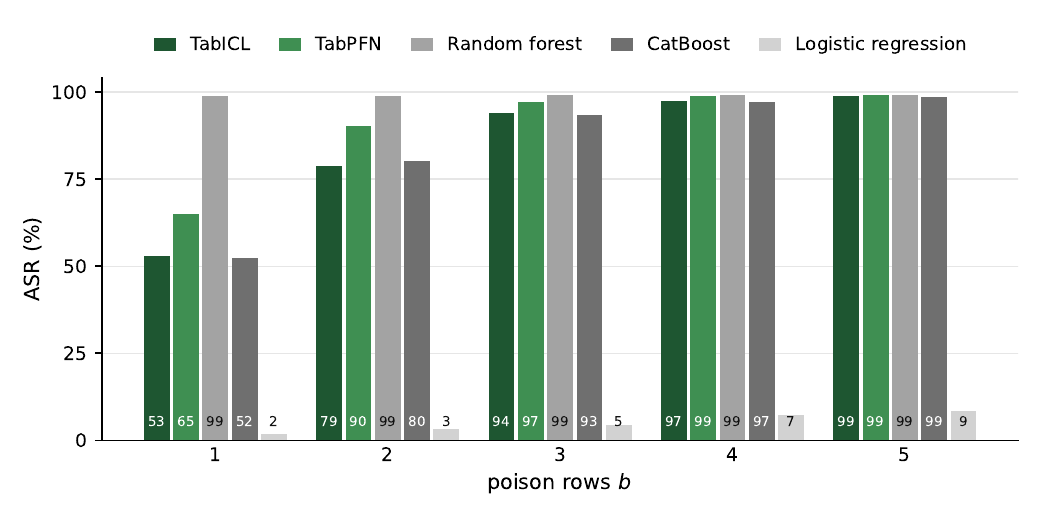}
\caption{ASR of all models per budget $b$: first the two TFMs (green), then random forest, CatBoost and logistic regression (grey). Bars: mean over the \zDS\ datasets, in per cent as the number in the bar. Values per dataset in Tables~\ref{tab:app_asr_ds_b} and~\ref{tab:app_trainiert_asr}.}
\label{fig:asr}
\end{figure}

A single poison row already flips the majority of the attacked target
rows in both TFMs, on average \zICLasrEins\,\% for TabICL and
\zPFNasrEins\,\% for TabPFN. The share rises with every further poison
row, and from three poison rows on almost all flip
(Figure~\ref{fig:asr}). The few target rows that hold even at five
poison rows mostly have a genuine duplicate with the true label in the
context (Table~\ref{tab:app_asr_ds_b}). The more duplicates there are,
the further $\pi_T$ stays above the decision threshold. Without
duplicates or with only a few, it stays only just above it.

The effect of the attack is due to the wrong label, not to the copy.
Across all datasets, models and budgets, the same copy with the true
label flips only one target row in total. The test confirms this: in both TFMs
and at every budget, poison rows flip statistically significantly more
predictions than control rows (research questions~(1a) and~(1b),
$p \zHolmArmMax$, Table~\ref{tab:test_a}). In the control arm, the copy
even has a corrective effect: a single control row already corrects most
of the predictions that are wrong without attack (\zICLkorrigiertEins\,\%
for TabICL and \zPFNkorrigiertEins\,\% for TabPFN,
Table~\ref{tab:app_kontrolle_ds_b}). The label of the copy thus determines in which direction the prediction
moves.

The effect of a single poison row varies strongly across the datasets. On
most datasets, it flips at least half of the target rows. On a few, it
flips almost none (Table~\ref{tab:app_asr_ds_b}). At one and two poison
rows, about \zDiffPFNICL\ percentage points more target rows flip in
TabPFN than in TabICL, but the datasets on which many predictions flip are similar for
both: at one poison row, the ASR of the two TFMs correlates across the
datasets with Spearman's $\rho = \zRhoTFMEins$. With more poison rows
the gap shrinks, at three to \zDiffPFNICLDrei\ percentage points,
because almost all target rows then flip in both TFMs.

The attack stays targeted. Averaged over the datasets, the AUC on the
remaining test rows changes by at most \zDaucMittelMax. This holds for
both TFMs and all budgets. On no single dataset does it change by more
than about \zTFMaucMax\ (Appendix~\ref{app:ausserhalb_auc}). The attack
thus does not show in the overall performance of the TFMs.

Whether flipping is a peculiarity of TFMs is shown by the comparison
with three models that are trained on the same poisoned context. Their
predictions flip as well, but how often differs greatly between the
models (Figure~\ref{fig:asr}, Table~\ref{tab:app_trainiert_asr}): one
poison row flips almost all attacked target rows for the random forest
(\zRFasrEins\,\%), a good half for CatBoost (\zCATasrEins\,\%) and almost
none for the logistic regression (\zLRasrEins\,\%). In the control arm,
no target row flips for any of the three models. Because the datasets
are selected by the performance of the TFMs, these shares do not show
whether predictions of trained models in general flip more or less often
than those of TFMs.

It is more informative to look at the datasets on which many predictions
flip. Despite a similar ASR, CatBoost at one poison row
differs clearly from the TFMs in this respect. On \texttt{Pulsar-Dataset} and \texttt{wilt},
one poison row flips more than half of the attacked target rows for
CatBoost and almost none for TabICL and TabPFN. On \texttt{phoneme} and
\texttt{electricity} it is the other way round: half or more flip there
for the TFMs, at most a quarter for CatBoost
(Tables~\ref{tab:app_asr_ds_b} and~\ref{tab:app_trainiert_asr}). Across
all \zDS\ datasets, the ranking by ASR for CatBoost is hardly related to
that for TabICL or TabPFN (Spearman's $\rho = \zRhoCATICLEins$ and
\zRhoCATPFNEins).

\subsection{Inside the models}
\label{sec:innen}

\subsubsection{TabICL: the position of the target row}
\label{sec:icl}

\begin{figure}[tb]
\centering
\includegraphics[width=0.75\linewidth]{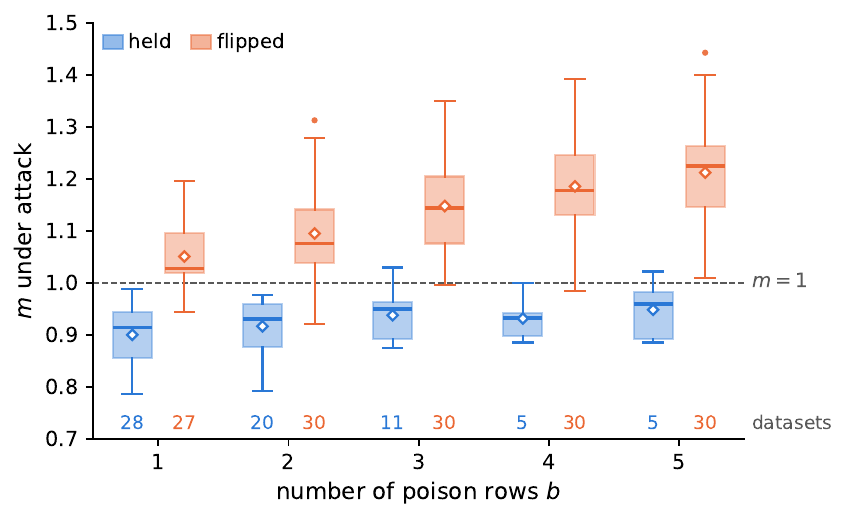}
\caption{TabICL, the relative closeness $m$ under attack per budget $b$, for the held (blue) and the flipped (orange) target rows. The basis is, per dataset, the mean of $m$ over its target rows of the group. Box: quartiles over the datasets, line: median, diamond: mean, whiskers: up to 1.5 times the interquartile range, points: values beyond. Number at the bottom: number of datasets with the group. Dashed line: the threshold $m = 1$ of the readout rule. Values per dataset in Table~\ref{tab:app_m_liste_m}.}
\label{fig:m_arme}
\end{figure}

Research questions (2a) and (2b) test whether the readout rule reported for
TabICL also holds under attack. Figure~\ref{fig:m_arme} shows the mean
$m$ per dataset for the flipped and the held target rows. On average, the rule holds for both
groups: flipped target rows lie above the threshold
$m = 1$ of the readout rule, held ones below it
(Table~\ref{tab:test_d}). At every budget, flipped target rows have
$m > 1$ with statistical significance, so they lie closer to the
centroid $\mu_F$ of the wrong class ($p < 0.001$). At one poison row they lie only just above the
threshold, and further above it with every additional poison row. Held target rows have $m < 1$, so they lie closer to the
centroid $\mu_T$ of the true class, significantly at one to three poison
rows ($p < 0.01$). From four poison rows on, only five datasets still
have held target rows, too few for the test to become significant,
because with five datasets even the uncorrected $p$ does not fall below
0.063. At four and five poison rows, too, the held target rows lie
below the threshold on almost all of these datasets.

The relative closeness $m$ also orders the target rows well: within a
dataset, it separates flipped from held target rows with an average AUC
of \zMauc\ (Table~\ref{tab:app_m_liste_innen}). For the individual target
row, however, the threshold $m = 1$ does not always agree with the
decision of TabICL: at one poison row, about a quarter of the flipped
target rows still lie below it (Table~\ref{tab:app_m_liste_regel}). Even
without attack, \zMnullUeberZeilen\ of the \zICLangegriffen\ attacked
target rows already lie above it, although TabICL predicts the true class
for them (Table~\ref{tab:app_d_alle}). The reason lies in the measure:
$m$ derives only two cosine values from $z$ and the context, whereas the
MLP decides from all dimensions of $z$.

The target row crosses the threshold through a rotation of its
embedding, while the class centroids hardly shift. For flipped target
rows, $\cos(z, z_0)$ drops to \zMcosZgekippt\ on average already at one
poison row and to \zMcosZgekipptFuenf\ at five. The class centroids, in
contrast, barely move: $\cos(\mu_T, \mu_{T,0})$ and
$\cos(\mu_F, \mu_{F,0})$ stay at \zMcosMuAlleMin\ or above on average
(Tables~\ref{tab:app_d_zz}
and~\ref{tab:app_d_zentroide}), because the poison row enters $\mu_F$ as
only one of many rows.

Flipped and held target rows differ in their relative closeness without
attack and in how much the attack changes it: at one poison row, flipped
target rows lie closer to the threshold even without attack
($m_0 = \zMstartGekippt$ against \zMstartGehalten) and the attack
increases it more ($m - m_0 = \zMschubGekippt$ against \zMschubGehalten), both on almost
all datasets (Tables~\ref{tab:app_m} and~\ref{tab:app_m_liste_m0}).

Under attack, the target row moves into a different neighbourhood at the
input of the MLP. Without attack, hardly any of its 25 nearest genuine
context rows carries the wrong label ($\nu_F$ on average
\zKoppICLohneMittel). Under one poison row, the number is
\zKoppICLgekipptMittel\ for flipped target rows and
\zKoppICLgehaltenMittel\ for held ones
(Table~\ref{tab:app_kopp_ds_icl}), so most flipped target rows end up in
a neighbourhood in which the wrong class holds the majority
($\nu_F > 12.5$). This is a consequence of the
rotation and not the reason for the flip: when the embedding of the target row rotates towards the wrong
class, it is mainly rows of that class that lie near it afterwards. The
MLP does not see these neighbours, it reads only $z$. $\nu_F$ thus
describes where $z$ ends up, in the same representation that the MLP
reads.

In the control arm, the control row pulls the target row in the opposite
direction, towards the centroid of the true class: with a single control
row, $m$ is already smaller than without attack on all \zDS\ datasets
(Table~\ref{tab:app_d_alle}), and $\nu_F$ drops from
\zKoppICLohneMittel\ to \zKoppICLkontrMittel\
(Table~\ref{tab:app_kopp_ds_icl}). As for the prediction
(Section~\ref{sec:budget}), the label of the copy thus determines in
which direction the embedding of the target row moves.

\subsubsection{TabPFN: the weight of the poison rows}
\label{sec:pfn}

\begin{figure}[!t]
\centering
\abbildung{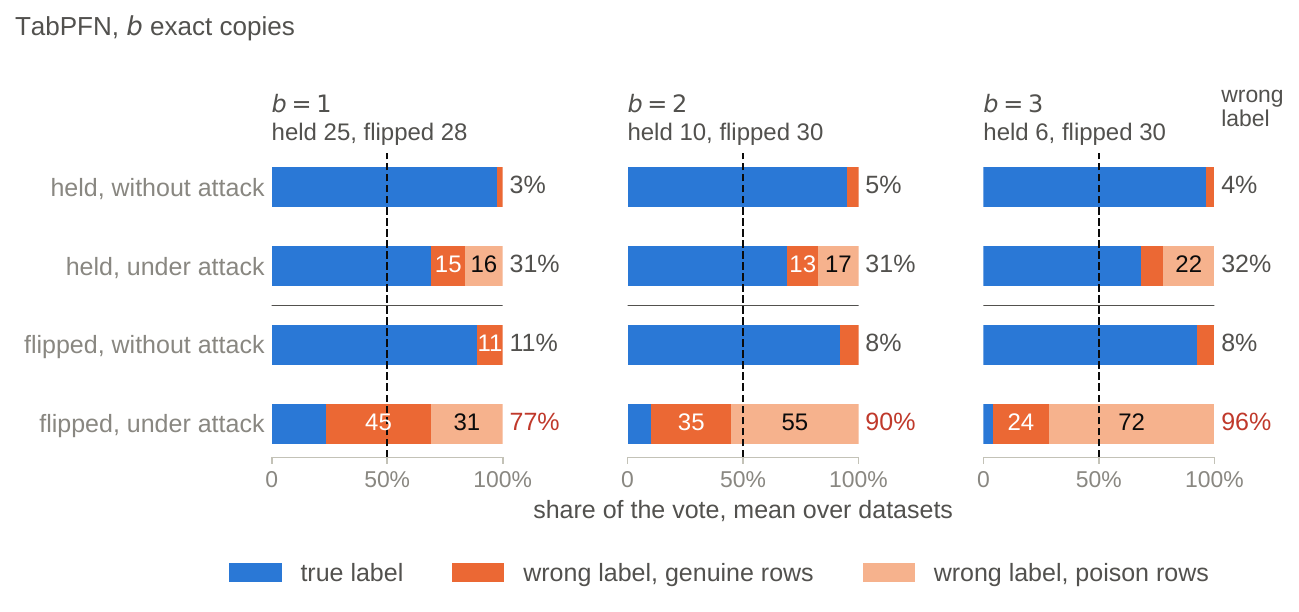}
\caption{TabPFN, the vote at one to three poison rows, one panel per $b$. Per panel, two bars each for the target rows that hold under attack and for those that flip: the same target rows without attack and under attack, per group over all datasets with target rows in the group (number of datasets above the panel). Because different target rows flip at every budget, the bars without attack differ between the panels. Shares of the vote are averaged per dataset, then over the datasets. Right: the share of the wrong label. The control arm and the values per dataset are shown in Tables~\ref{tab:app_e_vote_1} to~\ref{tab:app_e_vote_3}.}
\label{fig:vote}
\end{figure}

In TabPFN, the decision can be read off directly from the vote. Because
$\pf = a_F$, the prediction flips as soon as the rows with the wrong
label together receive more than half of the weight: $a_F > a_T$.
Without attack, the wrong class is far from this: for the target rows
that flip under one poison row, it receives on average
\zPFNaFnullGekipptPct\,\% of the weight without attack, for those that
hold \zPFNaFnullGehaltenPct\,\% (Figure~\ref{fig:vote}). Even without
attack, flipped target rows thus start closer to the decision
threshold. Because every
poison row agrees with the target row in all features, it is natural to
expect that the poison rows alone receive more than half of the weight.
Research question~(3) tests this for the flipped target rows.

For flipped target rows, a single poison row receives significantly less
than half of the weight, on average \zPFNgiftEins\,\%. This holds on
almost all datasets (research question~(3), $p \zHolmAPEins$,
Table~\ref{tab:test_e}). It is nevertheless the strongest single row: for
an average of \zPFNstaerksteEins\,\% of the flipped target rows, it
carries the largest weight in the context
(Table~\ref{tab:app_e_staerkste}). At the same time, the genuine rows of
the wrong class gain weight: their joint weight rises to
\zPFNechtEins\,\% on average, about four times as much as without attack.
Flipping is therefore not due to the poison row alone: only together with
the genuine rows of the wrong class does it lift the wrong class above the
decision threshold. Section~\ref{sec:antworten} returns to this.

With more poison rows, the split between poison rows and genuine rows
reverses (Figure~\ref{fig:vote}). Two poison rows receive on average just
over half, but only on two thirds of the datasets, and the result
($p \zHolmAPZwei$) does not hold if one corrects over all 25 tests
(Section~\ref{sec:statistik}). From three poison rows on, their weight lies significantly above the decision threshold on
almost all datasets ($p < 0.001$)
and, for most flipped target rows, suffices alone for flipping. The
poison rows then account for \zPFNgiftDrei\,\% of the weight, the genuine
rows of the wrong class for only \zPFNechtDrei\,\%. At one poison row, it
is thus mainly the genuine rows that carry the flip, at three the poison
rows. Even at three poison rows, the genuine rows of the wrong class
still receive about three times as much as without attack. For flipped
target rows, less and less weight remains for the true label:
\zPFNwahrEins\,\% at one poison row, \zPFNwahrDrei\,\% at three. The
flipped prediction is therefore not narrowly but confidently wrong. For
held target rows, the weight of the wrong label stays at about one third
at each of these budgets, because there the poison rows receive only
about one fifth of the weight even at three poison rows.

Under attack, the genuine rows of the wrong class gain weight on every
dataset, for flipped as for held target rows
(Table~\ref{tab:app_e_vote_1}). In the control arm it is the other way
round: with one control row they lose weight on all \zDS\ datasets, and
at three control rows the wrong label receives only
\zPFNkontrFalschDrei\,\% of the weight instead of \zPFNohneFalsch\,\%
without attack
(Tables~\ref{tab:app_e_vote_1} to~\ref{tab:app_e_vote_3}). In both arms,
the copy thus shifts the weight towards the class of its label.

\section{Discussion}
\label{sec:diskussion}

\subsection{Answers to the research questions}
\label{sec:antworten}

A single poison row already flips the majority of the predictions that
are correct without attack in both TFMs, and three poison rows flip
almost all. The same copies with the true label flip almost no
prediction, so in both TFMs the effect is due to the wrong label
(research questions~(1a) and~(1b)). In TabICL, flipped target rows lie
on average closer to the centroid of the wrong class under attack, and
held ones closer to the centroid of the true class, significantly up to
three poison rows. The readout rule reported for TabICL thus also holds
under attack (research questions~(2a) and~(2b)). In TabPFN, a single
poison row receives less than half of the weight for flipped target rows
on average. From three poison rows on, the poison rows receive more than
half, and for most flipped target rows their weight alone suffices for
the flip (research question~(3)).

The poison row flips the prediction in TabICL and TabPFN by different
routes. In TabICL, the
copy rotates the embedding of the target row towards the mean of the
wrong class, while the class centroids keep their direction. The target
row thus moves into a neighbourhood in which the wrong class holds the
majority (Section~\ref{sec:icl}). In TabPFN, the label of the copy enters
the vote of the decoder directly. The copy usually receives the largest
weight, and together with it the genuine rows of the wrong class gain
weight. \citet{nguyen2026} call it \enquote{joint processing} when a
label in the context changes the effect of the other labels in this way.
They show it for regression on five TFMs, among them TabICL in the
version examined here and TabPFN in version~3, and find it mainly in the
attention weights. Our measurements show joint processing for
classification as well. At one poison row, the copy thus acts less as a
vote than as a catalyst: it gives the genuine rows of the wrong class the
weight that it lacks itself to flip the prediction. From three poison
rows on, the copies carry the flip themselves.

At one and two poison rows, more predictions flip in TabPFN than in
TabICL. This difference does not protect TabICL, because the gap shrinks
with every poison row, and two poison rows in TabICL already flip more
predictions than one in TabPFN. Since the attacker chooses the budget,
the difference costs the attacker only one additional row. The
difference between the TFMs fits their two routes: in
TabICL, the copy must first rotate the embedding of the target row,
whereas in TabPFN its label enters the vote directly
(Table~\ref{tab:app_kopp}). The two TFMs share the datasets on which many
predictions flip, and CatBoost does not show this ranking
(Section~\ref{sec:budget}). The ranking therefore does not follow from
the datasets alone.

Four studies that attack or exploit the context help to place these
results. For
single-layer linear transformers, \citet{anwar2025} prove that a single
altered context row can force an arbitrary prediction. Our results show
something similar for current TFMs, and without optimisation, because
the exact copy is already the nearest evidence. \citet{bilos2026} attack
the context in an untargeted way and measure how much the overall
performance drops. Flipping the labels of the most central context rows
lowers it by a few percentage points, and a single row with the wrong
label at the centroid of a class hardly changes it. Here, a single copy that agrees with the target row in all
features flips the majority of the target rows
and leaves the other predictions almost unchanged
(Section~\ref{sec:budget}): what matters is where a row lies, not how
many there are.

\citet{wahdany2026} use the same copy for membership inference, assuming
that rows in the context resist relabelling. Our duplicates confirm
this: no single copy flips a target row against a genuine duplicate with
the true label (Table~\ref{tab:app_asr_ds_b}). Finally,
\citet{djilani2026} use the context for
defence against evasion attacks by gradually replacing context rows with
adversarially perturbed examples. Our control arm shows the flip side of the attack:
copies with the true label correct most wrong predictions
(Section~\ref{sec:budget}).

\subsection{Not a faulty inference but forged evidence}
\label{sec:einezeile}

Since a TFM approximates the posterior predictive distribution given its
context (Section~\ref{sec:einleitung}), every context row counts as an
observation, including a poison row: the model cannot check whether a
label is correct. In feature space, an exact
copy of the target row is the nearest evidence about it that a single
context row can provide. Even one such row can flip the prediction, in TabPFN
through its weight in the vote and in TabICL by rotating the embedding of
the target row. Hence, the prediction of a TFM does not flip
because of a faulty inference, but because the model infers soundly from
an observation that is forged in this case.

A faulty inference would have no reason to follow the label, to grow
stronger with the number of copies or to weigh counter-evidence. The
results show all three properties. First, the same copy with the true
label flips almost no prediction and corrects most wrong ones. Second,
more evidence has a stronger effect, since more poison rows flip more
predictions. Third, counter-evidence is weighed. The target rows that
hold lie deeper in the true class even without attack, in TabICL with a
smaller $m_0$, in TabPFN with less weight on the wrong label. A poison row never prevails against a genuine duplicate
with the true label, and five poison rows prevail against only a few
duplicates. The TFM thus infers from the poison rows as from any other
observation.

Two further findings indicate that the influence of a poison row depends
on its nearness to the target row. The effect of the attack stays confined to the target row, and
the other predictions of its fold hardly change, whereas a faulty
inference in the model would affect them as well. In TabPFN, moreover,
the poison row usually receives the largest weight
(Section~\ref{sec:antworten}), which is how its decoder is designed: it
predicts the class through a soft nearest-neighbour retrieval over the
context~\citep{grinsztajn2026}.

Following the forged label is not peculiar to TFMs. Trained models do so
as well: with one poison row, almost all predictions flip for the random
forest and, for CatBoost, about as many as for TabICL. Hardly any flip
for the logistic regression, which, however, also has the lowest
performance without attack (mean AUC \zLRauc\ against \zICLauc\ and
\zPFNauc\ for the TFMs). What is peculiar to the TFM is that no training
lies between context and prediction: the poison row acts in a single
forward pass. TFMs are thus in the same position as trained models, for
which label flipping is considered effective~\citep{vassilev2025}.

Language models show a similar weakness. Indirect prompt injection
exploits the fact that their input changes their behaviour, which
\citet{schneier2024} calls a feature rather than a bug. A language model
is steered by injected instructions, a TFM by forged evidence. In both
cases, the model cannot tell a manipulated input from a genuine one, so
the defence lies at the trust boundary, here at the source of the
context.

\subsection{Implications for operation}
\label{sec:trainiert}

For the protection of the source of the context, the EU Artificial Intelligence Act (EU AI Act) and the NIST each provide
a framework. For high-risk AI systems, the EU AI Act requires, for example,
measures to prevent and detect attacks on the training data and
input data
\citep[Art.~15(5)]{euaiact2024}.
Against targeted poisoning, the NIST names, for example, access controls,
data validation and attestation of the provenance and integrity of the
data. Against indirect prompt injection, the NIST proposes connecting untrusted sources only through defined
interfaces~\citep{vassilev2025}.

Technically, the context of a TFM lies between training data and input
data in the sense of the EU AI Act. The Act defines training data as data
used to fit learnable parameters, and input data as data on the basis of
which the system produces an output~\citep[Art.~3(29) and~(33)]{euaiact2024}.
Like input data, the context is passed only at prediction time and does
not fit any parameters. Like training data, it provides the observations
from which the TFM infers. Its legal categorisation is left to legal assessment. Regardless of this categorisation, the context differs from training
data in when a poison row takes effect. A poison row reaches a trained
model only before training, through data that are typically compiled,
validated and versioned~\citep{baylor2017, breck2019, kreuzberger2023}.
The same holds for a TFM with a \enquote{frozen} context: the context is fixed once
and then used for all subsequent predictions, so here, too, a poison row
can only get in beforehand.

A \enquote{live} context is different. The TFM reads it at prediction time from a
table that is written to in operation, so a new row takes effect at the
very next prediction. It is what \citet{sculley2015} call an unstable
data dependency, an input whose content can change at any time. The
attacker can write the poison row as soon as the target row is known.
The easiest position is that of someone who records cases with their
outcome and submits an application of their own: a single copy of the
application with the desired outcome often flips the prediction, a few
copies almost always. A regularly retrained model~\citep{kreuzberger2023}
approaches this case the shorter the interval between training and
prediction. With a live context, there is no such interval: a
poison row takes effect at the very next prediction.

Three modes of operation that are currently offered or studied for TFMs
read the context at prediction time. If their source is written to in
operation, the context is live. A TFM can run as a function in a
database and form its context directly from a table, so that predictions
can be made with a database query, without a separate training
step~\citep{kongdas2026, googlecloud2026}. Agents can assemble the
context themselves in response to a request in natural language, so that
predictions can be requested without writing
code~\citep{priorlabsmcp2026}. Retrieval-based TFMs draw the nearest
neighbours of each query row from a larger pool, so that TFMs scale to
datasets too large to fit into the context~\citep{ma2025, thomas2024}. With retrieval, the exact copy is closest to the query
row and is therefore selected for certain. At the same time, it faces
fewer genuine rows, though the most similar ones.

The results show that three established security requirements must cover
the context of a TFM: access control, traceability and validation.

\begin{enumerate}
\item \textbf{Access control.} Write and overwrite access to the source
of the context is access to the model and its predictions. It should be
granted in the sense of the access controls named by the NIST, following
the principle of least privilege~\citep{saltzer1975, nist2024csf}, for
example as read-only access for everyone who does not need to record
cases. The same applies to the pipelines that read data from the source
and assemble the context: whoever can change them also determines which
rows the model sees. In the experimental setup, in which the whole table
forms the context, three poison rows flip almost all attacked predictions
with this access and knowledge of the target row.

\item \textbf{Traceability.} The context on which a prediction rested
should be logged together with its provenance, that is, with the source
and the state of the pipeline that assembled it~\citep{buneman2001}. In a
TFM with a live context, the poison row can be deleted or overwritten
after the prediction, and with it the trace of the attack. Moreover, a
prediction can then be neither understood nor reproduced from the query
row and the output alone, because it depends on the state of the context
at the moment of prediction, and a single row can reverse it. Anyone who
wants to understand such a decision needs the context on which it rests.
Such a log can also deter: whoever knows that a poison row can be traced
even after deletion must expect to be found out.

\item \textbf{Validation.} A check of the context can be added at
prediction time. Since the operator does not know which row an attack
targets, every query row is a potential target, and the check must run
before every prediction. Monitoring the overall performance of the model
does not replace it, because such monitoring hardly notices the attack:
the AUC on the remaining test rows stays almost unchanged. Against the
attack studied here, a simple check suffices: an exact copy agrees with
the query row in all features and can therefore be found as a duplicate.
Applied before every prediction to duplicates of the query row, the check
finds all poison rows in the experiment, at every budget, without a model
and without a threshold. Without attack, it raises an alarm for
\zDupSauber\,\% of all test rows on average (Appendix~\ref{app:dup}). An
alarm is therefore a reason to check the prediction, not proof of an
attack.
\end{enumerate}

The three requirements interlock. Access control is meant to prevent manipulation of the
context. Validation is meant to find poison rows that get in
nonetheless. Traceability is meant to establish what a prediction
rested on, even if a poison row was deleted or overwritten afterwards. This corresponds to the prevention
and detection required by the EU AI Act~\citep[Art.~15(5)]{euaiact2024},
extended by investigation after the fact.

\subsection{Limitations and open questions}
\label{sec:grenzen}

Three limitations determine how far the results reach.

First, the study deliberately examines the most conspicuous form of the
attack, the exact copy of the target row with the wrong label. A poison
row that merely resembles the target row would escape a duplicate check.
Harder to find would be a row that is close to the target row only in
the representation of the model or that was constructed using its
gradients. The latter, however, requires knowledge of the model and goes
beyond the threat model considered here.

Second, the scope is narrow: binary classification with numerical
features, contexts of at most 2,400 rows, two TFMs whose results do not
necessarily carry over to other TFMs, and datasets selected by the
performance of the two TFMs examined. The datasets are broadly collected
but not a sample from a defined population: the results describe these
datasets, not a rate for tabular data in general. All models run at
their default settings, and more strongly regularised trained models
might flip less often. The target rows are balanced by class, so the ASR
describes this sample and not the queries in operation.

Third, the study describes what happens inside the models and does not
explain it causally. The measures are approximations. $m$ reduces the
embedding to two cosine values, although the MLP reads the whole
embedding. The weights in TabPFN do not show whether the query of the
target row or the keys of the context rows shift. $\nu_F$ counts the
neighbours in TabICL with a set threshold.

The measurements of this study open up several questions that can be
pursued with the same setup. Interventions in the
attention~\citep{vig2020, meng2022} can clarify whether TabICL decides by
nearest neighbours or by the class centroids, whether the context rows
shift together with the target row, and whether the gap between TabPFN
and TabICL follows from their two routes. Further TFMs will show whether
the shared ranking of the datasets is a property of TFMs in general.  With
retrieval as the context, the effect of the poison row against the
nearest neighbours of the target row can be measured. It can also be examined whether experts, alone or supported by a language
model, recognise poison rows. Finally, it can be examined why, on
average, about as many predictions flip for CatBoost as for TabICL,
although they flip on different datasets.

\section{Conclusion}
\label{sec:fazit}

Targeted in-context poisoning works even in its simplest form. An exact
copy of the target row with the wrong label suffices to flip the majority
of the predictions that are correct without attack in both TFMs, and a
few copies flip almost all. The same copy with the true label flips
almost none: the effect is due to the label. The performance on the
remaining rows stays almost unchanged, so monitoring the overall
performance hardly notices the attack.

The TFMs do not err in their inference: they infer soundly from an
observation that is forged in this case. Inside the TFMs, the poison row
flips the prediction by two routes. In TabICL, it rotates the embedding
of the target row towards the mean of the wrong class. In TabPFN, a
single poison row receives on average less than half of the weight and
pulls genuine rows of the wrong class along. From three poison rows on,
the poison rows themselves receive more than half.

What is new with TFMs is not the kind of vulnerability, since
predictions of trained models such as random forest and CatBoost flip as
well. What is new is the timing: the data take effect at the moment of
prediction, and with a live context a poison row can be deleted or
overwritten afterwards. Three established requirements for secure
operation, which agree with the recommendations of the NIST against
poisoning, must therefore cover the context of a TFM: access control,
traceability and validation.

Whoever can write the data from which a model infers can redirect
individual predictions. In a TFM, this happens at the moment of
prediction, and that is exactly where it can be checked.

\backmatter

\section*{Declarations}
\hypersetup{bookmarksdepth=1}

\bmhead{Funding} This study received no funding.

\bmhead{Data availability} All datasets are publicly available on
OpenML; their identifiers are given in Table~\ref{tab:app_daten_guete}.
The measurements behind every table and figure and the code for the
experiments, the tables and the figures are available at \url{https://github.com/adem-alparslan/ticp}.
Packages, checkpoints and seeds are listed in Appendix~\ref{app:repro}.

\bmhead{Use of AI tools} The study was conducted in October and November 2026. The author wrote the code himself, assisted by Claude Code (Anthropic; models Claude Opus 5.5 and Claude Fable 5.2). The AI assistant also checked the text for typos, translated the author’s German drafts into English, and checked numbers, the description of the methods and citations against the data, the code and the cited sources.

\hypersetup{bookmarksdepth=5}
\newcommand{\nurde}[1]{}
\begin{appendices}
\providecommand{\nurde}[1]{#1}%
\counterwithin{table}{section}
\renewcommand{\thetable}{\Alph{section}\arabic{table}}
\counterwithin{figure}{section}
\renewcommand{\thefigure}{\Alph{section}\arabic{figure}}
\newcommand{\tabelle}[2]{\begin{table}[H]\centering\scriptsize\caption{#2}\label{tab:#1}\input{#1}\end{table}}
\newcommand{\tabelletfm}[2]{\begin{table}[H]\centering\scriptsize\caption{#2}\label{tab:#1}\input{#1_tfm}\end{table}}

\section*{Conventions of the appendix}
A target row is attacked if the model predicts it correctly without attack. Under attack it flips if the model then
predicts it wrongly, and holds otherwise, at the number of poison rows $b$ of the column or table. Poison rows are exact
copies of the target row with the wrong label, control rows the same copies with the true label. Every quantity is
formed per dataset as the mean over its target rows from all five folds and then averaged over the datasets (Section~\ref{sec:statistik}).
Last rows give mean $\pm$ SD over the datasets with a value, with fewer than five datasets the mean alone. A dash (--)
marks an empty group. The ASR is the share of attacked target rows that flip. DS is the number of datasets behind a
value, and higher/equal/lower counts the datasets on which a value lies above, at and below the reference. In
Appendix~\ref{app:tests}, $p$ is two-sided and Holm-corrected per test over the five numbers of poison rows; all other
tables are descriptive. Symbols as in Table~\ref{tab:masse}.

\section{Data and performance without attack}\label{app:daten}\label{app:datensaetze}\label{app:guete}
\noindent Two further candidates, \texttt{higgs} (OpenML 23512) and
\texttt{numerai28.6} (OpenML 23517), were excluded because in a
preliminary run both TFMs stayed clearly below an AUC of 0.8 on them,
on \texttt{numerai28.6} close to chance (Section~\ref{sec:daten}).

\noindent{\sloppy On the 13 datasets with a clearly smaller class, a minority share below 0.4 (Table~\ref{tab:app_daten_guete}), TabICL and TabPFN predict 234 and 239 of the 305 target rows of the minority class correctly, but 318 of the 325 of the majority class.
\par}

\noindent In \texttt{creditcard} each test fold holds a single row of the minority class, so its per-fold AUC rests on one positive case and the AUC criterion of Section~\ref{sec:daten} says little there. Without \texttt{creditcard}, the mean ASR of either TFM changes by at most \zOhneCreditMax\ at any number of poison rows.

\begin{landscape}
\begin{table}[H]\centering\scriptsize\caption{The \zDS\ datasets. Rows: before the reduction to at most 3,000 rows; context: context size after it; minority: share of the smaller class; with duplicate: target rows with an exact copy carrying the true label in the clean context (most copies of one row in brackets). Per TFM without attack: the AUC on the test rows, mean over the five folds, and the number of correctly predicted, that is attacked, target rows. Last rows: mean $\pm$ SD of the AUC and sums of the target rows.}\label{tab:app_daten_guete}\centering\scriptsize\maxbreite{\begin{tabular}{@{}lrrrrrcccccc@{}}\toprule
 & & & & & & & & \multicolumn{2}{c}{TabICL} & \multicolumn{2}{c}{TabPFN} \\ \cmidrule(lr){9-10} \cmidrule(lr){11-12}
Dataset & OpenML & rows & context & features & minority & target rows & with duplicate (max) & AUC & correct & AUC & correct \\ \midrule
\texttt{banknote} & 1462 & 1,372 & 1,097 & 4 & 0.445 & 50 & 3 (3) & $1.000$ & 50 & $1.000$ & 50 \\
\texttt{california} & 44090 & 20,634 & 2,400 & 8 & 0.500 & 50 & 0 & $0.969$ & 44 & $0.971$ & 44 \\
\texttt{climate-model} & 40994 & 540 & 432 & 18 & 0.085 & 50 & 0 & $0.960$ & 40 & $0.948$ & 40 \\
\texttt{codrna} & 351 & 488,565 & 2,400 & 8 & 0.333 & 50 & 0 & $0.994$ & 50 & $0.994$ & 50 \\
\texttt{covertype} & 44121 & 566,602 & 2,400 & 10 & 0.500 & 50 & 0 & $0.881$ & 44 & $0.899$ & 42 \\
\texttt{credit} & 44089 & 16,714 & 2,400 & 10 & 0.500 & 50 & 0 & $0.844$ & 40 & $0.845$ & 41 \\
\texttt{creditcard} & 1597 & 284,807 & 2,400 & 29 & 0.002 & 30 & 0 & $0.993$ & 29 & $0.997$ & 27 \\
\texttt{delta\_ailerons} & 803 & 7,129 & 2,400 & 5 & 0.469 & 50 & 0 & $0.979$ & 49 & $0.980$ & 47 \\
\texttt{eeg-eye-state} & 1471 & 14,980 & 2,400 & 14 & 0.449 & 50 & 0 & $0.996$ & 49 & $0.997$ & 49 \\
\texttt{electrical\_grid} & 46298 & 10,000 & 2,400 & 12 & 0.362 & 50 & 0 & $0.993$ & 47 & $0.994$ & 48 \\
\texttt{electricity} & 44120 & 38,474 & 2,400 & 7 & 0.500 & 50 & 0 & $0.891$ & 40 & $0.937$ & 45 \\
\texttt{elevators} & 846 & 16,599 & 2,400 & 18 & 0.309 & 50 & 0 & $0.944$ & 42 & $0.948$ & 42 \\
\texttt{hazelnut-spread} & 46930 & 2,400 & 1,920 & 30 & 0.500 & 50 & 0 & $0.996$ & 48 & $0.997$ & 49 \\
\texttt{hill-valley} & 1479 & 1,212 & 969 & 100 & 0.500 & 50 & 0 & $0.991$ & 47 & $1.000$ & 50 \\
\texttt{house\_16H} & 44123 & 13,488 & 2,400 & 16 & 0.500 & 50 & 0 & $0.958$ & 45 & $0.960$ & 45 \\
\texttt{jannis} & 45021 & 57,580 & 2,400 & 54 & 0.500 & 50 & 0 & $0.853$ & 32 & $0.859$ & 32 \\
\texttt{kin8nm} & 807 & 8,192 & 2,400 & 8 & 0.491 & 50 & 0 & $0.972$ & 48 & $0.975$ & 47 \\
\texttt{MagicTelescope} & 1120 & 19,020 & 2,400 & 10 & 0.352 & 50 & 0 & $0.943$ & 47 & $0.948$ & 47 \\
\texttt{mammography} & 310 & 11,183 & 2,400 & 6 & 0.023 & 50 & 5 (708) & $0.943$ & 35 & $0.942$ & 38 \\
\texttt{MiniBooNE} & 44128 & 72,998 & 2,400 & 50 & 0.500 & 50 & 0 & $0.983$ & 45 & $0.984$ & 46 \\
\texttt{ozone-level-8hr} & 1487 & 2,534 & 2,027 & 72 & 0.063 & 50 & 0 & $0.941$ & 31 & $0.944$ & 31 \\
\texttt{phoneme} & 44127 & 3,172 & 2,400 & 5 & 0.500 & 50 & 0 & $0.959$ & 43 & $0.964$ & 42 \\
\texttt{polish\_companies} & 46950 & 5,910 & 2,400 & 64 & 0.069 & 50 & 0 & $0.975$ & 43 & $0.991$ & 47 \\
\texttt{Pulsar-Dataset} & 45558 & 17,898 & 2,400 & 8 & 0.092 & 50 & 0 & $0.980$ & 46 & $0.980$ & 46 \\
\texttt{puma8NH} & 816 & 8,192 & 2,400 & 8 & 0.498 & 50 & 0 & $0.916$ & 42 & $0.916$ & 42 \\
\texttt{spambase} & 44 & 4,601 & 2,400 & 57 & 0.394 & 50 & 6 (36) & $0.992$ & 47 & $0.992$ & 47 \\
\texttt{stock} & 841 & 950 & 760 & 9 & 0.486 & 50 & 0 & $0.997$ & 49 & $0.996$ & 49 \\
\texttt{wdbc} & 1510 & 569 & 455 & 30 & 0.373 & 50 & 0 & $0.997$ & 49 & $0.997$ & 48 \\
\texttt{wilt} & 40983 & 4,839 & 2,400 & 5 & 0.054 & 50 & 0 & $0.995$ & 46 & $0.994$ & 46 \\
\texttt{wine} & 43980 & 2,554 & 2,043 & 11 & 0.500 & 50 & 3 (2) & $0.920$ & 40 & $0.920$ & 39 \\
\midrule mean $\pm$ SD &  &  &  &  &  &  &  & $0.958 \pm 0.044$ &  & $0.962 \pm 0.041$ &  \\
sum &  &  &  &  &  & 1,480 & &  & 1,307 &  & 1,316 \\
\bottomrule\end{tabular}}
\end{table}
\end{landscape}

\begin{landscape}
\section{Effect of the attack}\label{app:asr}
\subsection{ASR per dataset and number of poison rows}\label{app:asr_raster}
\begin{table}[H]\centering\scriptsize\caption{ASR of the two TFMs per dataset and number of poison rows $b$. Last rows: mean $\pm$ SD, the number of datasets with ASR $\geq 0.5$ and $< 0.1$, the mean difference TabICL $-$ TabPFN and Spearman's $\rho$ of the two columns. At $b = 5$, \zICLhaltenFuenf\ (TabICL) and \zPFNhaltenFuenf\ (TabPFN) attacked target rows hold, \zICLhaltenDupFuenf\ and \zPFNhaltenDupFuenf\ of them with an exact duplicate carrying the true label (Table~\ref{tab:app_daten_guete}). Their $\pi_T$ rises with the number of duplicates: without one it is at most \zHaltenPiTOhneMax, with several hundred (\texttt{mammography}) at least \zHaltenPiTVielMin. With one poison row, none of the \zDupZiele\ target rows with a duplicate flips in either TFM; at $b = 5$, those that flip have at most \zDupGekipptFuenfMax\ duplicates.}\label{tab:app_asr_ds_b}\scalebox{0.93}{\parbox{\linewidth}{\renewcommand{\arraystretch}{0.93}\centering\scriptsize\maxbreite{\begin{tabular}{@{}lcccccccccc@{}}\toprule
Dataset & \multicolumn{2}{c}{ASR, $b = 1$} & \multicolumn{2}{c}{ASR, $b = 2$} & \multicolumn{2}{c}{ASR, $b = 3$} & \multicolumn{2}{c}{ASR, $b = 4$} & \multicolumn{2}{c}{ASR, $b = 5$} \\ \cmidrule(lr){2-3} \cmidrule(lr){4-5} \cmidrule(lr){6-7} \cmidrule(lr){8-9} \cmidrule(lr){10-11}
 & TabICL & TabPFN & TabICL & TabPFN & TabICL & TabPFN & TabICL & TabPFN & TabICL & TabPFN \\ \midrule
\texttt{banknote} & $0.42$ & $0.68$ & $0.92$ & $0.92$ & $0.94$ & $0.94$ & $0.94$ & $0.94$ & $0.96$ & $0.98$ \\
\texttt{california} & $0.59$ & $0.61$ & $0.98$ & $1.00$ & $1.00$ & $1.00$ & $1.00$ & $1.00$ & $1.00$ & $1.00$ \\
\texttt{climate-model} & $0.25$ & $0.95$ & $1.00$ & $1.00$ & $1.00$ & $1.00$ & $1.00$ & $1.00$ & $1.00$ & $1.00$ \\
\texttt{codrna} & $0.34$ & $0.74$ & $0.96$ & $1.00$ & $1.00$ & $1.00$ & $1.00$ & $1.00$ & $1.00$ & $1.00$ \\
\texttt{covertype} & $0.95$ & $1.00$ & $1.00$ & $1.00$ & $1.00$ & $1.00$ & $1.00$ & $1.00$ & $1.00$ & $1.00$ \\
\texttt{credit} & $0.03$ & $0.15$ & $0.20$ & $0.39$ & $0.90$ & $0.73$ & $1.00$ & $0.98$ & $1.00$ & $1.00$ \\
\texttt{creditcard} & $0.55$ & $0.63$ & $0.93$ & $1.00$ & $1.00$ & $1.00$ & $1.00$ & $1.00$ & $1.00$ & $1.00$ \\
\texttt{delta\_ailerons} & $0.08$ & $0.00$ & $0.20$ & $0.23$ & $0.78$ & $0.83$ & $1.00$ & $1.00$ & $1.00$ & $1.00$ \\
\texttt{eeg-eye-state} & $1.00$ & $1.00$ & $1.00$ & $1.00$ & $1.00$ & $1.00$ & $1.00$ & $1.00$ & $1.00$ & $1.00$ \\
\texttt{electrical\_grid} & $0.70$ & $0.73$ & $1.00$ & $1.00$ & $1.00$ & $1.00$ & $1.00$ & $1.00$ & $1.00$ & $1.00$ \\
\texttt{electricity} & $0.50$ & $0.87$ & $0.80$ & $1.00$ & $0.95$ & $1.00$ & $1.00$ & $1.00$ & $1.00$ & $1.00$ \\
\texttt{elevators} & $0.14$ & $0.26$ & $0.48$ & $0.93$ & $0.98$ & $1.00$ & $1.00$ & $1.00$ & $1.00$ & $1.00$ \\
\texttt{hazelnut-spread} & $0.98$ & $1.00$ & $1.00$ & $1.00$ & $1.00$ & $1.00$ & $1.00$ & $1.00$ & $1.00$ & $1.00$ \\
\texttt{hill-valley} & $0.62$ & $0.98$ & $0.77$ & $1.00$ & $0.94$ & $1.00$ & $0.96$ & $1.00$ & $0.96$ & $1.00$ \\
\texttt{house\_16H} & $0.38$ & $0.64$ & $0.91$ & $1.00$ & $1.00$ & $1.00$ & $1.00$ & $1.00$ & $1.00$ & $1.00$ \\
\texttt{jannis} & $0.72$ & $1.00$ & $1.00$ & $1.00$ & $1.00$ & $1.00$ & $1.00$ & $1.00$ & $1.00$ & $1.00$ \\
\texttt{kin8nm} & $0.65$ & $0.53$ & $0.96$ & $1.00$ & $1.00$ & $1.00$ & $1.00$ & $1.00$ & $1.00$ & $1.00$ \\
\texttt{MagicTelescope} & $0.26$ & $0.49$ & $0.74$ & $0.87$ & $1.00$ & $1.00$ & $1.00$ & $1.00$ & $1.00$ & $1.00$ \\
\texttt{mammography} & $0.20$ & $0.29$ & $0.63$ & $0.66$ & $0.86$ & $0.84$ & $0.86$ & $0.87$ & $0.86$ & $0.87$ \\
\texttt{MiniBooNE} & $0.91$ & $0.93$ & $1.00$ & $1.00$ & $1.00$ & $1.00$ & $1.00$ & $1.00$ & $1.00$ & $1.00$ \\
\texttt{ozone-level-8hr} & $1.00$ & $1.00$ & $1.00$ & $1.00$ & $1.00$ & $1.00$ & $1.00$ & $1.00$ & $1.00$ & $1.00$ \\
\texttt{phoneme} & $0.98$ & $0.98$ & $1.00$ & $1.00$ & $1.00$ & $1.00$ & $1.00$ & $1.00$ & $1.00$ & $1.00$ \\
\texttt{polish\_companies} & $0.65$ & $0.70$ & $0.98$ & $1.00$ & $1.00$ & $1.00$ & $1.00$ & $1.00$ & $1.00$ & $1.00$ \\
\texttt{Pulsar-Dataset} & $0.00$ & $0.00$ & $0.26$ & $0.59$ & $1.00$ & $1.00$ & $1.00$ & $1.00$ & $1.00$ & $1.00$ \\
\texttt{puma8NH} & $0.00$ & $0.24$ & $0.02$ & $1.00$ & $0.12$ & $1.00$ & $0.62$ & $1.00$ & $0.98$ & $1.00$ \\
\texttt{spambase} & $0.72$ & $0.74$ & $0.81$ & $0.79$ & $0.83$ & $0.85$ & $0.85$ & $0.87$ & $0.91$ & $0.87$ \\
\texttt{stock} & $0.92$ & $0.53$ & $1.00$ & $1.00$ & $1.00$ & $1.00$ & $1.00$ & $1.00$ & $1.00$ & $1.00$ \\
\texttt{wdbc} & $0.43$ & $0.90$ & $0.98$ & $1.00$ & $1.00$ & $1.00$ & $1.00$ & $1.00$ & $1.00$ & $1.00$ \\
\texttt{wilt} & $0.00$ & $0.02$ & $0.20$ & $0.74$ & $0.98$ & $1.00$ & $1.00$ & $1.00$ & $1.00$ & $1.00$ \\
\texttt{wine} & $0.93$ & $0.92$ & $0.93$ & $0.97$ & $0.95$ & $0.97$ & $1.00$ & $1.00$ & $1.00$ & $1.00$ \\
\midrule mean $\pm$ SD & $0.53 \pm 0.34$ & $0.65 \pm 0.33$ & $0.79 \pm 0.31$ & $0.90 \pm 0.20$ & $0.94 \pm 0.17$ & $0.97 \pm 0.07$ & $0.97 \pm 0.08$ & $0.99 \pm 0.03$ & $0.99 \pm 0.03$ & $0.99 \pm 0.03$ \\
datasets with ASR $\geq 0.5$ & 17 & 22 & 24 & 28 & 29 & 30 & 30 & 30 & 30 & 30 \\
datasets with ASR $< 0.1$ & 5 & 3 & 1 & 0 & 0 & 0 & 0 & 0 & 0 & 0 \\
TabICL $-$ TabPFN, mean & \multicolumn{2}{c}{$-0.121$} & \multicolumn{2}{c}{$-0.115$} & \multicolumn{2}{c}{$-0.032$} & \multicolumn{2}{c}{$-0.014$} & \multicolumn{2}{c}{$-0.002$} \\
Spearman's $\rho$ & \multicolumn{2}{c}{$0.80$} & \multicolumn{2}{c}{$0.71$} & \multicolumn{2}{c}{$0.71$} & \multicolumn{2}{c}{$0.64$} & \multicolumn{2}{c}{$0.78$} \\
\bottomrule\end{tabular}}
}}\end{table}
\end{landscape}
\begin{landscape}
\subsection{Control arm}\label{app:kontrolle}
\begin{table}[H]\centering\scriptsize\caption{Control arm: of the target rows a TFM predicts wrongly without attack, the share that $b$ control rows correct. A dash marks datasets without such rows: \texttt{banknote} and \texttt{codrna} for both TFMs, \texttt{hill-valley} for TabPFN.}\label{tab:app_kontrolle_ds_b}\centering\scriptsize\maxbreite{\begin{tabular}{@{}lcccccccccc@{}}\toprule
Dataset & \multicolumn{5}{c}{share corrected, TabICL} & \multicolumn{5}{c}{share corrected, TabPFN} \\ \cmidrule(lr){2-6} \cmidrule(lr){7-11}
 & $b = 1$ & $b = 2$ & $b = 3$ & $b = 4$ & $b = 5$ & $b = 1$ & $b = 2$ & $b = 3$ & $b = 4$ & $b = 5$ \\ \midrule
\texttt{banknote} & -- & -- & -- & -- & -- & -- & -- & -- & -- & -- \\
\texttt{california} & $1.00$ & $1.00$ & $1.00$ & $1.00$ & $1.00$ & $0.83$ & $1.00$ & $1.00$ & $1.00$ & $1.00$ \\
\texttt{climate-model} & $1.00$ & $1.00$ & $1.00$ & $1.00$ & $1.00$ & $1.00$ & $1.00$ & $1.00$ & $1.00$ & $1.00$ \\
\texttt{codrna} & -- & -- & -- & -- & -- & -- & -- & -- & -- & -- \\
\texttt{covertype} & $1.00$ & $1.00$ & $1.00$ & $1.00$ & $1.00$ & $1.00$ & $1.00$ & $1.00$ & $1.00$ & $1.00$ \\
\texttt{credit} & $0.20$ & $0.50$ & $1.00$ & $1.00$ & $1.00$ & $0.11$ & $0.67$ & $0.89$ & $1.00$ & $1.00$ \\
\texttt{creditcard} & $1.00$ & $1.00$ & $1.00$ & $1.00$ & $1.00$ & $1.00$ & $1.00$ & $1.00$ & $1.00$ & $1.00$ \\
\texttt{delta\_ailerons} & $0.00$ & $0.00$ & $1.00$ & $1.00$ & $1.00$ & $0.67$ & $0.67$ & $1.00$ & $1.00$ & $1.00$ \\
\texttt{eeg-eye-state} & $1.00$ & $1.00$ & $1.00$ & $1.00$ & $1.00$ & $1.00$ & $1.00$ & $1.00$ & $1.00$ & $1.00$ \\
\texttt{electrical\_grid} & $1.00$ & $1.00$ & $1.00$ & $1.00$ & $1.00$ & $1.00$ & $1.00$ & $1.00$ & $1.00$ & $1.00$ \\
\texttt{electricity} & $0.80$ & $1.00$ & $1.00$ & $1.00$ & $1.00$ & $1.00$ & $1.00$ & $1.00$ & $1.00$ & $1.00$ \\
\texttt{elevators} & $0.50$ & $0.62$ & $1.00$ & $1.00$ & $1.00$ & $0.50$ & $1.00$ & $1.00$ & $1.00$ & $1.00$ \\
\texttt{hazelnut-spread} & $1.00$ & $1.00$ & $1.00$ & $1.00$ & $1.00$ & $1.00$ & $1.00$ & $1.00$ & $1.00$ & $1.00$ \\
\texttt{hill-valley} & $1.00$ & $1.00$ & $1.00$ & $1.00$ & $1.00$ & -- & -- & -- & -- & -- \\
\texttt{house\_16H} & $0.80$ & $1.00$ & $1.00$ & $1.00$ & $1.00$ & $0.80$ & $1.00$ & $1.00$ & $1.00$ & $1.00$ \\
\texttt{jannis} & $0.94$ & $1.00$ & $1.00$ & $1.00$ & $1.00$ & $1.00$ & $1.00$ & $1.00$ & $1.00$ & $1.00$ \\
\texttt{kin8nm} & $1.00$ & $1.00$ & $1.00$ & $1.00$ & $1.00$ & $1.00$ & $1.00$ & $1.00$ & $1.00$ & $1.00$ \\
\texttt{MagicTelescope} & $0.33$ & $1.00$ & $1.00$ & $1.00$ & $1.00$ & $0.67$ & $1.00$ & $1.00$ & $1.00$ & $1.00$ \\
\texttt{mammography} & $0.53$ & $0.93$ & $1.00$ & $1.00$ & $1.00$ & $0.58$ & $0.75$ & $0.92$ & $0.92$ & $1.00$ \\
\texttt{MiniBooNE} & $1.00$ & $1.00$ & $1.00$ & $1.00$ & $1.00$ & $1.00$ & $1.00$ & $1.00$ & $1.00$ & $1.00$ \\
\texttt{ozone-level-8hr} & $1.00$ & $1.00$ & $1.00$ & $1.00$ & $1.00$ & $1.00$ & $1.00$ & $1.00$ & $1.00$ & $1.00$ \\
\texttt{phoneme} & $1.00$ & $1.00$ & $1.00$ & $1.00$ & $1.00$ & $1.00$ & $1.00$ & $1.00$ & $1.00$ & $1.00$ \\
\texttt{polish\_companies} & $0.86$ & $1.00$ & $1.00$ & $1.00$ & $1.00$ & $1.00$ & $1.00$ & $1.00$ & $1.00$ & $1.00$ \\
\texttt{Pulsar-Dataset} & $0.50$ & $0.75$ & $1.00$ & $1.00$ & $1.00$ & $0.25$ & $1.00$ & $1.00$ & $1.00$ & $1.00$ \\
\texttt{puma8NH} & $0.00$ & $0.12$ & $0.25$ & $0.50$ & $0.88$ & $0.50$ & $1.00$ & $1.00$ & $1.00$ & $1.00$ \\
\texttt{spambase} & $1.00$ & $1.00$ & $1.00$ & $1.00$ & $1.00$ & $1.00$ & $1.00$ & $1.00$ & $1.00$ & $1.00$ \\
\texttt{stock} & $1.00$ & $1.00$ & $1.00$ & $1.00$ & $1.00$ & $1.00$ & $1.00$ & $1.00$ & $1.00$ & $1.00$ \\
\texttt{wdbc} & $1.00$ & $1.00$ & $1.00$ & $1.00$ & $1.00$ & $1.00$ & $1.00$ & $1.00$ & $1.00$ & $1.00$ \\
\texttt{wilt} & $0.75$ & $0.75$ & $1.00$ & $1.00$ & $1.00$ & $0.75$ & $0.75$ & $1.00$ & $1.00$ & $1.00$ \\
\texttt{wine} & $1.00$ & $1.00$ & $1.00$ & $1.00$ & $1.00$ & $1.00$ & $1.00$ & $1.00$ & $1.00$ & $1.00$ \\
\midrule mean $\pm$ SD & $0.79 \pm 0.32$ & $0.88 \pm 0.26$ & $0.97 \pm 0.14$ & $0.98 \pm 0.09$ & $1.00 \pm 0.02$ & $0.84 \pm 0.25$ & $0.96 \pm 0.11$ & $0.99 \pm 0.03$ & $1.00 \pm 0.02$ & $1.00 \pm 0.00$ \\
\bottomrule\end{tabular}}
\end{table}
\end{landscape}
\begin{landscape}
\subsection{AUC of the remaining test rows}\label{app:ausserhalb_auc}
\begin{table}[H]\centering\scriptsize\caption{Change $\Delta$AUC of the AUC on the remaining test rows of the fold, predicted with the poisoned context of the target row, against the run without attack, per dataset and number of poison rows $b$; mean over the attacked target rows.}\label{tab:app_b_dauc}\centering\scriptsize\maxbreite{\begin{tabular}{@{}lcccccccccc@{}}\toprule
Dataset & \multicolumn{5}{c}{$\Delta$AUC, attack, TabICL} & \multicolumn{5}{c}{$\Delta$AUC, attack, TabPFN} \\ \cmidrule(lr){2-6}\cmidrule(lr){7-11}
 & $b = 1$ & $b = 2$ & $b = 3$ & $b = 4$ & $b = 5$ & $b = 1$ & $b = 2$ & $b = 3$ & $b = 4$ & $b = 5$ \\ \midrule
\texttt{banknote} & $0.0000$ & $0.0000$ & $0.0000$ & $0.0000$ & $0.0000$ & $0.0000$ & $0.0000$ & $0.0000$ & $0.0000$ & $0.0000$ \\
\texttt{california} & $0.0000$ & $-0.0000$ & $-0.0001$ & $-0.0001$ & $-0.0001$ & $-0.0000$ & $-0.0001$ & $-0.0001$ & $-0.0001$ & $-0.0001$ \\
\texttt{climate-model} & $-0.0043$ & $-0.0050$ & $-0.0039$ & $-0.0042$ & $-0.0049$ & $-0.0006$ & $-0.0005$ & $-0.0000$ & $0.0005$ & $0.0007$ \\
\texttt{codrna} & $-0.0000$ & $-0.0000$ & $-0.0000$ & $-0.0000$ & $-0.0001$ & $0.0000$ & $0.0000$ & $0.0000$ & $0.0000$ & $0.0000$ \\
\texttt{covertype} & $-0.0001$ & $-0.0001$ & $-0.0000$ & $-0.0002$ & $-0.0004$ & $-0.0001$ & $-0.0002$ & $-0.0003$ & $-0.0004$ & $-0.0005$ \\
\texttt{credit} & $-0.0000$ & $-0.0001$ & $-0.0002$ & $-0.0003$ & $-0.0005$ & $-0.0000$ & $-0.0001$ & $-0.0002$ & $-0.0002$ & $-0.0002$ \\
\texttt{creditcard} & $-0.0072$ & $-0.0084$ & $-0.0098$ & $-0.0104$ & $-0.0096$ & $-0.0066$ & $-0.0082$ & $-0.0077$ & $-0.0071$ & $-0.0064$ \\
\texttt{delta\_ailerons} & $0.0000$ & $-0.0001$ & $-0.0001$ & $-0.0001$ & $-0.0001$ & $-0.0000$ & $-0.0001$ & $-0.0001$ & $-0.0002$ & $-0.0001$ \\
\texttt{eeg-eye-state} & $-0.0001$ & $-0.0001$ & $-0.0001$ & $-0.0001$ & $-0.0001$ & $-0.0001$ & $-0.0001$ & $-0.0001$ & $-0.0001$ & $-0.0001$ \\
\texttt{electrical\_grid} & $-0.0001$ & $-0.0002$ & $-0.0002$ & $-0.0002$ & $-0.0002$ & $-0.0001$ & $-0.0001$ & $-0.0001$ & $-0.0001$ & $-0.0001$ \\
\texttt{electricity} & $-0.0000$ & $0.0001$ & $0.0002$ & $0.0002$ & $0.0002$ & $0.0000$ & $0.0000$ & $0.0001$ & $0.0000$ & $0.0000$ \\
\texttt{elevators} & $0.0000$ & $-0.0001$ & $-0.0001$ & $-0.0002$ & $-0.0003$ & $-0.0001$ & $-0.0001$ & $-0.0000$ & $-0.0000$ & $-0.0000$ \\
\texttt{hazelnut-spread} & $-0.0001$ & $-0.0001$ & $-0.0001$ & $-0.0001$ & $-0.0001$ & $-0.0001$ & $-0.0001$ & $-0.0001$ & $-0.0001$ & $-0.0001$ \\
\texttt{hill-valley} & $-0.0000$ & $0.0009$ & $0.0001$ & $-0.0003$ & $-0.0000$ & $0.0000$ & $0.0000$ & $-0.0000$ & $0.0000$ & $-0.0000$ \\
\texttt{house\_16H} & $-0.0000$ & $-0.0001$ & $-0.0001$ & $-0.0001$ & $-0.0001$ & $0.0000$ & $-0.0000$ & $0.0000$ & $0.0000$ & $0.0000$ \\
\texttt{jannis} & $-0.0001$ & $0.0000$ & $0.0003$ & $0.0003$ & $0.0004$ & $-0.0000$ & $-0.0001$ & $-0.0001$ & $-0.0001$ & $-0.0001$ \\
\texttt{kin8nm} & $-0.0002$ & $-0.0003$ & $-0.0004$ & $-0.0004$ & $-0.0004$ & $-0.0001$ & $-0.0001$ & $-0.0001$ & $-0.0000$ & $-0.0000$ \\
\texttt{MagicTelescope} & $-0.0001$ & $-0.0001$ & $-0.0002$ & $-0.0001$ & $-0.0001$ & $-0.0000$ & $-0.0000$ & $-0.0000$ & $-0.0000$ & $-0.0000$ \\
\texttt{mammography} & $-0.0001$ & $-0.0021$ & $-0.0045$ & $-0.0057$ & $-0.0029$ & $0.0009$ & $0.0007$ & $0.0009$ & $0.0003$ & $-0.0004$ \\
\texttt{MiniBooNE} & $-0.0000$ & $-0.0001$ & $-0.0001$ & $-0.0001$ & $-0.0001$ & $0.0000$ & $-0.0000$ & $-0.0000$ & $-0.0000$ & $-0.0000$ \\
\texttt{ozone-level-8hr} & $0.0002$ & $-0.0000$ & $-0.0002$ & $-0.0004$ & $-0.0004$ & $-0.0001$ & $-0.0001$ & $0.0001$ & $0.0001$ & $0.0001$ \\
\texttt{phoneme} & $-0.0001$ & $-0.0001$ & $-0.0002$ & $-0.0002$ & $-0.0002$ & $-0.0001$ & $-0.0002$ & $-0.0003$ & $-0.0003$ & $-0.0003$ \\
\texttt{polish\_companies} & $-0.0005$ & $-0.0007$ & $-0.0008$ & $-0.0009$ & $-0.0009$ & $0.0000$ & $-0.0001$ & $-0.0001$ & $-0.0000$ & $0.0000$ \\
\texttt{Pulsar-Dataset} & $-0.0002$ & $-0.0001$ & $0.0001$ & $0.0001$ & $0.0001$ & $-0.0005$ & $-0.0006$ & $-0.0004$ & $-0.0004$ & $-0.0005$ \\
\texttt{puma8NH} & $-0.0000$ & $-0.0000$ & $-0.0001$ & $-0.0002$ & $-0.0004$ & $-0.0000$ & $-0.0001$ & $-0.0001$ & $-0.0001$ & $-0.0000$ \\
\texttt{spambase} & $-0.0001$ & $-0.0002$ & $-0.0002$ & $-0.0002$ & $-0.0002$ & $-0.0002$ & $-0.0002$ & $-0.0002$ & $-0.0002$ & $-0.0002$ \\
\texttt{stock} & $-0.0001$ & $-0.0001$ & $-0.0001$ & $-0.0001$ & $-0.0001$ & $-0.0002$ & $-0.0002$ & $-0.0002$ & $-0.0002$ & $-0.0002$ \\
\texttt{wdbc} & $-0.0002$ & $-0.0002$ & $-0.0001$ & $-0.0001$ & $-0.0001$ & $-0.0004$ & $-0.0004$ & $-0.0003$ & $-0.0003$ & $-0.0003$ \\
\texttt{wilt} & $-0.0001$ & $-0.0004$ & $-0.0005$ & $-0.0005$ & $-0.0005$ & $-0.0003$ & $-0.0003$ & $-0.0002$ & $-0.0002$ & $-0.0002$ \\
\texttt{wine} & $-0.0004$ & $-0.0005$ & $-0.0005$ & $-0.0006$ & $-0.0007$ & $-0.0004$ & $-0.0005$ & $-0.0005$ & $-0.0005$ & $-0.0005$ \\
\midrule mean $\pm$ SD & $-0.0005 \pm 0.0015$ & $-0.0006 \pm 0.0018$ & $-0.0007 \pm 0.0020$ & $-0.0008 \pm 0.0022$ & $-0.0008 \pm 0.0020$ & $-0.0003 \pm 0.0012$ & $-0.0004 \pm 0.0015$ & $-0.0003 \pm 0.0014$ & $-0.0003 \pm 0.0013$ & $-0.0003 \pm 0.0012$ \\
\bottomrule\end{tabular}}
\end{table}
\end{landscape}
\noindent In single contexts, that is, for one attacked target row at one budget, the AUC on the remaining test rows shifts by more than 0.02 under attack in a share of 0.005 (TabICL) and 0.006 (TabPFN) of the contexts. All of these contexts belong to \texttt{climate-model}, \texttt{creditcard} and \texttt{mammography}, whose minority classes make up a share of 0.085, 0.002 and 0.023 of the rows, so that a handful of rows determine the AUC. The largest shift in a single context is 0.089 (TabICL) and 0.059 (TabPFN).

\section{Inside TabICL}\label{app:icl}
\noindent In every measurement and every ensemble member, $\cos(z, \mu_T)$ and $\cos(z, \mu_F)$ are positive (smallest values \zMcosTmin\ and \zMcosFmin), so $m$ is well defined and no member is left out of its mean. $m = 1$ means equal angle to both class centroids, not the midpoint between them.

\begin{landscape}
\subsection{Relative closeness \texorpdfstring{$m$}{m} under attack}\label{app:m_ds}
\begin{table}[H]\centering\scriptsize\caption{TabICL, the relative closeness $m = \cos(z, \mu_F)/\cos(z, \mu_T)$ under attack, for the flipped and the held target rows, per dataset and number of poison rows $b$; the values behind Figure~\ref{fig:m_arme}.}\label{tab:app_m_liste_m}\centering\scriptsize\maxbreite{\begin{tabular}{@{}lcccccccccc@{}}\toprule
Dataset & \multicolumn{5}{c}{$m$, flipped} & \multicolumn{5}{c}{$m$, held} \\ \cmidrule(lr){2-6}\cmidrule(lr){7-11}
 & $b = 1$ & $b = 2$ & $b = 3$ & $b = 4$ & $b = 5$ & $b = 1$ & $b = 2$ & $b = 3$ & $b = 4$ & $b = 5$ \\ \midrule
\texttt{banknote} & $0.946$ & $1.083$ & $1.182$ & $1.249$ & $1.298$ & $0.811$ & $0.878$ & $0.952$ & $1.000$ & $1.022$ \\
\texttt{california} & $1.039$ & $1.070$ & $1.125$ & $1.166$ & $1.193$ & $0.920$ & $0.957$ & -- & -- & -- \\
\texttt{climate-model} & $1.013$ & $1.101$ & $1.197$ & $1.231$ & $1.249$ & $0.945$ & -- & -- & -- & -- \\
\texttt{codrna} & $0.943$ & $0.984$ & $1.099$ & $1.179$ & $1.237$ & $0.805$ & $0.825$ & -- & -- & -- \\
\texttt{covertype} & $1.031$ & $1.057$ & $1.077$ & $1.087$ & $1.094$ & $0.966$ & -- & -- & -- & -- \\
\texttt{credit} & $1.003$ & $1.012$ & $1.013$ & $1.033$ & $1.042$ & $0.968$ & $0.976$ & $0.990$ & -- & -- \\
\texttt{creditcard} & $1.124$ & $1.270$ & $1.335$ & $1.391$ & $1.443$ & $0.932$ & $0.974$ & -- & -- & -- \\
\texttt{delta\_ailerons} & $1.026$ & $1.041$ & $1.023$ & $1.075$ & $1.126$ & $0.829$ & $0.872$ & $0.900$ & -- & -- \\
\texttt{eeg-eye-state} & $1.197$ & $1.313$ & $1.350$ & $1.375$ & $1.400$ & -- & -- & -- & -- & -- \\
\texttt{electrical\_grid} & $1.032$ & $1.123$ & $1.240$ & $1.288$ & $1.309$ & $0.862$ & -- & -- & -- & -- \\
\texttt{electricity} & $1.020$ & $1.029$ & $1.044$ & $1.060$ & $1.075$ & $0.951$ & $0.961$ & $0.961$ & -- & -- \\
\texttt{elevators} & $1.022$ & $1.006$ & $1.027$ & $1.084$ & $1.121$ & $0.880$ & $0.904$ & $0.875$ & -- & -- \\
\texttt{hazelnut-spread} & $1.167$ & $1.279$ & $1.318$ & $1.329$ & $1.339$ & $0.939$ & -- & -- & -- & -- \\
\texttt{hill-valley} & $1.082$ & $1.154$ & $1.187$ & $1.223$ & $1.250$ & $0.928$ & $0.955$ & $0.950$ & $0.932$ & $0.960$ \\
\texttt{house\_16H} & $1.017$ & $1.038$ & $1.104$ & $1.139$ & $1.164$ & $0.913$ & $0.946$ & -- & -- & -- \\
\texttt{jannis} & $1.023$ & $1.043$ & $1.074$ & $1.085$ & $1.090$ & $0.980$ & -- & -- & -- & -- \\
\texttt{kin8nm} & $1.021$ & $1.076$ & $1.150$ & $1.205$ & $1.224$ & $0.886$ & $0.927$ & -- & -- & -- \\
\texttt{MagicTelescope} & $1.022$ & $1.040$ & $1.091$ & $1.128$ & $1.144$ & $0.913$ & $0.932$ & -- & -- & -- \\
\texttt{mammography} & $0.962$ & $1.054$ & $1.143$ & $1.195$ & $1.226$ & $0.943$ & $0.975$ & $0.883$ & $0.885$ & $0.885$ \\
\texttt{MiniBooNE} & $1.047$ & $1.139$ & $1.187$ & $1.207$ & $1.219$ & $0.896$ & -- & -- & -- & -- \\
\texttt{ozone-level-8hr} & $1.086$ & $1.141$ & $1.163$ & $1.172$ & $1.177$ & -- & -- & -- & -- & -- \\
\texttt{phoneme} & $1.143$ & $1.195$ & $1.221$ & $1.240$ & $1.257$ & $0.974$ & -- & -- & -- & -- \\
\texttt{polish\_companies} & $1.107$ & $1.169$ & $1.223$ & $1.250$ & $1.266$ & $0.989$ & $0.974$ & -- & -- & -- \\
\texttt{Pulsar-Dataset} & -- & $0.988$ & $1.075$ & $1.171$ & $1.235$ & $0.789$ & $0.866$ & -- & -- & -- \\
\texttt{puma8NH} & -- & $1.006$ & $0.995$ & $0.985$ & $1.009$ & $0.915$ & $0.922$ & $0.931$ & $0.943$ & $0.982$ \\
\texttt{spambase} & $1.141$ & $1.213$ & $1.252$ & $1.278$ & $1.280$ & $0.835$ & $0.845$ & $0.874$ & $0.897$ & $0.891$ \\
\texttt{stock} & $1.000$ & $1.075$ & $1.121$ & $1.155$ & $1.185$ & $0.912$ & -- & -- & -- & -- \\
\texttt{wdbc} & $1.027$ & $1.084$ & $1.207$ & $1.267$ & $1.307$ & $0.807$ & $0.791$ & -- & -- & -- \\
\texttt{wilt} & -- & $0.921$ & $1.063$ & $1.177$ & $1.247$ & $0.786$ & $0.895$ & $1.029$ & -- & -- \\
\texttt{wine} & $1.121$ & $1.141$ & $1.145$ & $1.145$ & $1.154$ & $0.926$ & $0.951$ & $0.967$ & -- & -- \\
\midrule mean $\pm$ SD & $1.050 \pm 0.065$ & $1.095 \pm 0.093$ & $1.148 \pm 0.095$ & $1.186 \pm 0.097$ & $1.212 \pm 0.100$ & $0.900 \pm 0.062$ & $0.916 \pm 0.055$ & $0.937 \pm 0.050$ & $0.931 \pm 0.045$ & $0.948 \pm 0.059$ \\
\bottomrule\end{tabular}}
\end{table}
\end{landscape}
\begin{landscape}
\subsection{Relative closeness \texorpdfstring{$m_0$}{m0} without attack}\label{app:m_liste_m0}
\begin{table}[H]\centering\scriptsize\caption{TabICL, $m_0$, the relative closeness without attack, of the target rows that flip and of those that hold under attack at $b$.}\label{tab:app_m_liste_m0}\centering\scriptsize\maxbreite{\begin{tabular}{@{}lcccccccccc@{}}\toprule
Dataset & \multicolumn{5}{c}{$m_0$, flipped} & \multicolumn{5}{c}{$m_0$, held} \\ \cmidrule(lr){2-6}\cmidrule(lr){7-11}
 & $b = 1$ & $b = 2$ & $b = 3$ & $b = 4$ & $b = 5$ & $b = 1$ & $b = 2$ & $b = 3$ & $b = 4$ & $b = 5$ \\ \midrule
\texttt{banknote} & $0.482$ & $0.484$ & $0.484$ & $0.484$ & $0.485$ & $0.488$ & $0.494$ & $0.500$ & $0.500$ & $0.500$ \\
\texttt{california} & $0.854$ & $0.822$ & $0.820$ & $0.820$ & $0.820$ & $0.771$ & $0.733$ & -- & -- & -- \\
\texttt{climate-model} & $0.877$ & $0.800$ & $0.800$ & $0.800$ & $0.800$ & $0.774$ & -- & -- & -- & -- \\
\texttt{codrna} & $0.744$ & $0.696$ & $0.693$ & $0.693$ & $0.693$ & $0.667$ & $0.638$ & -- & -- & -- \\
\texttt{covertype} & $0.937$ & $0.935$ & $0.935$ & $0.935$ & $0.935$ & $0.889$ & -- & -- & -- & -- \\
\texttt{credit} & $0.994$ & $0.968$ & $0.956$ & $0.956$ & $0.956$ & $0.955$ & $0.953$ & $0.953$ & -- & -- \\
\texttt{creditcard} & $0.626$ & $0.592$ & $0.587$ & $0.587$ & $0.587$ & $0.538$ & $0.512$ & -- & -- & -- \\
\texttt{delta\_ailerons} & $0.964$ & $0.876$ & $0.797$ & $0.787$ & $0.787$ & $0.771$ & $0.763$ & $0.751$ & -- & -- \\
\texttt{eeg-eye-state} & $0.693$ & $0.693$ & $0.693$ & $0.693$ & $0.693$ & -- & -- & -- & -- & -- \\
\texttt{electrical\_grid} & $0.700$ & $0.671$ & $0.671$ & $0.671$ & $0.671$ & $0.603$ & -- & -- & -- & -- \\
\texttt{electricity} & $0.945$ & $0.930$ & $0.924$ & $0.922$ & $0.922$ & $0.899$ & $0.892$ & $0.896$ & -- & -- \\
\texttt{elevators} & $0.912$ & $0.834$ & $0.829$ & $0.828$ & $0.828$ & $0.814$ & $0.823$ & $0.799$ & -- & -- \\
\texttt{hazelnut-spread} & $0.704$ & $0.702$ & $0.702$ & $0.702$ & $0.702$ & $0.637$ & -- & -- & -- & -- \\
\texttt{hill-valley} & $0.777$ & $0.761$ & $0.786$ & $0.787$ & $0.787$ & $0.810$ & $0.884$ & $0.842$ & $0.842$ & $0.842$ \\
\texttt{house\_16H} & $0.908$ & $0.854$ & $0.848$ & $0.848$ & $0.848$ & $0.812$ & $0.791$ & -- & -- & -- \\
\texttt{jannis} & $0.968$ & $0.955$ & $0.955$ & $0.955$ & $0.955$ & $0.921$ & -- & -- & -- & -- \\
\texttt{kin8nm} & $0.826$ & $0.787$ & $0.782$ & $0.782$ & $0.782$ & $0.704$ & $0.688$ & -- & -- & -- \\
\texttt{MagicTelescope} & $0.883$ & $0.851$ & $0.850$ & $0.850$ & $0.850$ & $0.838$ & $0.846$ & -- & -- & -- \\
\texttt{mammography} & $0.862$ & $0.852$ & $0.849$ & $0.849$ & $0.849$ & $0.849$ & $0.850$ & $0.868$ & $0.868$ & $0.868$ \\
\texttt{MiniBooNE} & $0.786$ & $0.783$ & $0.783$ & $0.783$ & $0.783$ & $0.745$ & -- & -- & -- & -- \\
\texttt{ozone-level-8hr} & $0.932$ & $0.932$ & $0.932$ & $0.932$ & $0.932$ & -- & -- & -- & -- & -- \\
\texttt{phoneme} & $0.838$ & $0.836$ & $0.836$ & $0.836$ & $0.836$ & $0.754$ & -- & -- & -- & -- \\
\texttt{polish\_companies} & $0.817$ & $0.801$ & $0.799$ & $0.799$ & $0.799$ & $0.765$ & $0.731$ & -- & -- & -- \\
\texttt{Pulsar-Dataset} & -- & $0.717$ & $0.699$ & $0.699$ & $0.699$ & $0.699$ & $0.692$ & -- & -- & -- \\
\texttt{puma8NH} & -- & $0.991$ & $0.945$ & $0.904$ & $0.903$ & $0.905$ & $0.902$ & $0.899$ & $0.906$ & $0.953$ \\
\texttt{spambase} & $0.761$ & $0.753$ & $0.752$ & $0.751$ & $0.747$ & $0.702$ & $0.708$ & $0.710$ & $0.709$ & $0.712$ \\
\texttt{stock} & $0.705$ & $0.704$ & $0.704$ & $0.704$ & $0.704$ & $0.696$ & -- & -- & -- & -- \\
\texttt{wdbc} & $0.730$ & $0.663$ & $0.661$ & $0.661$ & $0.661$ & $0.610$ & $0.595$ & -- & -- & -- \\
\texttt{wilt} & -- & $0.704$ & $0.702$ & $0.702$ & $0.702$ & $0.702$ & $0.702$ & $0.705$ & -- & -- \\
\texttt{wine} & $0.919$ & $0.919$ & $0.917$ & $0.912$ & $0.912$ & $0.817$ & $0.817$ & $0.811$ & -- & -- \\
\midrule mean $\pm$ SD & $0.820 \pm 0.121$ & $0.795 \pm 0.119$ & $0.790 \pm 0.115$ & $0.788 \pm 0.113$ & $0.788 \pm 0.113$ & $0.755 \pm 0.115$ & $0.751 \pm 0.127$ & $0.794 \pm 0.126$ & $0.765 \pm 0.165$ & $0.775 \pm 0.176$ \\
\bottomrule\end{tabular}}
\end{table}
\end{landscape}
\subsection{Flipped against held target rows}\label{app:m_gekippt_gehalten}
\begin{table}[H]\centering\scriptsize\caption{TabICL, flipped against held target rows: $m$ under attack, $m_0$ without attack and the push $m - m_0$, per $b$. Mean $\pm$ SD over the datasets with the group (DS flipped/held), and within the datasets that have both groups (pairs), how often the flipped are higher, equal or lower than the held. For the push also over all attacked target rows (all).}\label{tab:app_m}\par\medskip\noindent\textit{$m$ under attack}\par\nopagebreak
\centering\scriptsize\begin{tabular}{@{}rccccc@{}}\toprule
$b$ & flipped & held & DS flipped/held & pairs & higher/equal/lower \\ \midrule
1 & $1.050 \pm 0.065$ & $0.900 \pm 0.062$ & 27/28 & 25 & 25/0/0 \\
2 & $1.095 \pm 0.093$ & $0.916 \pm 0.055$ & 30/20 & 20 & 20/0/0 \\
3 & $1.148 \pm 0.095$ & $0.937 \pm 0.050$ & 30/11 & 11 & 11/0/0 \\
4 & $1.186 \pm 0.097$ & $0.931 \pm 0.045$ & 30/5 & 5 & 5/0/0 \\
5 & $1.212 \pm 0.100$ & $0.948 \pm 0.059$ & 30/5 & 5 & 5/0/0 \\
\bottomrule\end{tabular}

\par\medskip\noindent\textit{$m_0$ without attack, the same target rows}\par\nopagebreak
\centering\scriptsize\begin{tabular}{@{}rccccc@{}}\toprule
$b$ & flipped & held & DS flipped/held & pairs & higher/equal/lower \\ \midrule
1 & $0.820 \pm 0.121$ & $0.755 \pm 0.115$ & 27/28 & 25 & 23/0/2 \\
2 & $0.795 \pm 0.119$ & $0.751 \pm 0.127$ & 30/20 & 20 & 18/0/2 \\
3 & $0.790 \pm 0.115$ & $0.794 \pm 0.126$ & 30/11 & 11 & 7/0/4 \\
4 & $0.788 \pm 0.113$ & $0.765 \pm 0.165$ & 30/5 & 5 & 1/0/4 \\
5 & $0.788 \pm 0.113$ & $0.775 \pm 0.176$ & 30/5 & 5 & 1/0/4 \\
\bottomrule\end{tabular}

\par\medskip\noindent\textit{push $m - m_0$}\par\nopagebreak
\centering\scriptsize\maxbreite{\begin{tabular}{@{}rccccccc@{}}\toprule
$b$ & all & flipped & held & DS flipped/held & above 0, all/flipped/held & pairs & higher/equal/lower \\ \midrule
1 & $0.195 \pm 0.135$ & $0.230 \pm 0.144$ & $0.145 \pm 0.095$ & 27/28 & 30/27/28 & 25 & 22/0/3 \\
2 & $0.284 \pm 0.173$ & $0.299 \pm 0.172$ & $0.166 \pm 0.111$ & 30/20 & 30/30/20 & 20 & 19/0/1 \\
3 & $0.352 \pm 0.185$ & $0.358 \pm 0.185$ & $0.143 \pm 0.135$ & 30/11 & 30/30/11 & 11 & 11/0/0 \\
4 & $0.393 \pm 0.193$ & $0.398 \pm 0.194$ & $0.166 \pm 0.198$ & 30/5 & 30/30/5 & 5 & 5/0/0 \\
5 & $0.421 \pm 0.202$ & $0.424 \pm 0.202$ & $0.173 \pm 0.206$ & 30/5 & 30/30/5 & 5 & 5/0/0 \\
\bottomrule\end{tabular}}
\end{table}
\begin{landscape}
\subsection{The rule \texorpdfstring{$m > 1$}{m > 1}}\label{app:m_regel}
\begin{table}[H]\centering\scriptsize\caption{TabICL, the rule $m > 1$: the share of the flipped target rows with $m > 1$ and of the held ones with $m \leq 1$, per dataset and $b$.}\label{tab:app_m_liste_regel}\centering\scriptsize\maxbreite{\begin{tabular}{@{}lcccccccccc@{}}\toprule
Dataset & \multicolumn{5}{c}{share with $m > 1$, flipped} & \multicolumn{5}{c}{share with $m \leq 1$, held} \\ \cmidrule(lr){2-6}\cmidrule(lr){7-11}
 & $b = 1$ & $b = 2$ & $b = 3$ & $b = 4$ & $b = 5$ & $b = 1$ & $b = 2$ & $b = 3$ & $b = 4$ & $b = 5$ \\ \midrule
\texttt{banknote} & $0.19$ & $0.85$ & $0.98$ & $1.00$ & $1.00$ & $1.00$ & $1.00$ & $1.00$ & $0.33$ & $0.00$ \\
\texttt{california} & $0.77$ & $0.79$ & $1.00$ & $1.00$ & $1.00$ & $1.00$ & $1.00$ & -- & -- & -- \\
\texttt{climate-model} & $0.60$ & $0.75$ & $1.00$ & $1.00$ & $1.00$ & $0.87$ & -- & -- & -- & -- \\
\texttt{codrna} & $0.24$ & $0.33$ & $0.80$ & $0.98$ & $1.00$ & $1.00$ & $1.00$ & -- & -- & -- \\
\texttt{covertype} & $0.88$ & $1.00$ & $1.00$ & $1.00$ & $1.00$ & $1.00$ & -- & -- & -- & -- \\
\texttt{credit} & $1.00$ & $0.75$ & $0.64$ & $0.88$ & $0.97$ & $0.92$ & $0.94$ & $1.00$ & -- & -- \\
\texttt{creditcard} & $0.94$ & $1.00$ & $1.00$ & $1.00$ & $1.00$ & $0.85$ & $0.50$ & -- & -- & -- \\
\texttt{delta\_ailerons} & $1.00$ & $0.90$ & $0.55$ & $0.90$ & $0.98$ & $1.00$ & $1.00$ & $1.00$ & -- & -- \\
\texttt{eeg-eye-state} & $0.96$ & $1.00$ & $1.00$ & $1.00$ & $1.00$ & -- & -- & -- & -- & -- \\
\texttt{electrical\_grid} & $0.61$ & $0.96$ & $1.00$ & $1.00$ & $1.00$ & $1.00$ & -- & -- & -- & -- \\
\texttt{electricity} & $0.75$ & $0.72$ & $0.92$ & $0.93$ & $1.00$ & $1.00$ & $1.00$ & $1.00$ & -- & -- \\
\texttt{elevators} & $0.83$ & $0.50$ & $0.59$ & $0.95$ & $1.00$ & $1.00$ & $1.00$ & $1.00$ & -- & -- \\
\texttt{hazelnut-spread} & $0.98$ & $1.00$ & $1.00$ & $1.00$ & $1.00$ & $1.00$ & -- & -- & -- & -- \\
\texttt{hill-valley} & $0.76$ & $0.97$ & $0.98$ & $1.00$ & $1.00$ & $0.94$ & $1.00$ & $1.00$ & $1.00$ & $1.00$ \\
\texttt{house\_16H} & $0.65$ & $0.66$ & $1.00$ & $1.00$ & $1.00$ & $1.00$ & $1.00$ & -- & -- & -- \\
\texttt{jannis} & $0.91$ & $0.91$ & $1.00$ & $1.00$ & $1.00$ & $0.89$ & -- & -- & -- & -- \\
\texttt{kin8nm} & $0.55$ & $0.89$ & $1.00$ & $1.00$ & $1.00$ & $1.00$ & $1.00$ & -- & -- & -- \\
\texttt{MagicTelescope} & $0.67$ & $0.80$ & $0.91$ & $1.00$ & $1.00$ & $1.00$ & $1.00$ & -- & -- & -- \\
\texttt{mammography} & $0.14$ & $0.59$ & $0.90$ & $0.97$ & $1.00$ & $0.82$ & $0.54$ & $1.00$ & $1.00$ & $1.00$ \\
\texttt{MiniBooNE} & $0.78$ & $1.00$ & $1.00$ & $1.00$ & $1.00$ & $1.00$ & -- & -- & -- & -- \\
\texttt{ozone-level-8hr} & $0.94$ & $1.00$ & $1.00$ & $1.00$ & $1.00$ & -- & -- & -- & -- & -- \\
\texttt{phoneme} & $0.98$ & $1.00$ & $1.00$ & $1.00$ & $1.00$ & $1.00$ & -- & -- & -- & -- \\
\texttt{polish\_companies} & $0.96$ & $1.00$ & $1.00$ & $1.00$ & $1.00$ & $0.53$ & $1.00$ & -- & -- & -- \\
\texttt{Pulsar-Dataset} & -- & $0.33$ & $0.63$ & $0.98$ & $1.00$ & $1.00$ & $1.00$ & -- & -- & -- \\
\texttt{puma8NH} & -- & $1.00$ & $0.60$ & $0.31$ & $0.56$ & $1.00$ & $1.00$ & $1.00$ & $1.00$ & $1.00$ \\
\texttt{spambase} & $0.88$ & $0.97$ & $0.97$ & $0.95$ & $0.93$ & $1.00$ & $1.00$ & $1.00$ & $1.00$ & $1.00$ \\
\texttt{stock} & $0.51$ & $0.88$ & $1.00$ & $1.00$ & $1.00$ & $1.00$ & -- & -- & -- & -- \\
\texttt{wdbc} & $0.48$ & $0.75$ & $1.00$ & $1.00$ & $1.00$ & $1.00$ & $1.00$ & -- & -- & -- \\
\texttt{wilt} & -- & $0.22$ & $0.58$ & $0.91$ & $1.00$ & $1.00$ & $0.81$ & $0.00$ & -- & -- \\
\texttt{wine} & $0.97$ & $0.97$ & $0.97$ & $0.95$ & $0.97$ & $1.00$ & $1.00$ & $1.00$ & -- & -- \\
\midrule mean $\pm$ SD & $0.74 \pm 0.25$ & $0.82 \pm 0.22$ & $0.90 \pm 0.16$ & $0.96 \pm 0.13$ & $0.98 \pm 0.08$ & $0.96 \pm 0.10$ & $0.94 \pm 0.15$ & $0.91 \pm 0.30$ & $0.87 \pm 0.30$ & $0.80 \pm 0.45$ \\
\bottomrule\end{tabular}}
\end{table}
\end{landscape}
\subsection{Relative closeness \texorpdfstring{$m$}{m} and the decision within the dataset}\label{app:m_innen}
\begin{table}[H]\centering\scriptsize\caption{TabICL, the AUC with which $m$ separates the flipped from the held target rows within a dataset, per $b$.}\label{tab:app_m_liste_innen}\centering\scriptsize\maxbreite{\begin{tabular}{@{}lccccc@{}}\toprule
Dataset & \multicolumn{5}{c}{AUC of $m$, flipped against held} \\ \cmidrule(lr){2-6}
 & $b = 1$ & $b = 2$ & $b = 3$ & $b = 4$ & $b = 5$ \\ \midrule
\texttt{banknote} & $0.96$ & $1.00$ & $1.00$ & $1.00$ & $1.00$ \\
\texttt{california} & $1.00$ & $0.98$ & -- & -- & -- \\
\texttt{climate-model} & $0.72$ & -- & -- & -- & -- \\
\texttt{codrna} & $0.90$ & $0.98$ & -- & -- & -- \\
\texttt{covertype} & $1.00$ & -- & -- & -- & -- \\
\texttt{credit} & $0.97$ & $0.98$ & $0.77$ & -- & -- \\
\texttt{creditcard} & $0.95$ & $1.00$ & -- & -- & -- \\
\texttt{delta\_ailerons} & $1.00$ & $1.00$ & $0.97$ & -- & -- \\
\texttt{eeg-eye-state} & -- & -- & -- & -- & -- \\
\texttt{electrical\_grid} & $0.97$ & -- & -- & -- & -- \\
\texttt{electricity} & $1.00$ & $1.00$ & $1.00$ & -- & -- \\
\texttt{elevators} & $1.00$ & $0.93$ & $1.00$ & -- & -- \\
\texttt{hazelnut-spread} & $1.00$ & -- & -- & -- & -- \\
\texttt{hill-valley} & $0.95$ & $1.00$ & $1.00$ & $1.00$ & $1.00$ \\
\texttt{house\_16H} & $1.00$ & $0.97$ & -- & -- & -- \\
\texttt{jannis} & $0.98$ & -- & -- & -- & -- \\
\texttt{kin8nm} & $0.99$ & $0.99$ & -- & -- & -- \\
\texttt{MagicTelescope} & $1.00$ & $1.00$ & -- & -- & -- \\
\texttt{mammography} & $0.58$ & $0.72$ & $1.00$ & $1.00$ & $1.00$ \\
\texttt{MiniBooNE} & $1.00$ & -- & -- & -- & -- \\
\texttt{ozone-level-8hr} & -- & -- & -- & -- & -- \\
\texttt{phoneme} & $1.00$ & -- & -- & -- & -- \\
\texttt{polish\_companies} & $0.91$ & $1.00$ & -- & -- & -- \\
\texttt{Pulsar-Dataset} & -- & $0.94$ & -- & -- & -- \\
\texttt{puma8NH} & -- & $1.00$ & $0.95$ & $0.78$ & $0.76$ \\
\texttt{spambase} & $1.00$ & $1.00$ & $1.00$ & $1.00$ & $1.00$ \\
\texttt{stock} & $0.93$ & -- & -- & -- & -- \\
\texttt{wdbc} & $0.99$ & $1.00$ & -- & -- & -- \\
\texttt{wilt} & -- & $0.57$ & $0.56$ & -- & -- \\
\texttt{wine} & $1.00$ & $1.00$ & $1.00$ & -- & -- \\
\midrule mean $\pm$ SD & $0.95 \pm 0.10$ & $0.95 \pm 0.11$ & $0.93 \pm 0.14$ & $0.96 \pm 0.10$ & $0.95 \pm 0.11$ \\
\bottomrule\end{tabular}}
\end{table}
\begin{landscape}
\subsection{All attacked target rows: change under attack and control arm}\label{app:m}\label{app:m_null}
\begin{table}[H]\centering\scriptsize\caption{TabICL, all attacked target rows of a dataset: $m_0$ and the share with $m_0 < 1$, the change $m - m_0$ under attack and $m$ in the control arm, per $b$.}\label{tab:app_d_alle}\centering\scriptsize\maxbreite{\begin{tabular}{@{}lcccccccccccc@{}}\toprule
Dataset & \multicolumn{2}{c}{without attack} & \multicolumn{5}{c}{$m - m_0$, under attack} & \multicolumn{5}{c}{$m$, control arm} \\ \cmidrule(lr){2-3}\cmidrule(lr){4-8}\cmidrule(lr){9-13}
 & $m_0$ & $m_0 < 1$, share & $b = 1$ & $b = 2$ & $b = 3$ & $b = 4$ & $b = 5$ & $b = 1$ & $b = 2$ & $b = 3$ & $b = 4$ & $b = 5$ \\ \midrule
\texttt{banknote} & $0.485$ & $1.000$ & $0.383$ & $0.582$ & $0.683$ & $0.748$ & $0.802$ & $0.485$ & $0.486$ & $0.486$ & $0.486$ & $0.487$ \\
\texttt{california} & $0.820$ & $1.000$ & $0.171$ & $0.248$ & $0.305$ & $0.346$ & $0.373$ & $0.784$ & $0.772$ & $0.763$ & $0.755$ & $0.749$ \\
\texttt{climate-model} & $0.800$ & $1.000$ & $0.162$ & $0.301$ & $0.397$ & $0.431$ & $0.449$ & $0.776$ & $0.761$ & $0.744$ & $0.739$ & $0.733$ \\
\texttt{codrna} & $0.693$ & $1.000$ & $0.159$ & $0.284$ & $0.406$ & $0.485$ & $0.544$ & $0.664$ & $0.651$ & $0.640$ & $0.635$ & $0.634$ \\
\texttt{covertype} & $0.935$ & $1.000$ & $0.093$ & $0.122$ & $0.142$ & $0.152$ & $0.159$ & $0.898$ & $0.886$ & $0.880$ & $0.878$ & $0.877$ \\
\texttt{credit} & $0.956$ & $1.000$ & $0.013$ & $0.027$ & $0.055$ & $0.077$ & $0.086$ & $0.951$ & $0.947$ & $0.937$ & $0.928$ & $0.919$ \\
\texttt{creditcard} & $0.587$ & $1.000$ & $0.451$ & $0.663$ & $0.749$ & $0.805$ & $0.856$ & $0.551$ & $0.542$ & $0.537$ & $0.535$ & $0.534$ \\
\texttt{delta\_ailerons} & $0.787$ & $1.000$ & $0.059$ & $0.120$ & $0.209$ & $0.289$ & $0.339$ & $0.776$ & $0.770$ & $0.763$ & $0.754$ & $0.745$ \\
\texttt{eeg-eye-state} & $0.693$ & $1.000$ & $0.504$ & $0.620$ & $0.657$ & $0.682$ & $0.707$ & $0.635$ & $0.633$ & $0.638$ & $0.641$ & $0.644$ \\
\texttt{electrical\_grid} & $0.671$ & $1.000$ & $0.310$ & $0.452$ & $0.570$ & $0.617$ & $0.638$ & $0.645$ & $0.637$ & $0.630$ & $0.627$ & $0.628$ \\
\texttt{electricity} & $0.922$ & $1.000$ & $0.063$ & $0.093$ & $0.118$ & $0.138$ & $0.153$ & $0.901$ & $0.893$ & $0.885$ & $0.880$ & $0.876$ \\
\texttt{elevators} & $0.828$ & $0.976$ & $0.072$ & $0.124$ & $0.195$ & $0.255$ & $0.293$ & $0.816$ & $0.809$ & $0.795$ & $0.779$ & $0.769$ \\
\texttt{hazelnut-spread} & $0.702$ & $1.000$ & $0.460$ & $0.577$ & $0.615$ & $0.627$ & $0.636$ & $0.643$ & $0.640$ & $0.639$ & $0.640$ & $0.642$ \\
\texttt{hill-valley} & $0.790$ & $0.979$ & $0.234$ & $0.318$ & $0.382$ & $0.421$ & $0.448$ & $0.744$ & $0.737$ & $0.733$ & $0.736$ & $0.736$ \\
\texttt{house\_16H} & $0.848$ & $1.000$ & $0.104$ & $0.181$ & $0.255$ & $0.291$ & $0.315$ & $0.816$ & $0.797$ & $0.775$ & $0.775$ & $0.776$ \\
\texttt{jannis} & $0.955$ & $1.000$ & $0.057$ & $0.089$ & $0.119$ & $0.131$ & $0.135$ & $0.924$ & $0.906$ & $0.891$ & $0.894$ & $0.893$ \\
\texttt{kin8nm} & $0.782$ & $1.000$ & $0.191$ & $0.288$ & $0.367$ & $0.422$ & $0.442$ & $0.756$ & $0.745$ & $0.734$ & $0.726$ & $0.720$ \\
\texttt{MagicTelescope} & $0.850$ & $1.000$ & $0.091$ & $0.163$ & $0.241$ & $0.278$ & $0.294$ & $0.824$ & $0.811$ & $0.796$ & $0.788$ & $0.793$ \\
\texttt{mammography} & $0.851$ & $0.971$ & $0.095$ & $0.173$ & $0.255$ & $0.299$ & $0.326$ & $0.837$ & $0.830$ & $0.817$ & $0.807$ & $0.806$ \\
\texttt{MiniBooNE} & $0.783$ & $1.000$ & $0.250$ & $0.356$ & $0.404$ & $0.424$ & $0.436$ & $0.738$ & $0.724$ & $0.718$ & $0.724$ & $0.728$ \\
\texttt{ozone-level-8hr} & $0.932$ & $0.871$ & $0.154$ & $0.209$ & $0.231$ & $0.240$ & $0.245$ & $0.877$ & $0.858$ & $0.853$ & $0.856$ & $0.857$ \\
\texttt{phoneme} & $0.836$ & $1.000$ & $0.303$ & $0.360$ & $0.385$ & $0.405$ & $0.421$ & $0.751$ & $0.748$ & $0.748$ & $0.748$ & $0.747$ \\
\texttt{polish\_companies} & $0.799$ & $1.000$ & $0.267$ & $0.366$ & $0.423$ & $0.451$ & $0.466$ & $0.758$ & $0.751$ & $0.748$ & $0.748$ & $0.750$ \\
\texttt{Pulsar-Dataset} & $0.699$ & $1.000$ & $0.090$ & $0.199$ & $0.376$ & $0.472$ & $0.536$ & $0.693$ & $0.689$ & $0.684$ & $0.677$ & $0.671$ \\
\texttt{puma8NH} & $0.905$ & $1.000$ & $0.010$ & $0.020$ & $0.034$ & $0.064$ & $0.104$ & $0.903$ & $0.901$ & $0.899$ & $0.897$ & $0.894$ \\
\texttt{spambase} & $0.744$ & $1.000$ & $0.312$ & $0.399$ & $0.443$ & $0.477$ & $0.503$ & $0.699$ & $0.701$ & $0.702$ & $0.702$ & $0.702$ \\
\texttt{stock} & $0.704$ & $1.000$ & $0.289$ & $0.371$ & $0.417$ & $0.451$ & $0.481$ & $0.690$ & $0.686$ & $0.682$ & $0.680$ & $0.679$ \\
\texttt{wdbc} & $0.661$ & $1.000$ & $0.240$ & $0.417$ & $0.546$ & $0.606$ & $0.646$ & $0.630$ & $0.621$ & $0.615$ & $0.613$ & $0.611$ \\
\texttt{wilt} & $0.702$ & $1.000$ & $0.084$ & $0.198$ & $0.360$ & $0.474$ & $0.544$ & $0.691$ & $0.684$ & $0.671$ & $0.655$ & $0.650$ \\
\texttt{wine} & $0.912$ & $0.950$ & $0.194$ & $0.215$ & $0.224$ & $0.234$ & $0.242$ & $0.817$ & $0.821$ & $0.822$ & $0.820$ & $0.819$ \\
\midrule mean $\pm$ SD & $0.788 \pm 0.113$ & $0.992 \pm 0.025$ & $0.195 \pm 0.135$ & $0.284 \pm 0.173$ & $0.352 \pm 0.185$ & $0.393 \pm 0.193$ & $0.421 \pm 0.202$ & $0.756 \pm 0.112$ & $0.748 \pm 0.110$ & $0.741 \pm 0.108$ & $0.737 \pm 0.107$ & $0.736 \pm 0.107$ \\
\bottomrule\end{tabular}}
\end{table}
\end{landscape}
\begin{landscape}
\subsection{Movement of the target row}\label{app:m_bewegung}
\begin{table}[H]\centering\scriptsize\caption{TabICL, the movement of the target row: $\cos(z, z_0)$ between its embedding under and without attack (1: no movement), for the flipped and the held target rows and in the control arm, per dataset and $b$.}\label{tab:app_d_zz}\centering\scriptsize\maxbreite{\begin{tabular}{@{}lccccccccccccccc@{}}\toprule
Dataset & \multicolumn{5}{c}{$\cos(z, z_0)$, flipped} & \multicolumn{5}{c}{$\cos(z, z_0)$, held} & \multicolumn{5}{c}{$\cos(z, z_0)$, control arm} \\ \cmidrule(lr){2-6}\cmidrule(lr){7-11}\cmidrule(lr){12-16}
 & $b = 1$ & $b = 2$ & $b = 3$ & $b = 4$ & $b = 5$ & $b = 1$ & $b = 2$ & $b = 3$ & $b = 4$ & $b = 5$ & $b = 1$ & $b = 2$ & $b = 3$ & $b = 4$ & $b = 5$ \\ \midrule
\texttt{banknote} & $0.507$ & $0.406$ & $0.354$ & $0.322$ & $0.304$ & $0.585$ & $0.622$ & $0.594$ & $0.571$ & $0.568$ & $0.995$ & $0.991$ & $0.986$ & $0.982$ & $0.978$ \\
\texttt{california} & $0.843$ & $0.760$ & $0.676$ & $0.613$ & $0.580$ & $0.830$ & $0.733$ & -- & -- & -- & $0.976$ & $0.959$ & $0.932$ & $0.897$ & $0.868$ \\
\texttt{climate-model} & $0.907$ & $0.653$ & $0.471$ & $0.451$ & $0.446$ & $0.856$ & -- & -- & -- & -- & $0.991$ & $0.974$ & $0.927$ & $0.919$ & $0.900$ \\
\texttt{codrna} & $0.796$ & $0.667$ & $0.526$ & $0.461$ & $0.425$ & $0.832$ & $0.690$ & -- & -- & -- & $0.975$ & $0.946$ & $0.899$ & $0.868$ & $0.857$ \\
\texttt{covertype} & $0.831$ & $0.750$ & $0.664$ & $0.632$ & $0.622$ & $0.827$ & -- & -- & -- & -- & $0.944$ & $0.897$ & $0.829$ & $0.783$ & $0.777$ \\
\texttt{credit} & $0.997$ & $0.916$ & $0.774$ & $0.691$ & $0.690$ & $0.986$ & $0.970$ & $0.942$ & -- & -- & $0.996$ & $0.989$ & $0.934$ & $0.839$ & $0.790$ \\
\texttt{creditcard} & $0.670$ & $0.496$ & $0.434$ & $0.402$ & $0.382$ & $0.726$ & $0.664$ & -- & -- & -- & $0.967$ & $0.950$ & $0.937$ & $0.927$ & $0.920$ \\
\texttt{delta\_ailerons} & $0.980$ & $0.897$ & $0.778$ & $0.693$ & $0.648$ & $0.963$ & $0.917$ & $0.853$ & -- & -- & $0.997$ & $0.992$ & $0.985$ & $0.966$ & $0.943$ \\
\texttt{eeg-eye-state} & $0.532$ & $0.458$ & $0.412$ & $0.383$ & $0.366$ & -- & -- & -- & -- & -- & $0.920$ & $0.886$ & $0.853$ & $0.835$ & $0.829$ \\
\texttt{electrical\_grid} & $0.726$ & $0.597$ & $0.452$ & $0.328$ & $0.281$ & $0.668$ & -- & -- & -- & -- & $0.985$ & $0.970$ & $0.942$ & $0.876$ & $0.798$ \\
\texttt{electricity} & $0.904$ & $0.862$ & $0.800$ & $0.761$ & $0.736$ & $0.937$ & $0.907$ & $0.911$ & -- & -- & $0.978$ & $0.961$ & $0.936$ & $0.913$ & $0.899$ \\
\texttt{elevators} & $0.915$ & $0.794$ & $0.685$ & $0.560$ & $0.523$ & $0.927$ & $0.906$ & $0.893$ & -- & -- & $0.992$ & $0.983$ & $0.944$ & $0.855$ & $0.797$ \\
\texttt{hazelnut-spread} & $0.568$ & $0.466$ & $0.409$ & $0.395$ & $0.394$ & $0.601$ & -- & -- & -- & -- & $0.921$ & $0.884$ & $0.851$ & $0.837$ & $0.835$ \\
\texttt{hill-valley} & $0.744$ & $0.648$ & $0.621$ & $0.579$ & $0.555$ & $0.887$ & $0.925$ & $0.923$ & $0.923$ & $0.922$ & $0.952$ & $0.936$ & $0.916$ & $0.896$ & $0.882$ \\
\texttt{house\_16H} & $0.914$ & $0.788$ & $0.612$ & $0.577$ & $0.565$ & $0.889$ & $0.807$ & -- & -- & -- & $0.979$ & $0.942$ & $0.832$ & $0.833$ & $0.853$ \\
\texttt{jannis} & $0.915$ & $0.782$ & $0.559$ & $0.537$ & $0.561$ & $0.908$ & -- & -- & -- & -- & $0.966$ & $0.890$ & $0.714$ & $0.735$ & $0.760$ \\
\texttt{kin8nm} & $0.843$ & $0.732$ & $0.608$ & $0.455$ & $0.392$ & $0.775$ & $0.659$ & -- & -- & -- & $0.986$ & $0.972$ & $0.943$ & $0.894$ & $0.852$ \\
\texttt{MagicTelescope} & $0.877$ & $0.785$ & $0.635$ & $0.568$ & $0.561$ & $0.919$ & $0.896$ & -- & -- & -- & $0.981$ & $0.959$ & $0.887$ & $0.837$ & $0.849$ \\
\texttt{mammography} & $0.925$ & $0.801$ & $0.636$ & $0.554$ & $0.525$ & $0.927$ & $0.891$ & $0.967$ & $0.975$ & $0.968$ & $0.991$ & $0.981$ & $0.956$ & $0.917$ & $0.908$ \\
\texttt{MiniBooNE} & $0.752$ & $0.607$ & $0.503$ & $0.501$ & $0.512$ & $0.792$ & -- & -- & -- & -- & $0.949$ & $0.906$ & $0.864$ & $0.882$ & $0.893$ \\
\texttt{ozone-level-8hr} & $0.737$ & $0.621$ & $0.537$ & $0.525$ & $0.520$ & -- & -- & -- & -- & -- & $0.922$ & $0.838$ & $0.791$ & $0.799$ & $0.800$ \\
\texttt{phoneme} & $0.622$ & $0.571$ & $0.525$ & $0.498$ & $0.482$ & $0.722$ & -- & -- & -- & -- & $0.879$ & $0.844$ & $0.812$ & $0.793$ & $0.785$ \\
\texttt{polish\_companies} & $0.724$ & $0.640$ & $0.572$ & $0.540$ & $0.526$ & $0.782$ & $0.703$ & -- & -- & -- & $0.941$ & $0.922$ & $0.907$ & $0.900$ & $0.896$ \\
\texttt{Pulsar-Dataset} & -- & $0.779$ & $0.610$ & $0.521$ & $0.480$ & $0.930$ & $0.837$ & -- & -- & -- & $0.998$ & $0.995$ & $0.988$ & $0.972$ & $0.955$ \\
\texttt{puma8NH} & -- & $0.995$ & $0.906$ & $0.755$ & $0.676$ & $0.995$ & $0.987$ & $0.963$ & $0.929$ & $0.977$ & $1.000$ & $0.999$ & $0.998$ & $0.996$ & $0.985$ \\
\texttt{spambase} & $0.636$ & $0.578$ & $0.555$ & $0.539$ & $0.535$ & $0.815$ & $0.823$ & $0.794$ & $0.779$ & $0.802$ & $0.918$ & $0.910$ & $0.905$ & $0.904$ & $0.904$ \\
\texttt{stock} & $0.728$ & $0.666$ & $0.602$ & $0.558$ & $0.526$ & $0.796$ & -- & -- & -- & -- & $0.989$ & $0.975$ & $0.956$ & $0.939$ & $0.927$ \\
\texttt{wdbc} & $0.771$ & $0.562$ & $0.426$ & $0.386$ & $0.367$ & $0.743$ & $0.653$ & -- & -- & -- & $0.980$ & $0.965$ & $0.947$ & $0.942$ & $0.934$ \\
\texttt{wilt} & -- & $0.822$ & $0.654$ & $0.543$ & $0.495$ & $0.945$ & $0.849$ & $0.732$ & -- & -- & $0.996$ & $0.991$ & $0.972$ & $0.936$ & $0.913$ \\
\texttt{wine} & $0.618$ & $0.594$ & $0.588$ & $0.582$ & $0.573$ & $0.739$ & $0.716$ & $0.659$ & -- & -- & $0.773$ & $0.759$ & $0.749$ & $0.743$ & $0.738$ \\
\midrule mean $\pm$ SD & $0.777 \pm 0.137$ & $0.690 \pm 0.144$ & $0.586 \pm 0.127$ & $0.530 \pm 0.112$ & $0.508 \pm 0.111$ & $0.832 \pm 0.111$ & $0.808 \pm 0.117$ & $0.839 \pm 0.128$ & $0.836 \pm 0.165$ & $0.847 \pm 0.171$ & $0.961 \pm 0.047$ & $0.939 \pm 0.056$ & $0.903 \pm 0.072$ & $0.880 \pm 0.067$ & $0.867 \pm 0.066$ \\
\bottomrule\end{tabular}}
\end{table}
\end{landscape}
\begin{landscape}
\subsection{Movement of the class centroids}\label{app:m_zentroide}
\begin{table}[H]\centering\scriptsize\caption{TabICL, the movement of the class centroids under attack: $\cos(\mu_T, \mu_{T,0})$ and $\cos(\mu_F, \mu_{F,0})$, mean over all attacked target rows, per dataset and $b$.}\label{tab:app_d_zentroide}\centering\scriptsize\maxbreite{\begin{tabular}{@{}lcccccccccc@{}}\toprule
Dataset & \multicolumn{5}{c}{$\cos(\mu_T, \mu_{T,0})$} & \multicolumn{5}{c}{$\cos(\mu_F, \mu_{F,0})$} \\ \cmidrule(lr){2-6}\cmidrule(lr){7-11}
 & $b = 1$ & $b = 2$ & $b = 3$ & $b = 4$ & $b = 5$ & $b = 1$ & $b = 2$ & $b = 3$ & $b = 4$ & $b = 5$ \\ \midrule
\texttt{banknote} & $0.9890$ & $0.9896$ & $0.9901$ & $0.9905$ & $0.9908$ & $0.9962$ & $0.9958$ & $0.9956$ & $0.9955$ & $0.9955$ \\
\texttt{california} & $1.0000$ & $1.0000$ & $0.9999$ & $0.9999$ & $0.9999$ & $1.0000$ & $1.0000$ & $1.0000$ & $0.9999$ & $0.9999$ \\
\texttt{climate-model} & $0.9985$ & $0.9977$ & $0.9970$ & $0.9967$ & $0.9966$ & $0.9990$ & $0.9978$ & $0.9966$ & $0.9962$ & $0.9959$ \\
\texttt{codrna} & $0.9996$ & $0.9994$ & $0.9994$ & $0.9993$ & $0.9993$ & $0.9999$ & $0.9998$ & $0.9997$ & $0.9996$ & $0.9996$ \\
\texttt{covertype} & $1.0000$ & $1.0000$ & $0.9999$ & $0.9998$ & $0.9997$ & $1.0000$ & $1.0000$ & $0.9999$ & $0.9998$ & $0.9998$ \\
\texttt{credit} & $1.0000$ & $1.0000$ & $0.9998$ & $0.9995$ & $0.9993$ & $1.0000$ & $1.0000$ & $0.9998$ & $0.9996$ & $0.9994$ \\
\texttt{creditcard} & $0.9914$ & $0.9892$ & $0.9895$ & $0.9900$ & $0.9905$ & $0.9867$ & $0.9789$ & $0.9753$ & $0.9724$ & $0.9704$ \\
\texttt{delta\_ailerons} & $1.0000$ & $0.9999$ & $0.9997$ & $0.9993$ & $0.9990$ & $1.0000$ & $1.0000$ & $0.9997$ & $0.9994$ & $0.9992$ \\
\texttt{eeg-eye-state} & $0.9999$ & $1.0000$ & $0.9999$ & $0.9999$ & $0.9999$ & $0.9999$ & $1.0000$ & $1.0000$ & $0.9999$ & $0.9999$ \\
\texttt{electrical\_grid} & $0.9988$ & $0.9992$ & $0.9994$ & $0.9993$ & $0.9992$ & $0.9993$ & $0.9997$ & $0.9997$ & $0.9995$ & $0.9994$ \\
\texttt{electricity} & $1.0000$ & $1.0000$ & $0.9999$ & $0.9999$ & $0.9998$ & $1.0000$ & $1.0000$ & $0.9999$ & $0.9999$ & $0.9998$ \\
\texttt{elevators} & $0.9996$ & $0.9995$ & $0.9995$ & $0.9994$ & $0.9992$ & $1.0000$ & $0.9999$ & $0.9998$ & $0.9996$ & $0.9994$ \\
\texttt{hazelnut-spread} & $0.9998$ & $0.9999$ & $0.9999$ & $0.9998$ & $0.9998$ & $1.0000$ & $1.0000$ & $0.9999$ & $0.9999$ & $0.9999$ \\
\texttt{hill-valley} & $0.9994$ & $0.9993$ & $0.9993$ & $0.9991$ & $0.9990$ & $0.9995$ & $0.9994$ & $0.9993$ & $0.9992$ & $0.9990$ \\
\texttt{house\_16H} & $1.0000$ & $0.9999$ & $0.9999$ & $0.9998$ & $0.9997$ & $1.0000$ & $1.0000$ & $0.9999$ & $0.9998$ & $0.9997$ \\
\texttt{jannis} & $1.0000$ & $1.0000$ & $0.9999$ & $0.9999$ & $0.9998$ & $1.0000$ & $1.0000$ & $0.9999$ & $0.9999$ & $0.9998$ \\
\texttt{kin8nm} & $0.9998$ & $0.9999$ & $0.9999$ & $0.9998$ & $0.9997$ & $0.9999$ & $0.9999$ & $0.9999$ & $0.9998$ & $0.9997$ \\
\texttt{MagicTelescope} & $1.0000$ & $0.9999$ & $0.9998$ & $0.9997$ & $0.9997$ & $1.0000$ & $1.0000$ & $0.9999$ & $0.9998$ & $0.9997$ \\
\texttt{mammography} & $0.9993$ & $0.9992$ & $0.9988$ & $0.9986$ & $0.9985$ & $0.9996$ & $0.9993$ & $0.9987$ & $0.9983$ & $0.9980$ \\
\texttt{MiniBooNE} & $1.0000$ & $0.9999$ & $0.9999$ & $0.9998$ & $0.9998$ & $1.0000$ & $0.9999$ & $0.9999$ & $0.9999$ & $0.9998$ \\
\texttt{ozone-level-8hr} & $0.9999$ & $0.9998$ & $0.9997$ & $0.9996$ & $0.9996$ & $0.9999$ & $0.9998$ & $0.9995$ & $0.9994$ & $0.9994$ \\
\texttt{phoneme} & $1.0000$ & $1.0000$ & $0.9999$ & $0.9999$ & $0.9999$ & $1.0000$ & $1.0000$ & $1.0000$ & $0.9999$ & $0.9999$ \\
\texttt{polish\_companies} & $0.9999$ & $0.9998$ & $0.9998$ & $0.9997$ & $0.9997$ & $1.0000$ & $0.9999$ & $0.9998$ & $0.9997$ & $0.9996$ \\
\texttt{Pulsar-Dataset} & $0.9994$ & $0.9988$ & $0.9978$ & $0.9974$ & $0.9973$ & $0.9999$ & $0.9997$ & $0.9989$ & $0.9986$ & $0.9985$ \\
\texttt{puma8NH} & $1.0000$ & $1.0000$ & $0.9999$ & $0.9992$ & $0.9980$ & $1.0000$ & $1.0000$ & $0.9999$ & $0.9993$ & $0.9982$ \\
\texttt{spambase} & $0.9999$ & $0.9998$ & $0.9998$ & $0.9998$ & $0.9998$ & $0.9999$ & $0.9999$ & $0.9999$ & $0.9999$ & $0.9999$ \\
\texttt{stock} & $0.9986$ & $0.9989$ & $0.9990$ & $0.9990$ & $0.9990$ & $0.9995$ & $0.9997$ & $0.9997$ & $0.9997$ & $0.9996$ \\
\texttt{wdbc} & $0.9941$ & $0.9943$ & $0.9946$ & $0.9946$ & $0.9946$ & $0.9968$ & $0.9966$ & $0.9964$ & $0.9962$ & $0.9962$ \\
\texttt{wilt} & $0.9991$ & $0.9985$ & $0.9979$ & $0.9971$ & $0.9966$ & $0.9998$ & $0.9995$ & $0.9986$ & $0.9977$ & $0.9973$ \\
\texttt{wine} & $1.0000$ & $1.0000$ & $1.0000$ & $0.9999$ & $0.9999$ & $1.0000$ & $1.0000$ & $1.0000$ & $1.0000$ & $1.0000$ \\
\midrule mean $\pm$ SD & $0.9989 \pm 0.0026$ & $0.9987 \pm 0.0028$ & $0.9987 \pm 0.0027$ & $0.9985 \pm 0.0026$ & $0.9985 \pm 0.0025$ & $0.9992 \pm 0.0025$ & $0.9988 \pm 0.0039$ & $0.9985 \pm 0.0045$ & $0.9983 \pm 0.0050$ & $0.9981 \pm 0.0054$ \\
\bottomrule\end{tabular}}
\end{table}
\end{landscape}

\subsection{Neighbourhood of the target row}\label{app:kopp}
\begin{table}[H]\centering\scriptsize\caption{TabICL, the neighbourhood of the attacked target row under attack: $\nu_F$ counts the rows with the wrong label among its 25 nearest genuine context rows by cosine at the input of the MLP (copies left out), mean over the members; majority: $\nu_F > 12.5$. Per $b$: flipped target rows with and without such a majority, held ones with it, and the AUC of $\nu_F$ for flipped against held target rows (datasets with both groups).}\label{tab:app_kopp}\centering\scriptsize\maxbreite{\begin{tabular}{@{}rcccc@{}}\toprule
$b$ & \multicolumn{2}{c}{flipped} & held & AUC \\
& majority & no majority & majority & of $\nu_F$ \\ \midrule
1 & 643 & 47 & 92 & $0.97 \pm 0.03$ \\
2 & 966 & 63 & 50 & $0.96 \pm 0.05$ \\
3 & 1164 & 66 & 11 & $0.96 \pm 0.09$ \\
4 & 1221 & 53 & 8 & $0.94 \pm 0.14$ \\
5 & 1249 & 44 & 1 & $0.92 \pm 0.18$ \\
\bottomrule\end{tabular}}
\end{table}

\subsection{Neighbourhood of the target row per dataset}\label{app:kopp_ds}
\begin{table}[H]\centering\scriptsize\caption{TabICL at one poison row, per dataset: the mean of $\nu_F$ (Table~\ref{tab:app_kopp}) and the share of target rows with $\nu_F > 12.5$, without attack, under attack for the held and the flipped target rows, and in the control arm; $n$: number of target rows, summed in the last row.}\label{tab:app_kopp_ds_icl}\centering\scriptsize\maxbreite{\begin{tabular}{@{}lccccccccccc@{}}\toprule
& & \multicolumn{2}{c}{without attack} & \multicolumn{3}{c}{held} & \multicolumn{3}{c}{flipped} & \multicolumn{2}{c}{control arm} \\ \cmidrule(lr){3-4}\cmidrule(lr){5-7}\cmidrule(lr){8-10}\cmidrule(lr){11-12}
Dataset & $n$ & $\nu_F$ & majority & $n$ & $\nu_F$ & majority & $n$ & $\nu_F$ & majority & $\nu_F$ & majority \\ \midrule
\texttt{banknote} & 50 & $0.0$ & $0.00$ & 29 & $0.1$ & $0.00$ & 21 & $10.2$ & $0.38$ & $0.0$ & $0.00$ \\
\texttt{california} & 44 & $1.0$ & $0.02$ & 18 & $9.2$ & $0.28$ & 26 & $22.0$ & $1.00$ & $0.0$ & $0.00$ \\
\texttt{climate-model} & 40 & $2.5$ & $0.05$ & 30 & $5.5$ & $0.20$ & 10 & $18.0$ & $0.70$ & $1.4$ & $0.00$ \\
\texttt{codrna} & 50 & $0.2$ & $0.00$ & 33 & $7.3$ & $0.24$ & 17 & $19.0$ & $0.88$ & $0.0$ & $0.00$ \\
\texttt{covertype} & 44 & $1.8$ & $0.02$ & 2 & $9.5$ & $0.50$ & 42 & $20.4$ & $1.00$ & $0.5$ & $0.00$ \\
\texttt{credit} & 40 & $1.7$ & $0.03$ & 39 & $6.6$ & $0.21$ & 1 & $17.0$ & $1.00$ & $0.7$ & $0.00$ \\
\texttt{creditcard} & 29 & $3.0$ & $0.14$ & 13 & $0.2$ & $0.00$ & 16 & $7.9$ & $0.25$ & $3.0$ & $0.14$ \\
\texttt{delta\_ailerons} & 49 & $0.4$ & $0.00$ & 45 & $1.6$ & $0.00$ & 4 & $17.3$ & $1.00$ & $0.1$ & $0.00$ \\
\texttt{eeg-eye-state} & 49 & $0.0$ & $0.00$ & 0 & -- & -- & 49 & $24.5$ & $1.00$ & $0.0$ & $0.00$ \\
\texttt{electrical\_grid} & 47 & $0.3$ & $0.00$ & 14 & $8.5$ & $0.29$ & 33 & $22.1$ & $1.00$ & $0.0$ & $0.00$ \\
\texttt{electricity} & 40 & $1.4$ & $0.00$ & 20 & $6.5$ & $0.20$ & 20 & $21.0$ & $1.00$ & $0.3$ & $0.00$ \\
\texttt{elevators} & 42 & $1.7$ & $0.05$ & 36 & $5.8$ & $0.22$ & 6 & $24.2$ & $1.00$ & $1.0$ & $0.02$ \\
\texttt{hazelnut-spread} & 48 & $0.0$ & $0.00$ & 1 & $14.1$ & $1.00$ & 47 & $24.8$ & $1.00$ & $0.0$ & $0.00$ \\
\texttt{hill-valley} & 47 & $1.8$ & $0.02$ & 18 & $7.7$ & $0.06$ & 29 & $19.7$ & $0.86$ & $1.1$ & $0.00$ \\
\texttt{house\_16H} & 45 & $1.0$ & $0.02$ & 28 & $10.1$ & $0.32$ & 17 & $23.8$ & $1.00$ & $0.0$ & $0.00$ \\
\texttt{jannis} & 32 & $2.1$ & $0.06$ & 9 & $15.1$ & $0.67$ & 23 & $23.8$ & $1.00$ & $0.0$ & $0.00$ \\
\texttt{kin8nm} & 48 & $2.4$ & $0.04$ & 17 & $7.7$ & $0.29$ & 31 & $17.2$ & $0.97$ & $1.0$ & $0.00$ \\
\texttt{MagicTelescope} & 47 & $0.5$ & $0.00$ & 35 & $8.0$ & $0.31$ & 12 & $23.2$ & $1.00$ & $0.0$ & $0.00$ \\
\texttt{mammography} & 35 & $0.9$ & $0.00$ & 28 & $2.0$ & $0.04$ & 7 & $21.7$ & $1.00$ & $0.7$ & $0.00$ \\
\texttt{MiniBooNE} & 45 & $0.3$ & $0.00$ & 4 & $20.2$ & $1.00$ & 41 & $24.7$ & $1.00$ & $0.0$ & $0.00$ \\
\texttt{ozone-level-8hr} & 31 & $2.4$ & $0.10$ & 0 & -- & -- & 31 & $22.2$ & $1.00$ & $0.0$ & $0.00$ \\
\texttt{phoneme} & 43 & $1.1$ & $0.02$ & 1 & $11.0$ & $0.00$ & 42 & $24.3$ & $1.00$ & $0.0$ & $0.00$ \\
\texttt{polish\_companies} & 43 & $0.7$ & $0.00$ & 15 & $7.6$ & $0.33$ & 28 & $23.5$ & $1.00$ & $0.1$ & $0.00$ \\
\texttt{Pulsar-Dataset} & 46 & $0.0$ & $0.00$ & 46 & $1.1$ & $0.00$ & 0 & -- & -- & $0.0$ & $0.00$ \\
\texttt{puma8NH} & 42 & $1.4$ & $0.02$ & 42 & $2.7$ & $0.05$ & 0 & -- & -- & $1.0$ & $0.00$ \\
\texttt{spambase} & 47 & $0.3$ & $0.00$ & 13 & $3.6$ & $0.08$ & 34 & $24.6$ & $1.00$ & $0.0$ & $0.00$ \\
\texttt{stock} & 49 & $0.2$ & $0.00$ & 4 & $6.8$ & $0.00$ & 45 & $18.8$ & $0.84$ & $0.0$ & $0.00$ \\
\texttt{wdbc} & 49 & $0.5$ & $0.00$ & 28 & $2.6$ & $0.00$ & 21 & $17.6$ & $0.76$ & $0.0$ & $0.00$ \\
\texttt{wilt} & 46 & $0.4$ & $0.00$ & 46 & $2.6$ & $0.04$ & 0 & -- & -- & $0.1$ & $0.00$ \\
\texttt{wine} & 40 & $2.0$ & $0.05$ & 3 & $0.5$ & $0.00$ & 37 & $24.5$ & $1.00$ & $0.0$ & $0.00$ \\
\midrule mean $\pm$ SD & 1307 & $1.1 \pm 0.9$ & $0.02 \pm 0.03$ & 617 & $6.6 \pm 4.8$ & $0.23 \pm 0.28$ & 690 & $20.7 \pm 4.3$ & $0.91 \pm 0.19$ & $0.4 \pm 0.6$ & $0.01 \pm 0.03$ \\
\bottomrule\end{tabular}}
\end{table}

\begin{landscape}
\section{Inside TabPFN}\label{app:pfn}
\subsection{The vote at one poison row}\label{app:pfn_b}
\begin{table}[H]\centering\scriptsize\caption{TabPFN at one poison row, the vote per dataset as in Figure~\ref{fig:vote}, split into parts that sum to 1. Without attack: true label ($a_{T,0}$) and wrong label ($a_{F,0}$). Control arm: true label of genuine rows ($a_{T,G}$) and of control rows ($a_{T,C}$), wrong label ($a_{F,G}$). Under attack, for the flipped and the held target rows: true label ($a_T$), wrong label of genuine rows ($a_{F,G}$) and of poison rows ($a_{F,P}$), and $a_{F,0}$ of the same target rows without attack. $\pi_F = a_F$ holds per target row up to \zPFNidentMax.}\label{tab:app_e_vote_1}\centering\scriptsize\maxbreite{\begin{tabular}{@{}lccccccccccccc@{}}\toprule
Dataset & \multicolumn{2}{c}{without attack} & \multicolumn{3}{c}{control arm} & \multicolumn{4}{c}{under attack, flipped} & \multicolumn{4}{c}{under attack, held} \\ \cmidrule(lr){2-3}\cmidrule(lr){4-6}\cmidrule(lr){7-10}\cmidrule(lr){11-14}
 & $a_{T,0}$ & $a_{F,0}$ & $a_{T,G}$ & $a_{T,C}$ & $a_{F,G}$ & $a_{F,0}$ & $a_T$ & $a_{F,G}$ & $a_{F,P}$ & $a_{F,0}$ & $a_T$ & $a_{F,G}$ & $a_{F,P}$ \\ \midrule
\texttt{banknote} & $1.000$ & $0.000$ & $0.893$ & $0.107$ & $0.000$ & $0.000$ & $0.246$ & $0.117$ & $0.637$ & $0.000$ & $0.684$ & $0.028$ & $0.288$ \\
\texttt{california} & $0.921$ & $0.079$ & $0.876$ & $0.102$ & $0.022$ & $0.126$ & $0.318$ & $0.445$ & $0.237$ & $0.005$ & $0.664$ & $0.145$ & $0.192$ \\
\texttt{climate-model} & $0.888$ & $0.112$ & $0.875$ & $0.103$ & $0.022$ & $0.114$ & $0.192$ & $0.438$ & $0.370$ & $0.077$ & $0.540$ & $0.304$ & $0.156$ \\
\texttt{codrna} & $0.983$ & $0.017$ & $0.913$ & $0.085$ & $0.002$ & $0.021$ & $0.289$ & $0.507$ & $0.204$ & $0.004$ & $0.609$ & $0.226$ & $0.165$ \\
\texttt{covertype} & $0.849$ & $0.151$ & $0.771$ & $0.210$ & $0.019$ & $0.151$ & $0.147$ & $0.527$ & $0.326$ & -- & -- & -- & -- \\
\texttt{credit} & $0.794$ & $0.206$ & $0.796$ & $0.016$ & $0.188$ & $0.408$ & $0.471$ & $0.478$ & $0.051$ & $0.172$ & $0.766$ & $0.214$ & $0.019$ \\
\texttt{creditcard} & $0.963$ & $0.037$ & $0.953$ & $0.035$ & $0.011$ & $0.058$ & $0.308$ & $0.311$ & $0.381$ & $0.000$ & $0.626$ & $0.129$ & $0.245$ \\
\texttt{delta\_ailerons} & $0.949$ & $0.051$ & $0.942$ & $0.018$ & $0.040$ & -- & -- & -- & -- & $0.051$ & $0.896$ & $0.085$ & $0.019$ \\
\texttt{eeg-eye-state} & $0.983$ & $0.017$ & $0.668$ & $0.331$ & $0.001$ & $0.017$ & $0.131$ & $0.325$ & $0.545$ & -- & -- & -- & -- \\
\texttt{electrical\_grid} & $0.968$ & $0.032$ & $0.867$ & $0.127$ & $0.006$ & $0.044$ & $0.227$ & $0.352$ & $0.421$ & $0.000$ & $0.722$ & $0.031$ & $0.247$ \\
\texttt{electricity} & $0.871$ & $0.129$ & $0.834$ & $0.131$ & $0.035$ & $0.145$ & $0.271$ & $0.544$ & $0.185$ & $0.023$ & $0.528$ & $0.317$ & $0.155$ \\
\texttt{elevators} & $0.925$ & $0.075$ & $0.893$ & $0.061$ & $0.046$ & $0.144$ & $0.339$ & $0.438$ & $0.224$ & $0.050$ & $0.750$ & $0.163$ & $0.087$ \\
\texttt{hazelnut-spread} & $0.980$ & $0.020$ & $0.711$ & $0.288$ & $0.000$ & $0.020$ & $0.047$ & $0.462$ & $0.492$ & -- & -- & -- & -- \\
\texttt{hill-valley} & $0.926$ & $0.074$ & $0.813$ & $0.173$ & $0.014$ & $0.074$ & $0.160$ & $0.510$ & $0.329$ & $0.047$ & $0.539$ & $0.335$ & $0.126$ \\
\texttt{house\_16H} & $0.902$ & $0.098$ & $0.841$ & $0.127$ & $0.032$ & $0.147$ & $0.310$ & $0.438$ & $0.252$ & $0.010$ & $0.632$ & $0.164$ & $0.204$ \\
\texttt{jannis} & $0.782$ & $0.218$ & $0.652$ & $0.327$ & $0.021$ & $0.218$ & $0.104$ & $0.489$ & $0.406$ & -- & -- & -- & -- \\
\texttt{kin8nm} & $0.951$ & $0.049$ & $0.893$ & $0.085$ & $0.023$ & $0.086$ & $0.348$ & $0.319$ & $0.333$ & $0.006$ & $0.679$ & $0.092$ & $0.229$ \\
\texttt{MagicTelescope} & $0.923$ & $0.077$ & $0.871$ & $0.095$ & $0.034$ & $0.088$ & $0.239$ & $0.516$ & $0.244$ & $0.067$ & $0.757$ & $0.174$ & $0.069$ \\
\texttt{mammography} & $0.912$ & $0.088$ & $0.906$ & $0.060$ & $0.034$ & $0.275$ & $0.272$ & $0.577$ & $0.151$ & $0.012$ & $0.821$ & $0.097$ & $0.082$ \\
\texttt{MiniBooNE} & $0.946$ & $0.054$ & $0.814$ & $0.182$ & $0.004$ & $0.058$ & $0.165$ & $0.454$ & $0.381$ & $0.001$ & $0.597$ & $0.129$ & $0.274$ \\
\texttt{ozone-level-8hr} & $0.871$ & $0.129$ & $0.822$ & $0.154$ & $0.024$ & $0.129$ & $0.221$ & $0.498$ & $0.281$ & -- & -- & -- & -- \\
\texttt{phoneme} & $0.928$ & $0.072$ & $0.785$ & $0.213$ & $0.002$ & $0.074$ & $0.086$ & $0.521$ & $0.393$ & $0.000$ & $0.592$ & $0.109$ & $0.300$ \\
\texttt{polish\_companies} & $0.960$ & $0.040$ & $0.852$ & $0.141$ & $0.007$ & $0.057$ & $0.212$ & $0.509$ & $0.279$ & $0.001$ & $0.789$ & $0.086$ & $0.125$ \\
\texttt{Pulsar-Dataset} & $0.989$ & $0.011$ & $0.970$ & $0.022$ & $0.008$ & -- & -- & -- & -- & $0.011$ & $0.912$ & $0.043$ & $0.046$ \\
\texttt{puma8NH} & $0.863$ & $0.137$ & $0.890$ & $0.015$ & $0.095$ & $0.294$ & $0.401$ & $0.556$ & $0.043$ & $0.088$ & $0.653$ & $0.273$ & $0.074$ \\
\texttt{spambase} & $0.965$ & $0.035$ & $0.855$ & $0.144$ & $0.001$ & $0.046$ & $0.093$ & $0.603$ & $0.304$ & $0.001$ & $0.863$ & $0.058$ & $0.079$ \\
\texttt{stock} & $0.978$ & $0.022$ & $0.938$ & $0.054$ & $0.008$ & $0.040$ & $0.361$ & $0.338$ & $0.301$ & $0.002$ & $0.620$ & $0.139$ & $0.241$ \\
\texttt{wdbc} & $0.962$ & $0.038$ & $0.900$ & $0.099$ & $0.001$ & $0.043$ & $0.164$ & $0.370$ & $0.466$ & $0.000$ & $0.579$ & $0.113$ & $0.308$ \\
\texttt{wilt} & $0.972$ & $0.028$ & $0.947$ & $0.035$ & $0.019$ & $0.049$ & $0.431$ & $0.320$ & $0.249$ & $0.028$ & $0.815$ & $0.087$ & $0.098$ \\
\texttt{wine} & $0.863$ & $0.137$ & $0.853$ & $0.145$ & $0.002$ & $0.149$ & $0.023$ & $0.772$ & $0.205$ & $0.001$ & $0.641$ & $0.135$ & $0.223$ \\
\midrule mean $\pm$ SD & $0.926 \pm 0.057$ & $0.074 \pm 0.057$ & $0.853 \pm 0.078$ & $0.123 \pm 0.086$ & $0.024 \pm 0.037$ & $0.110 \pm 0.094$ & $0.235 \pm 0.116$ & $0.455 \pm 0.122$ & $0.310 \pm 0.135$ & $0.026 \pm 0.041$ & $0.691 \pm 0.113$ & $0.147 \pm 0.087$ & $0.162 \pm 0.090$ \\
\bottomrule\end{tabular}}
\end{table}
\end{landscape}
\begin{landscape}
\subsection{The vote at two poison rows}\label{app:pfn_vote_2}
\begin{table}[H]\centering\scriptsize\caption{TabPFN at two poison rows, as in Table~\ref{tab:app_e_vote_1}.}\label{tab:app_e_vote_2}\centering\scriptsize\maxbreite{\begin{tabular}{@{}lccccccccccccc@{}}\toprule
Dataset & \multicolumn{2}{c}{without attack} & \multicolumn{3}{c}{control arm} & \multicolumn{4}{c}{under attack, flipped} & \multicolumn{4}{c}{under attack, held} \\ \cmidrule(lr){2-3}\cmidrule(lr){4-6}\cmidrule(lr){7-10}\cmidrule(lr){11-14}
 & $a_{T,0}$ & $a_{F,0}$ & $a_{T,G}$ & $a_{T,C}$ & $a_{F,G}$ & $a_{F,0}$ & $a_T$ & $a_{F,G}$ & $a_{F,P}$ & $a_{F,0}$ & $a_T$ & $a_{F,G}$ & $a_{F,P}$ \\ \midrule
\texttt{banknote} & $1.000$ & $0.000$ & $0.813$ & $0.187$ & $0.000$ & $0.000$ & $0.101$ & $0.144$ & $0.756$ & $0.000$ & $0.679$ & $0.038$ & $0.283$ \\
\texttt{california} & $0.921$ & $0.079$ & $0.779$ & $0.211$ & $0.011$ & $0.079$ & $0.159$ & $0.358$ & $0.483$ & -- & -- & -- & -- \\
\texttt{climate-model} & $0.888$ & $0.112$ & $0.414$ & $0.582$ & $0.004$ & $0.112$ & $0.008$ & $0.305$ & $0.687$ & -- & -- & -- & -- \\
\texttt{codrna} & $0.983$ & $0.017$ & $0.723$ & $0.276$ & $0.001$ & $0.017$ & $0.071$ & $0.369$ & $0.560$ & -- & -- & -- & -- \\
\texttt{covertype} & $0.849$ & $0.151$ & $0.535$ & $0.453$ & $0.012$ & $0.151$ & $0.069$ & $0.357$ & $0.574$ & -- & -- & -- & -- \\
\texttt{credit} & $0.794$ & $0.206$ & $0.779$ & $0.062$ & $0.159$ & $0.246$ & $0.322$ & $0.366$ & $0.312$ & $0.181$ & $0.713$ & $0.250$ & $0.037$ \\
\texttt{creditcard} & $0.963$ & $0.037$ & $0.928$ & $0.067$ & $0.006$ & $0.037$ & $0.093$ & $0.271$ & $0.636$ & -- & -- & -- & -- \\
\texttt{delta\_ailerons} & $0.949$ & $0.051$ & $0.922$ & $0.046$ & $0.032$ & $0.081$ & $0.186$ & $0.215$ & $0.599$ & $0.042$ & $0.791$ & $0.119$ & $0.091$ \\
\texttt{eeg-eye-state} & $0.983$ & $0.017$ & $0.518$ & $0.481$ & $0.001$ & $0.017$ & $0.051$ & $0.310$ & $0.639$ & -- & -- & -- & -- \\
\texttt{electrical\_grid} & $0.968$ & $0.032$ & $0.660$ & $0.337$ & $0.003$ & $0.032$ & $0.035$ & $0.302$ & $0.664$ & -- & -- & -- & -- \\
\texttt{electricity} & $0.871$ & $0.129$ & $0.709$ & $0.270$ & $0.021$ & $0.129$ & $0.146$ & $0.486$ & $0.367$ & -- & -- & -- & -- \\
\texttt{elevators} & $0.925$ & $0.075$ & $0.756$ & $0.220$ & $0.024$ & $0.070$ & $0.151$ & $0.331$ & $0.518$ & $0.142$ & $0.588$ & $0.306$ & $0.106$ \\
\texttt{hazelnut-spread} & $0.980$ & $0.020$ & $0.483$ & $0.517$ & $0.000$ & $0.020$ & $0.009$ & $0.435$ & $0.555$ & -- & -- & -- & -- \\
\texttt{hill-valley} & $0.926$ & $0.074$ & $0.634$ & $0.354$ & $0.011$ & $0.074$ & $0.059$ & $0.516$ & $0.425$ & -- & -- & -- & -- \\
\texttt{house\_16H} & $0.902$ & $0.098$ & $0.550$ & $0.444$ & $0.006$ & $0.098$ & $0.038$ & $0.275$ & $0.687$ & -- & -- & -- & -- \\
\texttt{jannis} & $0.782$ & $0.218$ & $0.317$ & $0.680$ & $0.003$ & $0.218$ & $0.004$ & $0.288$ & $0.709$ & -- & -- & -- & -- \\
\texttt{kin8nm} & $0.951$ & $0.049$ & $0.725$ & $0.265$ & $0.010$ & $0.049$ & $0.081$ & $0.231$ & $0.688$ & -- & -- & -- & -- \\
\texttt{MagicTelescope} & $0.923$ & $0.077$ & $0.724$ & $0.261$ & $0.016$ & $0.078$ & $0.124$ & $0.436$ & $0.441$ & $0.070$ & $0.620$ & $0.198$ & $0.182$ \\
\texttt{mammography} & $0.912$ & $0.088$ & $0.874$ & $0.105$ & $0.021$ & $0.133$ & $0.257$ & $0.418$ & $0.325$ & $0.002$ & $0.839$ & $0.058$ & $0.103$ \\
\texttt{MiniBooNE} & $0.946$ & $0.054$ & $0.578$ & $0.421$ & $0.001$ & $0.054$ & $0.005$ & $0.407$ & $0.588$ & -- & -- & -- & -- \\
\texttt{ozone-level-8hr} & $0.871$ & $0.129$ & $0.580$ & $0.411$ & $0.009$ & $0.129$ & $0.044$ & $0.394$ & $0.562$ & -- & -- & -- & -- \\
\texttt{phoneme} & $0.928$ & $0.072$ & $0.576$ & $0.423$ & $0.001$ & $0.072$ & $0.045$ & $0.374$ & $0.582$ & -- & -- & -- & -- \\
\texttt{polish\_companies} & $0.960$ & $0.040$ & $0.750$ & $0.249$ & $0.002$ & $0.040$ & $0.087$ & $0.539$ & $0.375$ & -- & -- & -- & -- \\
\texttt{Pulsar-Dataset} & $0.989$ & $0.011$ & $0.950$ & $0.044$ & $0.006$ & $0.011$ & $0.247$ & $0.124$ & $0.629$ & $0.011$ & $0.695$ & $0.097$ & $0.208$ \\
\texttt{puma8NH} & $0.863$ & $0.137$ & $0.724$ & $0.234$ & $0.043$ & $0.137$ & $0.125$ & $0.285$ & $0.590$ & -- & -- & -- & -- \\
\texttt{spambase} & $0.965$ & $0.035$ & $0.730$ & $0.269$ & $0.001$ & $0.044$ & $0.040$ & $0.478$ & $0.481$ & $0.001$ & $0.756$ & $0.066$ & $0.177$ \\
\texttt{stock} & $0.978$ & $0.022$ & $0.888$ & $0.108$ & $0.005$ & $0.022$ & $0.222$ & $0.331$ & $0.447$ & -- & -- & -- & -- \\
\texttt{wdbc} & $0.962$ & $0.038$ & $0.661$ & $0.339$ & $0.000$ & $0.038$ & $0.014$ & $0.342$ & $0.645$ & -- & -- & -- & -- \\
\texttt{wilt} & $0.972$ & $0.028$ & $0.915$ & $0.071$ & $0.013$ & $0.024$ & $0.255$ & $0.199$ & $0.546$ & $0.040$ & $0.611$ & $0.137$ & $0.252$ \\
\texttt{wine} & $0.863$ & $0.137$ & $0.529$ & $0.470$ & $0.001$ & $0.141$ & $0.028$ & $0.517$ & $0.455$ & $0.002$ & $0.628$ & $0.064$ & $0.308$ \\
\midrule mean $\pm$ SD & $0.926 \pm 0.057$ & $0.074 \pm 0.057$ & $0.691 \pm 0.161$ & $0.295 \pm 0.169$ & $0.014 \pm 0.029$ & $0.078 \pm 0.061$ & $0.102 \pm 0.088$ & $0.347 \pm 0.106$ & $0.551 \pm 0.118$ & $0.049 \pm 0.064$ & $0.692 \pm 0.084$ & $0.133 \pm 0.090$ & $0.175 \pm 0.090$ \\
\bottomrule\end{tabular}}
\end{table}
\end{landscape}
\begin{landscape}
\subsection{The vote at three poison rows}\label{app:pfn_vote_3}
\begin{table}[H]\centering\scriptsize\caption{TabPFN at three poison rows, as in Table~\ref{tab:app_e_vote_1}.}\label{tab:app_e_vote_3}\centering\scriptsize\maxbreite{\begin{tabular}{@{}lccccccccccccc@{}}\toprule
Dataset & \multicolumn{2}{c}{without attack} & \multicolumn{3}{c}{control arm} & \multicolumn{4}{c}{under attack, flipped} & \multicolumn{4}{c}{under attack, held} \\ \cmidrule(lr){2-3}\cmidrule(lr){4-6}\cmidrule(lr){7-10}\cmidrule(lr){11-14}
 & $a_{T,0}$ & $a_{F,0}$ & $a_{T,G}$ & $a_{T,C}$ & $a_{F,G}$ & $a_{F,0}$ & $a_T$ & $a_{F,G}$ & $a_{F,P}$ & $a_{F,0}$ & $a_T$ & $a_{F,G}$ & $a_{F,P}$ \\ \midrule
\texttt{banknote} & $1.000$ & $0.000$ & $0.732$ & $0.268$ & $0.000$ & $0.000$ & $0.043$ & $0.153$ & $0.804$ & $0.000$ & $0.603$ & $0.034$ & $0.363$ \\
\texttt{california} & $0.921$ & $0.079$ & $0.644$ & $0.351$ & $0.005$ & $0.079$ & $0.061$ & $0.258$ & $0.681$ & -- & -- & -- & -- \\
\texttt{climate-model} & $0.888$ & $0.112$ & $0.100$ & $0.900$ & $0.000$ & $0.112$ & $0.001$ & $0.109$ & $0.891$ & -- & -- & -- & -- \\
\texttt{codrna} & $0.983$ & $0.017$ & $0.550$ & $0.449$ & $0.001$ & $0.017$ & $0.022$ & $0.206$ & $0.772$ & -- & -- & -- & -- \\
\texttt{covertype} & $0.849$ & $0.151$ & $0.396$ & $0.595$ & $0.009$ & $0.151$ & $0.040$ & $0.265$ & $0.695$ & -- & -- & -- & -- \\
\texttt{credit} & $0.794$ & $0.206$ & $0.679$ & $0.205$ & $0.116$ & $0.211$ & $0.194$ & $0.249$ & $0.557$ & $0.194$ & $0.656$ & $0.271$ & $0.073$ \\
\texttt{creditcard} & $0.963$ & $0.037$ & $0.889$ & $0.107$ & $0.004$ & $0.037$ & $0.012$ & $0.176$ & $0.812$ & -- & -- & -- & -- \\
\texttt{delta\_ailerons} & $0.949$ & $0.051$ & $0.889$ & $0.087$ & $0.023$ & $0.056$ & $0.188$ & $0.158$ & $0.653$ & $0.026$ & $0.637$ & $0.138$ & $0.224$ \\
\texttt{eeg-eye-state} & $0.983$ & $0.017$ & $0.416$ & $0.584$ & $0.000$ & $0.017$ & $0.021$ & $0.313$ & $0.666$ & -- & -- & -- & -- \\
\texttt{electrical\_grid} & $0.968$ & $0.032$ & $0.349$ & $0.651$ & $0.001$ & $0.032$ & $0.005$ & $0.151$ & $0.844$ & -- & -- & -- & -- \\
\texttt{electricity} & $0.871$ & $0.129$ & $0.590$ & $0.395$ & $0.015$ & $0.129$ & $0.081$ & $0.410$ & $0.509$ & -- & -- & -- & -- \\
\texttt{elevators} & $0.925$ & $0.075$ & $0.478$ & $0.514$ & $0.008$ & $0.075$ & $0.027$ & $0.196$ & $0.777$ & -- & -- & -- & -- \\
\texttt{hazelnut-spread} & $0.980$ & $0.020$ & $0.346$ & $0.654$ & $0.000$ & $0.020$ & $0.002$ & $0.399$ & $0.599$ & -- & -- & -- & -- \\
\texttt{hill-valley} & $0.926$ & $0.074$ & $0.403$ & $0.588$ & $0.009$ & $0.074$ & $0.025$ & $0.356$ & $0.619$ & -- & -- & -- & -- \\
\texttt{house\_16H} & $0.902$ & $0.098$ & $0.265$ & $0.734$ & $0.002$ & $0.098$ & $0.003$ & $0.099$ & $0.899$ & -- & -- & -- & -- \\
\texttt{jannis} & $0.782$ & $0.218$ & $0.138$ & $0.862$ & $0.000$ & $0.218$ & $0.000$ & $0.113$ & $0.886$ & -- & -- & -- & -- \\
\texttt{kin8nm} & $0.951$ & $0.049$ & $0.346$ & $0.653$ & $0.001$ & $0.049$ & $0.006$ & $0.099$ & $0.896$ & -- & -- & -- & -- \\
\texttt{MagicTelescope} & $0.923$ & $0.077$ & $0.516$ & $0.475$ & $0.009$ & $0.077$ & $0.028$ & $0.248$ & $0.724$ & -- & -- & -- & -- \\
\texttt{mammography} & $0.912$ & $0.088$ & $0.836$ & $0.152$ & $0.013$ & $0.104$ & $0.128$ & $0.319$ & $0.554$ & $0.003$ & $0.914$ & $0.025$ & $0.061$ \\
\texttt{MiniBooNE} & $0.946$ & $0.054$ & $0.313$ & $0.687$ & $0.000$ & $0.054$ & $0.001$ & $0.243$ & $0.757$ & -- & -- & -- & -- \\
\texttt{ozone-level-8hr} & $0.871$ & $0.129$ & $0.419$ & $0.577$ & $0.005$ & $0.129$ & $0.018$ & $0.263$ & $0.719$ & -- & -- & -- & -- \\
\texttt{phoneme} & $0.928$ & $0.072$ & $0.434$ & $0.566$ & $0.001$ & $0.072$ & $0.030$ & $0.298$ & $0.672$ & -- & -- & -- & -- \\
\texttt{polish\_companies} & $0.960$ & $0.040$ & $0.652$ & $0.347$ & $0.001$ & $0.040$ & $0.018$ & $0.541$ & $0.440$ & -- & -- & -- & -- \\
\texttt{Pulsar-Dataset} & $0.989$ & $0.011$ & $0.915$ & $0.081$ & $0.004$ & $0.011$ & $0.059$ & $0.068$ & $0.873$ & -- & -- & -- & -- \\
\texttt{puma8NH} & $0.863$ & $0.137$ & $0.317$ & $0.670$ & $0.013$ & $0.137$ & $0.004$ & $0.032$ & $0.964$ & -- & -- & -- & -- \\
\texttt{spambase} & $0.965$ & $0.035$ & $0.662$ & $0.338$ & $0.001$ & $0.040$ & $0.049$ & $0.416$ & $0.535$ & $0.001$ & $0.763$ & $0.050$ & $0.187$ \\
\texttt{stock} & $0.978$ & $0.022$ & $0.840$ & $0.157$ & $0.003$ & $0.022$ & $0.106$ & $0.333$ & $0.561$ & -- & -- & -- & -- \\
\texttt{wdbc} & $0.962$ & $0.038$ & $0.474$ & $0.526$ & $0.000$ & $0.038$ & $0.002$ & $0.178$ & $0.820$ & -- & -- & -- & -- \\
\texttt{wilt} & $0.972$ & $0.028$ & $0.823$ & $0.170$ & $0.007$ & $0.028$ & $0.052$ & $0.164$ & $0.785$ & -- & -- & -- & -- \\
\texttt{wine} & $0.863$ & $0.137$ & $0.396$ & $0.603$ & $0.001$ & $0.141$ & $0.019$ & $0.427$ & $0.553$ & $0.002$ & $0.508$ & $0.070$ & $0.422$ \\
\midrule mean $\pm$ SD & $0.926 \pm 0.057$ & $0.074 \pm 0.057$ & $0.527 \pm 0.227$ & $0.465 \pm 0.232$ & $0.008 \pm 0.021$ & $0.076 \pm 0.057$ & $0.041 \pm 0.051$ & $0.241 \pm 0.122$ & $0.717 \pm 0.138$ & $0.038 \pm 0.077$ & $0.680 \pm 0.141$ & $0.098 \pm 0.094$ & $0.222 \pm 0.148$ \\
\bottomrule\end{tabular}}
\end{table}
\end{landscape}
\begin{landscape}
\subsection{The strongest row}\label{app:rang_pfn}
\begin{table}[H]\centering\scriptsize\caption{TabPFN, the share of target rows for which one of the poison rows carries the largest weight in every ensemble member, for all attacked, the flipped and the held target rows, per dataset and $b$.}\label{tab:app_e_staerkste}\centering\scriptsize\maxbreite{\begin{tabular}{@{}lccccccccccccccc@{}}\toprule
Dataset & \multicolumn{5}{c}{share strongest, all attacked} & \multicolumn{5}{c}{share strongest, flipped} & \multicolumn{5}{c}{share strongest, held} \\ \cmidrule(lr){2-6}\cmidrule(lr){7-11}\cmidrule(lr){12-16}
 & $b = 1$ & $b = 2$ & $b = 3$ & $b = 4$ & $b = 5$ & $b = 1$ & $b = 2$ & $b = 3$ & $b = 4$ & $b = 5$ & $b = 1$ & $b = 2$ & $b = 3$ & $b = 4$ & $b = 5$ \\ \midrule
\texttt{banknote} & $0.92$ & $0.96$ & $0.98$ & $0.96$ & $0.94$ & $1.00$ & $1.00$ & $1.00$ & $1.00$ & $0.96$ & $0.75$ & $0.50$ & $0.67$ & $0.33$ & $0.00$ \\
\texttt{california} & $0.95$ & $0.98$ & $0.95$ & $0.98$ & $0.98$ & $1.00$ & $0.98$ & $0.95$ & $0.98$ & $0.98$ & $0.88$ & -- & -- & -- & -- \\
\texttt{climate-model} & $0.88$ & $0.75$ & $0.93$ & $0.97$ & $1.00$ & $0.89$ & $0.75$ & $0.93$ & $0.97$ & $1.00$ & $0.50$ & -- & -- & -- & -- \\
\texttt{codrna} & $0.44$ & $0.80$ & $0.90$ & $0.96$ & $1.00$ & $0.43$ & $0.80$ & $0.90$ & $0.96$ & $1.00$ & $0.46$ & -- & -- & -- & -- \\
\texttt{covertype} & $0.74$ & $1.00$ & $1.00$ & $1.00$ & $0.98$ & $0.74$ & $1.00$ & $1.00$ & $1.00$ & $0.98$ & -- & -- & -- & -- & -- \\
\texttt{credit} & $0.49$ & $0.63$ & $0.83$ & $0.93$ & $0.98$ & $0.83$ & $0.88$ & $1.00$ & $0.95$ & $0.98$ & $0.43$ & $0.48$ & $0.36$ & $0.00$ & -- \\
\texttt{creditcard} & $0.96$ & $0.96$ & $0.81$ & $0.59$ & $0.44$ & $0.94$ & $0.96$ & $0.81$ & $0.59$ & $0.44$ & $1.00$ & -- & -- & -- & -- \\
\texttt{delta\_ailerons} & $0.19$ & $0.49$ & $0.94$ & $1.00$ & $1.00$ & -- & $0.91$ & $1.00$ & $1.00$ & $1.00$ & $0.19$ & $0.36$ & $0.62$ & -- & -- \\
\texttt{eeg-eye-state} & $1.00$ & $1.00$ & $0.86$ & $0.78$ & $0.78$ & $1.00$ & $1.00$ & $0.86$ & $0.78$ & $0.78$ & -- & -- & -- & -- & -- \\
\texttt{electrical\_grid} & $0.96$ & $0.92$ & $0.98$ & $1.00$ & $0.98$ & $1.00$ & $0.92$ & $0.98$ & $1.00$ & $0.98$ & $0.85$ & -- & -- & -- & -- \\
\texttt{electricity} & $0.73$ & $0.89$ & $0.93$ & $0.96$ & $0.98$ & $0.74$ & $0.89$ & $0.93$ & $0.96$ & $0.98$ & $0.67$ & -- & -- & -- & -- \\
\texttt{elevators} & $0.86$ & $0.79$ & $0.83$ & $0.90$ & $0.95$ & $1.00$ & $0.77$ & $0.83$ & $0.90$ & $0.95$ & $0.81$ & $1.00$ & -- & -- & -- \\
\texttt{hazelnut-spread} & $0.96$ & $0.69$ & $0.39$ & $0.35$ & $0.37$ & $0.96$ & $0.69$ & $0.39$ & $0.35$ & $0.37$ & -- & -- & -- & -- & -- \\
\texttt{hill-valley} & $0.94$ & $0.84$ & $0.86$ & $0.92$ & $1.00$ & $0.94$ & $0.84$ & $0.86$ & $0.92$ & $1.00$ & $1.00$ & -- & -- & -- & -- \\
\texttt{house\_16H} & $1.00$ & $0.96$ & $0.96$ & $0.98$ & $0.98$ & $1.00$ & $0.96$ & $0.96$ & $0.98$ & $0.98$ & $1.00$ & -- & -- & -- & -- \\
\texttt{jannis} & $0.97$ & $0.62$ & $0.88$ & $0.94$ & $0.94$ & $0.97$ & $0.62$ & $0.88$ & $0.94$ & $0.94$ & -- & -- & -- & -- & -- \\
\texttt{kin8nm} & $0.96$ & $0.96$ & $0.98$ & $0.98$ & $0.98$ & $1.00$ & $0.96$ & $0.98$ & $0.98$ & $0.98$ & $0.91$ & -- & -- & -- & -- \\
\texttt{MagicTelescope} & $0.60$ & $0.72$ & $0.74$ & $0.94$ & $0.89$ & $0.78$ & $0.68$ & $0.74$ & $0.94$ & $0.89$ & $0.42$ & $1.00$ & -- & -- & -- \\
\texttt{mammography} & $0.66$ & $0.76$ & $0.82$ & $0.84$ & $0.84$ & $0.82$ & $0.88$ & $0.94$ & $0.97$ & $0.97$ & $0.59$ & $0.54$ & $0.17$ & $0.00$ & $0.00$ \\
\texttt{MiniBooNE} & $0.96$ & $0.65$ & $0.67$ & $0.65$ & $0.61$ & $0.95$ & $0.65$ & $0.67$ & $0.65$ & $0.61$ & $1.00$ & -- & -- & -- & -- \\
\texttt{ozone-level-8hr} & $1.00$ & $0.71$ & $0.71$ & $0.77$ & $0.87$ & $1.00$ & $0.71$ & $0.71$ & $0.77$ & $0.87$ & -- & -- & -- & -- & -- \\
\texttt{phoneme} & $0.76$ & $0.93$ & $0.95$ & $0.95$ & $0.93$ & $0.76$ & $0.93$ & $0.95$ & $0.95$ & $0.93$ & $1.00$ & -- & -- & -- & -- \\
\texttt{polish\_companies} & $0.81$ & $0.74$ & $0.47$ & $0.32$ & $0.30$ & $0.94$ & $0.74$ & $0.47$ & $0.32$ & $0.30$ & $0.50$ & -- & -- & -- & -- \\
\texttt{Pulsar-Dataset} & $0.35$ & $0.89$ & $1.00$ & $1.00$ & $1.00$ & -- & $1.00$ & $1.00$ & $1.00$ & $1.00$ & $0.35$ & $0.74$ & -- & -- & -- \\
\texttt{puma8NH} & $0.21$ & $0.81$ & $1.00$ & $1.00$ & $1.00$ & $0.00$ & $0.81$ & $1.00$ & $1.00$ & $1.00$ & $0.28$ & -- & -- & -- & -- \\
\texttt{spambase} & $0.40$ & $0.30$ & $0.23$ & $0.28$ & $0.30$ & $0.51$ & $0.38$ & $0.28$ & $0.32$ & $0.34$ & $0.08$ & $0.00$ & $0.00$ & $0.00$ & $0.00$ \\
\texttt{stock} & $0.98$ & $1.00$ & $0.98$ & $0.94$ & $0.96$ & $0.96$ & $1.00$ & $0.98$ & $0.94$ & $0.96$ & $1.00$ & -- & -- & -- & -- \\
\texttt{wdbc} & $0.98$ & $0.85$ & $0.98$ & $1.00$ & $0.98$ & $0.98$ & $0.85$ & $0.98$ & $1.00$ & $0.98$ & $1.00$ & -- & -- & -- & -- \\
\texttt{wilt} & $0.37$ & $0.89$ & $0.96$ & $0.91$ & $0.93$ & $1.00$ & $0.91$ & $0.96$ & $0.91$ & $0.93$ & $0.36$ & $0.83$ & -- & -- & -- \\
\texttt{wine} & $0.13$ & $0.51$ & $0.69$ & $0.72$ & $0.62$ & $0.14$ & $0.53$ & $0.71$ & $0.72$ & $0.62$ & $0.00$ & $0.00$ & $0.00$ & -- & -- \\
\midrule mean $\pm$ SD & $0.74 \pm 0.28$ & $0.80 \pm 0.17$ & $0.84 \pm 0.19$ & $0.85 \pm 0.21$ & $0.85 \pm 0.22$ & $0.83 \pm 0.26$ & $0.83 \pm 0.16$ & $0.85 \pm 0.19$ & $0.86 \pm 0.21$ & $0.86 \pm 0.22$ & $0.64 \pm 0.32$ & $0.54 \pm 0.36$ & $0.30 \pm 0.30$ & $0.08$ & $0.00$ \\
\bottomrule\end{tabular}}
\end{table}
\end{landscape}
\section{Tests}\label{app:tests}
\noindent The tests of Table~\ref{tab:tests}, at every number of poison rows (Section~\ref{sec:statistik}). With Holm over all 25 tests instead, only test (3) at $b = 2$ changes, to $p = \zHolmGesamtAPZwei$.

\begin{table}[H]
\caption{The tests of the study. Each test is run at every number of poison rows $b$ from one to five, over the datasets that have target rows of the given group.}
\label{tab:tests}
\centering\scriptsize
\maxbreite{\begin{tabular}{@{}lllll@{}}\toprule
Test & Model & Quantity per dataset & Target rows & Comparison \\ \midrule
(1a) & TabICL & share of flipped target rows & all attacked & attack arm against control arm \\
(1b) & TabPFN & share of flipped target rows & all attacked & attack arm against control arm \\
(2a) & TabICL & relative closeness $m$ & flipped & against 1 \\
(2b) & TabICL & relative closeness $m$ & held & against 1 \\
(3) & TabPFN & weight $a_{F,P}$ of the poison rows & flipped & against 0.5 \\
\bottomrule\end{tabular}}
\end{table}

\begin{table}[H]
\caption{Tests (1a) and (1b): per TFM the ASR under attack against the share of flipped target rows in the control arm, over the same attacked target rows, per $b$: means over the \zDS\ datasets, higher/equal/lower, and $p$ of the paired Wilcoxon signed-rank test over the datasets. Values per dataset in Table~\ref{tab:app_asr_ds_b}.}
\label{tab:test_a}
\centering\footnotesize\setlength{\tabcolsep}{4pt}\begin{tabular}{@{}ccccccc@{}}\toprule
 & \multicolumn{3}{c}{TabICL} & \multicolumn{3}{c}{TabPFN} \\ \cmidrule(lr){2-4}\cmidrule(l){5-7}
$b$ & attack / control & higher/equal/lower & $p$ & attack / control & higher/equal/lower & $p$ \\ \midrule
1 & 0.530 / 0.000 & 27/3/0 & $<0.001$ & 0.651 / 0.000 & 28/2/0 & $<0.001$ \\
2 & 0.788 / 0.000 & 30/0/0 & $<0.001$ & 0.903 / 0.000 & 30/0/0 & $<0.001$ \\
3 & 0.940 / 0.000 & 30/0/0 & $<0.001$ & 0.972 / 0.000 & 30/0/0 & $<0.001$ \\
4 & 0.974 / 0.001 & 30/0/0 & $<0.001$ & 0.989 / 0.000 & 30/0/0 & $<0.001$ \\
5 & 0.989 / 0.000 & 30/0/0 & $<0.001$ & 0.991 / 0.000 & 30/0/0 & $<0.001$ \\
\bottomrule\end{tabular}

\end{table}

\begin{table}[H]
\caption{Tests (2a) and (2b): TabICL, $m - 1$ for the flipped and for the held target rows, per $b$: mean over the datasets with the group (DS), higher/equal/lower than 0, and $p$ of the Wilcoxon signed-rank test of $m$ against 1. Values per dataset in Table~\ref{tab:app_m_liste_m}.}
\label{tab:test_d}
\centering\footnotesize\setlength{\tabcolsep}{5pt}\begin{tabular}{@{}ccccccccc@{}}\toprule
 & \multicolumn{4}{c}{flipped, Test (2a)} & \multicolumn{4}{c}{held, Test (2b)} \\ \cmidrule(lr){2-5}\cmidrule(l){6-9}
$b$ & $m - 1$ & DS & higher/equal/lower & $p$ & $m - 1$ & DS & higher/equal/lower & $p$ \\ \midrule
1 & $+0.050$ & 27 & 24/0/3 & $<0.001$ & $-0.100$ & 28 & 0/0/28 & $<0.001$ \\
2 & $+0.095$ & 30 & 27/0/3 & $<0.001$ & $-0.084$ & 20 & 0/0/20 & $<0.001$ \\
3 & $+0.148$ & 30 & 29/0/1 & $<0.001$ & $-0.063$ & 11 & 1/0/10 & $0.009$ \\
4 & $+0.186$ & 30 & 29/0/1 & $<0.001$ & $-0.069$ & 5 & 0/0/5 & $0.125$ \\
5 & $+0.212$ & 30 & 30/0/0 & $<0.001$ & $-0.052$ & 5 & 1/0/4 & $0.188$ \\
\bottomrule\end{tabular}

\end{table}

\begin{table}[H]
\caption{Test (3): TabPFN, $a_{F,P} - 0.5$, the weight of the poison rows at the flipped target rows minus one half, per $b$: mean over the datasets with flipped target rows (DS), higher/equal/lower than 0, and $p$ of the Wilcoxon signed-rank test of $a_{F,P}$ against 0.5. Of the flipped target rows, the poison rows alone receive more than half of the weight for \zPFNaPueberZielEins\,\% at $b = 1$ and \zPFNaPueberZielDrei\,\% at $b = 3$. Values per dataset for $b \leq 3$ in Tables~\ref{tab:app_e_vote_1} to~\ref{tab:app_e_vote_3}.}
\label{tab:test_e}
\centering\footnotesize\begin{tabular}{@{}ccccc@{}}\toprule
$b$ & $a_{F,P} - 0.5$ & DS & higher/equal/lower & $p$ \\ \midrule
1 & $-0.190$ & 28 & 2/0/26 & $<0.001$ \\
2 & $+0.051$ & 30 & 20/0/10 & $0.022$ \\
3 & $+0.217$ & 30 & 29/0/1 & $<0.001$ \\
4 & $+0.309$ & 30 & 30/0/0 & $<0.001$ \\
5 & $+0.359$ & 30 & 30/0/0 & $<0.001$ \\
\bottomrule\end{tabular}

\end{table}

\section{Trained models}\label{app:referenz}
\noindent The random forest, CatBoost and the logistic regression are retrained on every context (Section~\ref{sec:modelle}). All three run at their defaults. CatBoost fits 1,000 trees of depth six with an L2 regularisation of 3, a learning rate set by the library and no early stopping. The random forest and the logistic regression use the defaults of scikit-learn: the random forest has 100 trees, each grown on a bootstrap sample until its leaves are pure, and the logistic regression has an L2 penalty with $C = 1$ and is fitted after mean imputation and standardisation on the context, with up to 5,000 iterations. Every fit converged. They are not part of the tests (Section~\ref{sec:statistik}).
\subsection{Performance without attack}\label{app:trainiert_guete}
\begin{table}[H]\centering\scriptsize\caption{The trained models without attack, per dataset: the AUC on the test rows, mean over the five folds, and the number of correctly predicted, that is attacked, target rows. Last rows: mean $\pm$ SD of the AUC and sums.}\label{tab:app_trainiert_guete}\maxbreite{\centering\scriptsize\begin{tabular}{@{}lcccccc@{}}\toprule
Dataset & \multicolumn{2}{c}{Random forest} & \multicolumn{2}{c}{CatBoost} & \multicolumn{2}{c}{Logistic regression} \\ \cmidrule(lr){2-3} \cmidrule(lr){4-5} \cmidrule(lr){6-7}
 & AUC & correct & AUC & correct & AUC & correct \\ \midrule
\texttt{banknote} & $1.000$ & 50 & $1.000$ & 50 & $1.000$ & 47 \\
\texttt{california} & $0.940$ & 44 & $0.955$ & 43 & $0.900$ & 42 \\
\texttt{climate-model} & $0.928$ & 31 & $0.951$ & 37 & $0.952$ & 40 \\
\texttt{codrna} & $0.983$ & 47 & $0.991$ & 48 & $0.988$ & 49 \\
\texttt{covertype} & $0.847$ & 42 & $0.849$ & 40 & $0.647$ & 33 \\
\texttt{credit} & $0.823$ & 43 & $0.835$ & 40 & $0.774$ & 37 \\
\texttt{creditcard} & $0.999$ & 25 & $0.995$ & 25 & $0.952$ & 25 \\
\texttt{delta\_ailerons} & $0.976$ & 48 & $0.977$ & 47 & $0.972$ & 46 \\
\texttt{eeg-eye-state} & $0.938$ & 45 & $0.942$ & 45 & $0.659$ & 27 \\
\texttt{electrical\_grid} & $0.967$ & 42 & $0.982$ & 46 & $0.892$ & 37 \\
\texttt{electricity} & $0.886$ & 41 & $0.891$ & 39 & $0.825$ & 40 \\
\texttt{elevators} & $0.867$ & 36 & $0.901$ & 37 & $0.934$ & 43 \\
\texttt{hazelnut-spread} & $0.960$ & 47 & $0.974$ & 46 & $0.952$ & 46 \\
\texttt{hill-valley} & $0.594$ & 23 & $0.546$ & 30 & $0.949$ & 34 \\
\texttt{house\_16H} & $0.943$ & 45 & $0.949$ & 45 & $0.898$ & 43 \\
\texttt{jannis} & $0.814$ & 34 & $0.829$ & 32 & $0.787$ & 36 \\
\texttt{kin8nm} & $0.903$ & 44 & $0.942$ & 44 & $0.803$ & 35 \\
\texttt{MagicTelescope} & $0.918$ & 44 & $0.921$ & 45 & $0.828$ & 41 \\
\texttt{mammography} & $0.921$ & 36 & $0.930$ & 35 & $0.871$ & 32 \\
\texttt{MiniBooNE} & $0.969$ & 47 & $0.977$ & 47 & $0.934$ & 47 \\
\texttt{ozone-level-8hr} & $0.910$ & 29 & $0.925$ & 28 & $0.915$ & 28 \\
\texttt{phoneme} & $0.948$ & 43 & $0.942$ & 44 & $0.807$ & 36 \\
\texttt{polish\_companies} & $0.923$ & 35 & $0.946$ & 38 & $0.800$ & 29 \\
\texttt{Pulsar-Dataset} & $0.964$ & 46 & $0.972$ & 46 & $0.976$ & 46 \\
\texttt{puma8NH} & $0.889$ & 43 & $0.907$ & 44 & $0.853$ & 42 \\
\texttt{spambase} & $0.984$ & 46 & $0.986$ & 48 & $0.970$ & 46 \\
\texttt{stock} & $0.997$ & 50 & $0.997$ & 50 & $0.934$ & 39 \\
\texttt{wdbc} & $0.991$ & 47 & $0.994$ & 47 & $0.995$ & 47 \\
\texttt{wilt} & $0.983$ & 44 & $0.991$ & 45 & $0.975$ & 27 \\
\texttt{wine} & $0.904$ & 38 & $0.889$ & 41 & $0.812$ & 39 \\
\midrule mean $\pm$ SD & $0.922 \pm 0.080$ &  & $0.930 \pm 0.087$ &  & $0.885 \pm 0.094$ &  \\
sum &  & 1,235 &  & 1,252 &  & 1,159 \\
\bottomrule\end{tabular}
}\end{table}
\begin{landscape}
\subsection{ASR per dataset and number of poison rows}\label{app:trainiert_asr}
\begin{table}[H]\centering\scriptsize\caption{ASR of the trained models per dataset and number of poison rows $b$. Last rows: mean $\pm$ SD and the number of datasets with ASR $\geq 0.5$ and $< 0.1$. In the control arm no target row flips for any of the three models. At $b = 1$, \zRFhaltenEins\ target rows hold for the random forest, all with an exact duplicate carrying the true label, and the logistic regression changes $\pi_T$ by a median of \zLRdeltaPiEins. Spearman's $\rho$ of the CatBoost column with TabICL and TabPFN (Table~\ref{tab:app_asr_ds_b}): \zRhoCATICLEins\ and \zRhoCATPFNEins\ at $b = 1$, \zRhoCATICLZwei\ and \zRhoCATPFNZwei\ at $b = 2$, \zRhoCATICLDrei\ and \zRhoCATPFNDrei\ at $b = 3$.}\label{tab:app_trainiert_asr}\centering\scriptsize\maxbreite{\begin{tabular}{@{}lccccccccccccccc@{}}\toprule
Dataset & \multicolumn{5}{c}{ASR, Random forest} & \multicolumn{5}{c}{ASR, CatBoost} & \multicolumn{5}{c}{ASR, Logistic regression} \\ \cmidrule(lr){2-6} \cmidrule(lr){7-11} \cmidrule(lr){12-16}
 & $b = 1$ & $b = 2$ & $b = 3$ & $b = 4$ & $b = 5$ & $b = 1$ & $b = 2$ & $b = 3$ & $b = 4$ & $b = 5$ & $b = 1$ & $b = 2$ & $b = 3$ & $b = 4$ & $b = 5$ \\ \midrule
\texttt{banknote} & $0.94$ & $0.94$ & $0.98$ & $1.00$ & $1.00$ & $0.94$ & $0.94$ & $0.94$ & $0.98$ & $1.00$ & $0.00$ & $0.00$ & $0.00$ & $0.00$ & $0.00$ \\
\texttt{california} & $1.00$ & $1.00$ & $1.00$ & $1.00$ & $1.00$ & $0.23$ & $0.67$ & $0.93$ & $0.98$ & $1.00$ & $0.00$ & $0.00$ & $0.00$ & $0.00$ & $0.02$ \\
\texttt{climate-model} & $1.00$ & $1.00$ & $1.00$ & $1.00$ & $1.00$ & $1.00$ & $1.00$ & $1.00$ & $1.00$ & $1.00$ & $0.05$ & $0.12$ & $0.17$ & $0.30$ & $0.35$ \\
\texttt{codrna} & $1.00$ & $1.00$ & $1.00$ & $1.00$ & $1.00$ & $0.27$ & $0.62$ & $0.96$ & $1.00$ & $1.00$ & $0.00$ & $0.00$ & $0.00$ & $0.00$ & $0.00$ \\
\texttt{covertype} & $1.00$ & $1.00$ & $1.00$ & $1.00$ & $1.00$ & $0.33$ & $0.65$ & $0.85$ & $0.97$ & $0.97$ & $0.00$ & $0.00$ & $0.03$ & $0.03$ & $0.03$ \\
\texttt{credit} & $1.00$ & $1.00$ & $1.00$ & $1.00$ & $1.00$ & $0.20$ & $0.62$ & $0.85$ & $0.95$ & $0.97$ & $0.03$ & $0.03$ & $0.03$ & $0.05$ & $0.05$ \\
\texttt{creditcard} & $1.00$ & $1.00$ & $1.00$ & $1.00$ & $1.00$ & $1.00$ & $1.00$ & $1.00$ & $1.00$ & $1.00$ & $0.04$ & $0.12$ & $0.28$ & $0.48$ & $0.56$ \\
\texttt{delta\_ailerons} & $1.00$ & $1.00$ & $1.00$ & $1.00$ & $1.00$ & $0.19$ & $0.45$ & $0.68$ & $0.87$ & $0.98$ & $0.02$ & $0.02$ & $0.02$ & $0.02$ & $0.02$ \\
\texttt{eeg-eye-state} & $1.00$ & $1.00$ & $1.00$ & $1.00$ & $1.00$ & $0.49$ & $0.93$ & $0.98$ & $1.00$ & $1.00$ & $0.07$ & $0.07$ & $0.07$ & $0.07$ & $0.07$ \\
\texttt{electrical\_grid} & $1.00$ & $1.00$ & $1.00$ & $1.00$ & $1.00$ & $0.59$ & $0.96$ & $1.00$ & $1.00$ & $1.00$ & $0.00$ & $0.00$ & $0.00$ & $0.00$ & $0.00$ \\
\texttt{electricity} & $1.00$ & $1.00$ & $1.00$ & $1.00$ & $1.00$ & $0.10$ & $0.36$ & $0.59$ & $0.77$ & $0.92$ & $0.00$ & $0.00$ & $0.00$ & $0.00$ & $0.00$ \\
\texttt{elevators} & $1.00$ & $1.00$ & $1.00$ & $1.00$ & $1.00$ & $0.35$ & $0.57$ & $0.89$ & $0.95$ & $0.95$ & $0.02$ & $0.05$ & $0.05$ & $0.05$ & $0.05$ \\
\texttt{hazelnut-spread} & $1.00$ & $1.00$ & $1.00$ & $1.00$ & $1.00$ & $0.52$ & $0.96$ & $1.00$ & $1.00$ & $1.00$ & $0.02$ & $0.02$ & $0.02$ & $0.02$ & $0.02$ \\
\texttt{hill-valley} & $1.00$ & $1.00$ & $1.00$ & $1.00$ & $1.00$ & $0.67$ & $0.80$ & $0.93$ & $1.00$ & $1.00$ & $0.06$ & $0.09$ & $0.12$ & $0.21$ & $0.24$ \\
\texttt{house\_16H} & $1.00$ & $1.00$ & $1.00$ & $1.00$ & $1.00$ & $0.33$ & $0.80$ & $0.98$ & $1.00$ & $1.00$ & $0.00$ & $0.00$ & $0.00$ & $0.00$ & $0.00$ \\
\texttt{jannis} & $1.00$ & $1.00$ & $1.00$ & $1.00$ & $1.00$ & $0.81$ & $1.00$ & $1.00$ & $1.00$ & $1.00$ & $0.03$ & $0.14$ & $0.14$ & $0.17$ & $0.22$ \\
\texttt{kin8nm} & $1.00$ & $1.00$ & $1.00$ & $1.00$ & $1.00$ & $0.43$ & $0.75$ & $0.98$ & $1.00$ & $1.00$ & $0.00$ & $0.00$ & $0.00$ & $0.00$ & $0.00$ \\
\texttt{MagicTelescope} & $1.00$ & $1.00$ & $1.00$ & $1.00$ & $1.00$ & $0.38$ & $0.80$ & $0.96$ & $1.00$ & $1.00$ & $0.00$ & $0.00$ & $0.00$ & $0.00$ & $0.00$ \\
\texttt{mammography} & $0.86$ & $0.86$ & $0.86$ & $0.86$ & $0.86$ & $0.43$ & $0.71$ & $0.86$ & $0.86$ & $0.86$ & $0.03$ & $0.03$ & $0.03$ & $0.03$ & $0.06$ \\
\texttt{MiniBooNE} & $1.00$ & $1.00$ & $1.00$ & $1.00$ & $1.00$ & $0.62$ & $1.00$ & $1.00$ & $1.00$ & $1.00$ & $0.04$ & $0.04$ & $0.04$ & $0.04$ & $0.04$ \\
\texttt{ozone-level-8hr} & $1.00$ & $1.00$ & $1.00$ & $1.00$ & $1.00$ & $0.68$ & $1.00$ & $1.00$ & $1.00$ & $1.00$ & $0.04$ & $0.14$ & $0.14$ & $0.29$ & $0.32$ \\
\texttt{phoneme} & $1.00$ & $1.00$ & $1.00$ & $1.00$ & $1.00$ & $0.23$ & $0.48$ & $0.86$ & $0.95$ & $1.00$ & $0.00$ & $0.00$ & $0.00$ & $0.00$ & $0.00$ \\
\texttt{polish\_companies} & $1.00$ & $1.00$ & $1.00$ & $1.00$ & $1.00$ & $0.71$ & $0.97$ & $1.00$ & $1.00$ & $1.00$ & $0.03$ & $0.03$ & $0.03$ & $0.07$ & $0.07$ \\
\texttt{Pulsar-Dataset} & $1.00$ & $1.00$ & $1.00$ & $1.00$ & $1.00$ & $0.59$ & $0.93$ & $1.00$ & $1.00$ & $1.00$ & $0.00$ & $0.00$ & $0.00$ & $0.00$ & $0.00$ \\
\texttt{puma8NH} & $1.00$ & $1.00$ & $1.00$ & $1.00$ & $1.00$ & $0.41$ & $0.75$ & $0.98$ & $1.00$ & $1.00$ & $0.00$ & $0.00$ & $0.00$ & $0.00$ & $0.00$ \\
\texttt{spambase} & $0.87$ & $0.91$ & $0.91$ & $0.91$ & $0.91$ & $0.40$ & $0.65$ & $0.88$ & $0.92$ & $0.92$ & $0.04$ & $0.04$ & $0.04$ & $0.11$ & $0.11$ \\
\texttt{stock} & $1.00$ & $1.00$ & $1.00$ & $1.00$ & $1.00$ & $0.66$ & $0.96$ & $1.00$ & $1.00$ & $1.00$ & $0.00$ & $0.00$ & $0.03$ & $0.03$ & $0.05$ \\
\texttt{wdbc} & $1.00$ & $1.00$ & $1.00$ & $1.00$ & $1.00$ & $1.00$ & $1.00$ & $1.00$ & $1.00$ & $1.00$ & $0.02$ & $0.04$ & $0.11$ & $0.19$ & $0.26$ \\
\texttt{wilt} & $1.00$ & $1.00$ & $1.00$ & $1.00$ & $1.00$ & $0.73$ & $0.98$ & $1.00$ & $1.00$ & $1.00$ & $0.00$ & $0.00$ & $0.00$ & $0.00$ & $0.00$ \\
\texttt{wine} & $0.95$ & $0.97$ & $1.00$ & $1.00$ & $1.00$ & $0.39$ & $0.71$ & $0.93$ & $0.98$ & $1.00$ & $0.00$ & $0.00$ & $0.00$ & $0.00$ & $0.00$ \\
\midrule mean $\pm$ SD & $0.99 \pm 0.04$ & $0.99 \pm 0.03$ & $0.99 \pm 0.03$ & $0.99 \pm 0.03$ & $0.99 \pm 0.03$ & $0.52 \pm 0.26$ & $0.80 \pm 0.19$ & $0.93 \pm 0.10$ & $0.97 \pm 0.05$ & $0.99 \pm 0.03$ & $0.02 \pm 0.02$ & $0.03 \pm 0.05$ & $0.05 \pm 0.07$ & $0.07 \pm 0.12$ & $0.09 \pm 0.14$ \\
datasets with ASR $\geq 0.5$ & 30 & 30 & 30 & 30 & 30 & 14 & 27 & 30 & 30 & 30 & 0 & 0 & 0 & 0 & 1 \\
datasets with ASR $< 0.1$ & 0 & 0 & 0 & 0 & 0 & 0 & 0 & 0 & 0 & 0 & 30 & 26 & 24 & 23 & 23 \\
\bottomrule\end{tabular}}
\end{table}
\end{landscape}

\section{Duplicate check}\label{app:dup}
\begin{table}[H]\centering\scriptsize\caption{Duplicate check without attack, over every test row of a fold in its clean context (queries): the share whose context holds a row with exactly its features (alarm) and the share whose duplicates carry both labels. Under attack and in the control arm the check raises an alarm for every target row.}\label{tab:app_dup}\centering\scriptsize\begin{tabular}{@{}lccc@{}}\toprule
Dataset & Queries & Alarm & Duplicates of both labels \\ \midrule
\texttt{banknote} & 1,372 & $0.026$ & $0.000$ \\
\texttt{california} & 3,000 & $0.000$ & $0.000$ \\
\texttt{climate-model} & 540 & $0.000$ & $0.000$ \\
\texttt{codrna} & 3,000 & $0.006$ & $0.000$ \\
\texttt{covertype} & 3,000 & $0.000$ & $0.000$ \\
\texttt{credit} & 3,000 & $0.000$ & $0.000$ \\
\texttt{creditcard} & 3,000 & $0.001$ & $0.000$ \\
\texttt{delta\_ailerons} & 3,000 & $0.000$ & $0.000$ \\
\texttt{eeg-eye-state} & 3,000 & $0.000$ & $0.000$ \\
\texttt{electrical\_grid} & 3,000 & $0.000$ & $0.000$ \\
\texttt{electricity} & 3,000 & $0.000$ & $0.000$ \\
\texttt{elevators} & 3,000 & $0.000$ & $0.000$ \\
\texttt{hazelnut-spread} & 2,400 & $0.000$ & $0.000$ \\
\texttt{hill-valley} & 1,212 & $0.000$ & $0.000$ \\
\texttt{house\_16H} & 3,000 & $0.000$ & $0.000$ \\
\texttt{jannis} & 3,000 & $0.000$ & $0.000$ \\
\texttt{kin8nm} & 3,000 & $0.000$ & $0.000$ \\
\texttt{MagicTelescope} & 3,000 & $0.001$ & $0.000$ \\
\texttt{mammography} & 3,000 & $0.293$ & $0.293$ \\
\texttt{MiniBooNE} & 3,000 & $0.000$ & $0.000$ \\
\texttt{ozone-level-8hr} & 2,534 & $0.004$ & $0.000$ \\
\texttt{phoneme} & 3,000 & $0.004$ & $0.000$ \\
\texttt{polish\_companies} & 3,000 & $0.007$ & $0.000$ \\
\texttt{Pulsar-Dataset} & 3,000 & $0.000$ & $0.000$ \\
\texttt{puma8NH} & 3,000 & $0.000$ & $0.000$ \\
\texttt{spambase} & 3,000 & $0.091$ & $0.000$ \\
\texttt{stock} & 950 & $0.000$ & $0.000$ \\
\texttt{wdbc} & 569 & $0.000$ & $0.000$ \\
\texttt{wilt} & 3,000 & $0.003$ & $0.000$ \\
\texttt{wine} & 2,554 & $0.187$ & $0.000$ \\
\midrule
Mean & & $0.021$ & $0.010$ \\
\bottomrule\end{tabular}
\end{table}

\section{Reproducibility}\label{app:repro}
\begin{table}[H]\centering\scriptsize\caption{Packages and checkpoints from the run records, for the random forest and the logistic regression from the environment, since their result file keeps no run record. Cells: predictions of the target rows, eleven per target row. Fits: training runs of the trained models, one per target row, arm and number of poison rows, and one per fold without attack.}\label{tab:app_repro}\centering\scriptsize\maxbreite{\begin{tabular}{@{}lllrrr@{}}\toprule
Model & package & checkpoint & torch & cells & fits \\ \midrule
TabICL & tabicl 2.1.1 & tabicl-classifier-v2-20260212.ckpt & 2.11.0.dev20260119+cu128 & 16,280 & -- \\
TabPFN & tabpfn 9.1.0 & tabpfn-v3.5-20260909.safetensors (TabPFNV3p5) & 2.10.0+cu130 & 16,280 & -- \\
CatBoost & catboost 1.2.8 & defaults, no checkpoint & -- & 16,280 & 14,950 \\
Random forest & scikit-learn 1.6.1 & defaults, no checkpoint & -- & 16,280 & 14,950 \\
Logistic regression & scikit-learn 1.6.1 & defaults, no checkpoint & -- & 16,280 & 14,950 \\
\bottomrule\end{tabular}}
\end{table}
\noindent The models ran in two environments, because TabICL 2.1.1 and TabPFN 3.5 did not run with a common PyTorch build on the GPU of the study, an NVIDIA GB10. The repository lists every environment package by package. TabICL and CatBoost ran under Python 3.10, TabPFN, the random forest and the logistic regression under Python 3.11, all with scikit-learn 1.6.1, and the tests with SciPy 1.15.3 and statsmodels 0.14.6. Seeds: \texttt{random\_state} of every model 0, folds and the reduction to 3,000 rows 42, selection of the target rows 20240711. The rows to replace are drawn per target row from a random generator seeded with the index of that target row (Section~\ref{sec:aufbau}). Code and data are listed in the Declarations\nurde{ (Erklärungen)} at the end of the main text.

\end{appendices}


\begin{thebibliography}{99}
\setlength{\itemsep}{2pt}\small

\bibitem[{Brown et~al.(2020)}]{brown2020}
Brown TB, Mann B, Ryder N, et~al (2020) Language models are few-shot learners. In: Advances in
Neural Information Processing Systems 33, pp~1877--1901

\bibitem[{Müller et~al.(2022)}]{muller2022}
Müller S, Hollmann N, Pineda~Arango S, Grabocka J, Hutter F (2022) Transformers can do Bayesian
inference. In: International Conference on Learning Representations (ICLR)

\bibitem[{Hollmann et~al.(2025)}]{hollmann2025}
Hollmann N, Müller S, Purucker L, et~al (2025) Accurate predictions on small data with a tabular
foundation model. Nature 637:319--326. \url{https://doi.org/10.1038/s41586-024-08328-6}

\bibitem[{Qu et~al.(2025)}]{qu2025}
Qu J, Holzmüller D, Varoquaux G, Le~Morvan M (2025) TabICL: a tabular foundation model for
in-context learning on large data. In: Proceedings of the 42nd International Conference on Machine
Learning, PMLR 267, pp~50817--50847

\bibitem[{van Breugel and van der Schaar(2024)}]{vanbreugel2024}
van Breugel B, van der Schaar M (2024) Position: why tabular foundation models should be a research
priority. In: Proceedings of the 41st International Conference on Machine Learning, PMLR 235

\bibitem[{Qu et~al.(2026)}]{qu2026}
Qu J, Holzmüller D, Varoquaux G, Le~Morvan M (2026) TabICLv2: a better, faster, scalable, and open
tabular foundation model. In: Proceedings of the 43rd International Conference on Machine Learning, PMLR 306, pp~102477--102555

\bibitem[{Grinsztajn et~al.(2026)}]{grinsztajn2026}
Grinsztajn L, Flöge K, Key O, et~al (2026) TabPFN-3: technical report.
arXiv:2605.13986

\bibitem[{Jäger et~al.(2026)}]{jaeger2026}
Jäger B, Erickson N, Grinsztajn L, et~al (2026) TabPFN-3.5: technical report.
arXiv:2609.17895

\bibitem[{Kurashkin et~al.(2026)}]{kurashkin2026}
Kurashkin SO, Tynchenko VS, Borodulin AS, Nelyub VA, Kalutsky NO, Connie T (2026) The current
generation of tabular foundation models: a critical review. Mach Learn Knowl Extr 8(8):244.
\url{https://doi.org/10.3390/make8080244}

\bibitem[{SAP(2026)}]{sap2026}
SAP (2026) TabPFN-3.5 Plus now available in SAP AI Core for instant business predictions. SAP News Center, 15~September 2026. \url{https://news.sap.com/2026/09/tabpfn-35-plus-now-available-sap-ai-core-instant-business-predictions/}.
Accessed 1~November 2026

\bibitem[{Kong and Das(2026)}]{kongdas2026}
Kong W, Das A (2026) Introducing TabFM: a zero-shot foundation model for tabular data. Google
Research Blog, 30~June 2026.
\url{https://research.google/blog/introducing-tabfm-a-zero-shot-foundation-model-for-tabular-data/}.
Accessed 1~November 2026

\bibitem[{Google Cloud(2026)}]{googlecloud2026}
Google Cloud (2026) The AI.PREDICT function. BigQuery documentation.
\url{https://docs.cloud.google.com/bigquery/docs/reference/standard-sql/bigqueryml-syntax-ai-predict}.
Accessed 13~October 2026

\bibitem[{Biggio et~al.(2012)}]{biggio2012}
Biggio B, Nelson B, Laskov P (2012) Poisoning attacks against support vector machines. In:
Proceedings of the 29th International Conference on Machine Learning, pp~1467--1474

\bibitem[{Xiao et~al.(2012)}]{xiao2012}
Xiao H, Xiao H, Eckert C (2012) Adversarial label flips attack on support vector machines. In:
Proceedings of the 20th European Conference on Artificial Intelligence (ECAI), pp~870--875. \url{https://doi.org/10.3233/978-1-61499-098-7-870}

\bibitem[{Shafahi et~al.(2018)}]{shafahi2018}
Shafahi A, Huang WR, Najibi M, Suciu O, Studer C, Dumitras T, Goldstein T (2018) Poison frogs!
Targeted clean-label poisoning attacks on neural networks. In: Advances in Neural Information
Processing Systems 31

\bibitem[{Goldblum et~al.(2023)}]{goldblum2023}
Goldblum M, Tsipras D, Xie C, Chen X, Schwarzschild A, Song D, Madry A, Li B, Goldstein T (2023)
Dataset security for machine learning: data poisoning, backdoor attacks, and defenses. IEEE Trans
Pattern Anal Mach Intell 45(2):1563--1580. \url{https://doi.org/10.1109/TPAMI.2022.3162397}

\bibitem[{Cinà et~al.(2023)}]{cina2023}
Cinà AE, Grosse K, Demontis A, Vascon S, Zellinger W, Moser BA, Oprea A, Biggio B, Pelillo M, Roli F
(2023) Wild patterns reloaded: a survey of machine learning security against training data
poisoning. ACM Comput Surv 55(13s):294. \url{https://doi.org/10.1145/3585385}

\bibitem[{Vassilev et~al.(2025)}]{vassilev2025}
Vassilev A, Oprea A, Fordyce A, Anderson H, Davies X, Hamin M (2025) Adversarial machine learning:
a taxonomy and terminology of attacks and mitigations. NIST AI 100-2e2025, National Institute of
Standards and Technology, Gaithersburg, MD. \url{https://doi.org/10.6028/NIST.AI.100-2e2025}

\bibitem[{Djilani et~al.(2026)}]{djilani2026}
Djilani M, Simonetto T, Tit K, Tambon F, Ghamizi S, Cordy M, Papadakis M (2026) On the robustness
of tabular foundation models: test-time attacks and in-context defenses. In: 2026 IEEE Conference on Secure and Trustworthy Machine Learning (SaTML). IEEE, pp~368--391. \url{https://doi.org/10.1109/SaTML68715.2026.00028}

\bibitem[{Biloš et~al.(2026)}]{bilos2026}
Biloš M, Wilson JT, Schneider A, Nevmyvaka Y (2026) A mechanistic study of tabular foundation
models. arXiv:2605.21288

\bibitem[{Anwar et~al.(2025)}]{anwar2025}
Anwar U, Von~Oswald J, Kirsch L, Krueger D, Frei S (2025) Understanding in-context learning of
linear models in transformers through an adversarial lens. Transactions on Machine Learning Research

\bibitem[{Wahdany et~al.(2026)}]{wahdany2026}
Wahdany D, Jagielski M, Cresswell JC, Dziedzic A, Boenisch F (2026) TabPATE: differentially private
tabular in-context learning without public data. arXiv:2606.31474

\bibitem[{He et~al.(2025)}]{he2025}
He P, Xu H, Xing Y, Liu H, Yamada M, Tang J (2025) Data poisoning for in-context learning. In: Findings of the Association for Computational Linguistics: NAACL 2025, pp~1680--1700. \url{https://doi.org/10.18653/v1/2025.findings-naacl.91}

\bibitem[{Zou et~al.(2025)}]{zou2025}
Zou W, Geng R, Wang B, Jia J (2025) PoisonedRAG: knowledge corruption attacks to retrieval-augmented
generation of large language models. In: Proceedings of the 34th USENIX Security Symposium, pp~3827--3844

\bibitem[{Greshake et~al.(2023)}]{greshake2023}
Greshake K, Abdelnabi S, Mishra S, Endres C, Holz T, Fritz M (2023) Not what you've signed up for:
compromising real-world LLM-integrated applications with indirect prompt injection. In: Proceedings
of the 16th ACM Workshop on Artificial Intelligence and Security, pp~79--90. \url{https://doi.org/10.1145/3605764.3623985}

\bibitem[{Prokhorenkova et~al.(2018)}]{prokhorenkova2018}
Prokhorenkova L, Gusev G, Vorobev A, Dorogush AV, Gulin A (2018) CatBoost: unbiased boosting with
categorical features. In: Advances in Neural Information Processing Systems 31

\bibitem[{Vanschoren et~al.(2013)}]{vanschoren2014}
Vanschoren J, van Rijn JN, Bischl B, Torgo L (2013) OpenML: networked science in machine learning.
ACM SIGKDD Explor Newsl 15(2):49--60. \url{https://doi.org/10.1145/2641190.2641198}

\bibitem[{Erickson et~al.(2025)}]{erickson2025}
Erickson N, Purucker L, Tschalzev A, Holzmüller D, Desai PM, Salinas D, Hutter F (2025) TabArena: a
living benchmark for machine learning on tabular data. In: Advances in Neural Information Processing
Systems 38, Datasets and Benchmarks Track

\bibitem[{Demšar(2006)}]{demsar2006}
Demšar J (2006) Statistical comparisons of classifiers over multiple data sets.
J Mach Learn Res 7:1--30

\bibitem[{Nguyen et~al.(2026)}]{nguyen2026}
Nguyen D, Chesneau N, Bhan M (2026) The standardization trap: certifying joint label processing in
tabular foundation models. arXiv:2610.08314

\bibitem[{Schneier(2024)}]{schneier2024}
Schneier B (2024) LLMs' data-control path insecurity. Commun ACM 67(9):31--32.
\url{https://doi.org/10.1145/3660674}

\bibitem[{European Union(2024)}]{euaiact2024}
European Union (2024) Regulation (EU) 2024/1689 of the European Parliament and of the Council of
13 June 2024 laying down harmonised rules on artificial intelligence (Artificial Intelligence Act).
Official Journal of the European Union, L series, 12~July 2024. \url{http://data.europa.eu/eli/reg/2024/1689/oj}

\bibitem[{Baylor et~al.(2017)}]{baylor2017}
Baylor D, Breck E, Cheng HT, Fiedel N, Foo CY, Haque Z, Haykal S, Ispir M, Jain V, Koc L, Koo CY,
Lew L, Mewald C, Modi AN, Polyzotis N, Ramesh S, Roy S, Whang SE, Wicke M, Wilkiewicz J, Zhang X,
Zinkevich M (2017) TFX: a TensorFlow-based production-scale machine learning platform. In:
Proceedings of the 23rd ACM SIGKDD International Conference on Knowledge Discovery and Data
Mining, pp~1387--1395. \url{https://doi.org/10.1145/3097983.3098021}

\bibitem[{Breck et~al.(2019)}]{breck2019}
Breck E, Zinkevich M, Polyzotis N, Whang S, Roy S (2019) Data validation for machine learning. In: Proceedings of Machine Learning and Systems (SysML)

\bibitem[{Kreuzberger et~al.(2023)}]{kreuzberger2023}
Kreuzberger D, Kühl N, Hirschl S (2023) Machine learning operations (MLOps): overview,
definition, and architecture. IEEE Access 11:31866--31879. \url{https://doi.org/10.1109/ACCESS.2023.3262138}

\bibitem[{Sculley et~al.(2015)}]{sculley2015}
Sculley D, Holt G, Golovin D, Davydov E, Phillips T, Ebner D, Chaudhary V, Young M, Crespo JF, Dennison D (2015) Hidden technical debt in machine learning systems. In: Advances in Neural Information Processing Systems 28

\bibitem[{Prior Labs(2026)}]{priorlabsmcp2026}
Prior Labs (2026) Model Context Protocol: TabPFN MCP. TabPFN documentation.
\url{https://docs.priorlabs.ai/agentic/mcp}. Accessed 7~November 2026

\bibitem[{Ma et~al.(2025)}]{ma2025}
Ma J, Thomas V, Hosseinzadeh R, Labach A, Cresswell JC, Golestan K, Yu G, Caterini AL, Volkovs M (2025) TabDPT: scaling tabular foundation models on real data. In: Advances in Neural Information Processing Systems 38

\bibitem[{Thomas et~al.(2024)}]{thomas2024}
Thomas V, Ma J, Hosseinzadeh R, Golestan K, Yu G, Volkovs M, Caterini A (2024) Retrieval \&
fine-tuning for in-context tabular models. In: Advances in Neural Information Processing Systems 37

\bibitem[{Saltzer and Schroeder(1975)}]{saltzer1975}
Saltzer JH, Schroeder MD (1975) The protection of information in computer systems. Proc IEEE
63(9):1278--1308. \url{https://doi.org/10.1109/PROC.1975.9939}

\bibitem[{NIST(2024)}]{nist2024csf}
National Institute of Standards and Technology (2024) The NIST Cybersecurity Framework (CSF) 2.0.
NIST CSWP 29. \url{https://doi.org/10.6028/NIST.CSWP.29}

\bibitem[{Buneman et~al.(2001)}]{buneman2001}
Buneman P, Khanna S, Tan WC (2001) Why and where: a characterization of data provenance. In:
Database Theory -- ICDT 2001. Lecture Notes in Computer Science 1973, pp~316--330. \url{https://doi.org/10.1007/3-540-44503-X_20}

\bibitem[{Vig et~al.(2020)}]{vig2020}
Vig J, Gehrmann S, Belinkov Y, Qian S, Nevo D, Singer Y, Shieber S (2020) Investigating gender bias
in language models using causal mediation analysis. In: Advances in Neural Information Processing
Systems 33

\bibitem[{Meng et~al.(2022)}]{meng2022}
Meng K, Bau D, Andonian A, Belinkov Y (2022) Locating and editing factual associations in GPT. In:
Advances in Neural Information Processing Systems 35

\end{thebibliography}
\end{document}